\documentclass[authoryear,preprint,10pt,3p]{elsarticle}

\usepackage{amssymb}
\usepackage{tikz}
\usepackage{mathtools}
\usepackage{natbib}
\usepackage{multirow}
\usepackage{color}
\usepackage{soul}
\usepackage{graphicx}
\usepackage{caption}
\usepackage{xcolor}
\usepackage{enumitem}
\usepackage{tablefootnote}

\usepackage[edges]{forest}

\journal{International Journal of Forecasting}

\begin{document}
\begin{frontmatter}

\title{The M6 forecasting competition:\\Bridging the gap between forecasting and investment decisions}

\author[label1]{Spyros Makridakis}
\address[label1]{Makridakis Open Forecasting Center, Institute For the Future, University of Nicosia, Cyprus}
\author[label2]{Evangelos Spiliotis\corref{cor2}}
\cortext[cor2]{Corresponding author}
\address[label2]{Forecasting and Strategy Unit, School of Electrical and Computer Engineering, National Technical University of Athens, Greece}
\ead{spiliotis@fsu.gr}
\author[label3]{Ross Hollyman}
\address[label3]{Business School, University of Exeter, Exeter, UK}
\author[label4,label1]{Fotios Petropoulos}
\address[label4]{School of Management, University of Bath, UK}
\author[label5]{Norman Swanson}
\address[label5]{Department of Economics, Rutgers University, NJ, USA}
\author[label6]{Anil Gaba}
\address[label6]{INSEAD, Singapore}

\begin{abstract}
The M6 forecasting competition, the sixth in the Makridakis’ competition sequence, is focused on financial forecasting. A key objective of the M6 competition was to contribute to the debate surrounding the Efficient Market Hypothesis (EMH) by examining how and why market participants make investment decisions. To address these objectives, the M6 competition investigated forecasting accuracy and investment performance on a universe of 100 publicly traded assets. The competition employed live evaluation on real data across multiple periods, a cross-sectional setting where participants predicted asset performance relative to that of other assets, and a direct evaluation of the utility of forecasts. In this way, we were able to measure the benefits of accurate forecasting and assess the importance of forecasting when making investment decisions. 
Our findings highlight the challenges that participants faced when attempting to accurately forecast the relative performance of assets, the great difficulty associated with trying to consistently outperform the market, the limited connection between submitted forecasts and investment decisions, the value added by information exchange and the “wisdom of crowds”, and the value of utilizing risk models when attempting to connect prediction and investing decisions.
%, and the dominance of relatively simple and standard methods employed in the investment challenge
\end{abstract}

\end{frontmatter}

\clearpage

\section{Introduction}

Investing involves allocating money or resources with the expectation of future profit. People invest for various reasons - saving for retirement, building wealth, funding education or achieving various financial goals, with different types of investments offering varying degrees of expected return and risk. The stock market is the asset class of choice for most investors, and is also a focal point of interest for academics and practitioners. The M6 forecasting competition requested participants to invest in 50 US stocks and 50 international Exchange Traded Funds (ETFs) with the winners sharing \$300,000 in prizes based on their forecasting and investment performance.

The objective of the M6 competition was to contribute to the debate around the Efficient Market Hypothesis (EMH), advanced by \cite{Fama1969-qj} and popularized a few years later by \cite{Malkiel1973-to}. The EMH simply states that asset prices fully reflect all available information, making it impossible to consistently outperform the market through stock picking or market timing. Empirical evidence on the validity of the EMH, comparing the performance of all types of active funds versus corresponding passively managed benchmarks, is provided yearly by \cite{Armour2023-aq} - the data suggest that the success of active funds diminishes considerably in comparison to corresponding market averages as the time horizon of evaluation increases.

\begin{figure}%[!ht]
    \centering
    \includegraphics[width=4.5in]{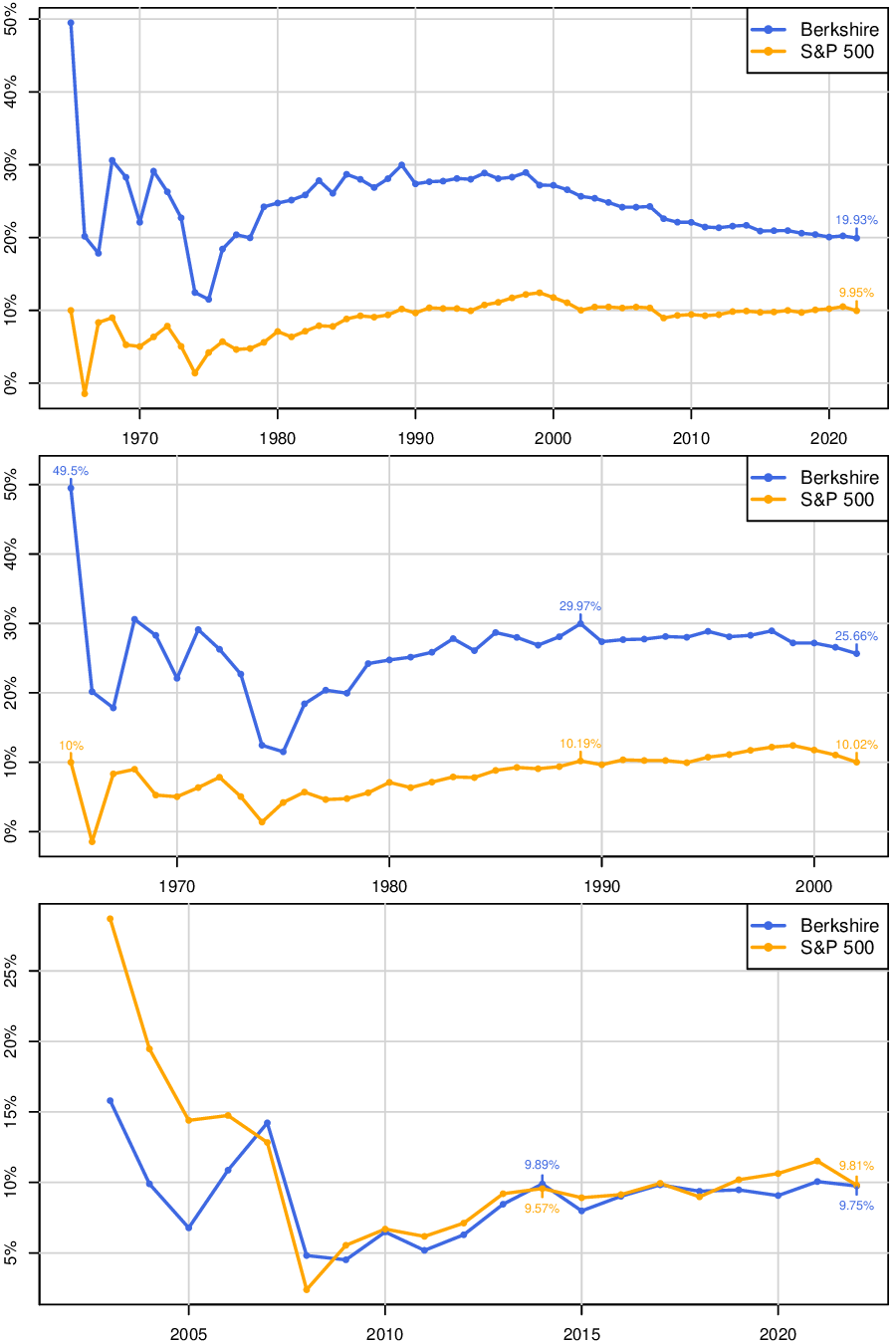}
    \caption{The performance of Berkshire versus that of S\&P500. Top panel: 1965-2022; Middle panel: 1965-2002; Bottom panel: 2003-2022.}
    \label{fig:intro}
\end{figure}

Yet there are a few investors and firms that seem to consistently outperform the market benchmarks, generating positive alphas, by stock picking or by exploiting a number of market inefficiencies \citep{Pedersen2015-on}. These investors, including the legendary Warren Buffett (who according to the 2022 letter to its Berkshires shareholders has generated compounded annual gains of about 10\% higher than the corresponding returns on the S\&P500 (see Figure \ref{fig:intro}, top panel)), Peter Lynch, George Soros, Carl Icahn and the firms Bridgewater Associates and Renaissance Technologies among others. Are these investors defying the EMH, and if they do for how long do they manage to do so and how do they achieve it, beyond being lucky? These are interesting questions requiring answers. In the case of Berkshire, its higher performance can be attributed to the application of Benjamin Graham’s value investing \citep{Graham1949-ra} by Buffett who was his student at Columbia University and applied a modified version to that of his mentor to stocks selection. According to the 2022 letter to Berkshire shareholders \citep{Buffett2023-dr} the firm produced a phenomenal overall compounded gain between 1964 and 2022 of 3,787,464\% versus 24,708\% for the corresponding S\&P500. Is Buffett a great stock picker beating the EMH over such a long period of 58 years?

The performance of Berkshire versus that of S\&P500 shown in the top panel of Figure \ref{fig:intro} produces, however, different results if separated into two sub periods, one between 1965 and 2002 (Figure \ref{fig:intro}, middle panel) and the other between 2003 and 2022 (Figure \ref{fig:intro}, bottom panel). In the middle panel of Figure \ref{fig:intro}, Berkshire’s gains are considerably higher than the S\&P500 ones, reaching a high of close to 20\% in 1989 and settling to 15.64\% higher at the end of 2002. In other words, Berkshire outperformed considerably the S\&P500 in the 1965 to 2002 period in compounded annual gains.

The bottom panel of Figure \ref{fig:intro} covers the 20 years period of 2003 to 2022 and displays a completely different picture to that of the middle panel. In this panel, we observe no significant differences in the annual compounded gains between Berkshire and S\&P500, bringing to an end the period of consistent out-performance observed during the previous 38 years. There can be many reasons for such fundamental change with a major one being increased competition, as Buffett's success and investment principles became widely known and more investors and funds began imitating his approaches. Other reasons could be the significant increase of the size of his investment portfolio, changes in the economic environment including low interest rates, quantitative easing, and technological advancements in stock selections. Additionally, Buffett's conservative investment philosophy focusing on long-term value investing may not have kept pace with the more aggressive, high-growth investment strategies being followed by others since the beginning of the 21st century. 

Figure \ref{fig:intro} provides three different ways of presenting the same data, raising questions about the fairest way of doing so. It is clear, however, that whilst Buffett consistently created significant positive alpha ($\alpha$) for 38 years (Figure \ref{fig:intro}, middle panel) that trend did not persist post 2002. The top panel of Figure \ref{fig:intro} shows consistent compounded gains of Berkshire over S\&P500 for the entire 1965-2022 period being greatly influenced by its considerable pre-2003 gains. The bottom panel of Figure \ref{fig:intro}, on the other hand, exhibits no differences in compounded gains between Berkshire and S\&P500 as the comparison starts in 2003 without taking into account its pre-2003 performance - there is indeed no guarantee that past out-performance  extends in to the future. This does not mean that Buffett has stopped being a great investor but that markets have become more efficient as information is disseminated faster and made instantaneously available to all players eliminating his advantages and proving the long-term validity of the EMH. What will happen in the future as Buffett’s stock selection has shifted - concentrating 78\% of its entire portfolio in just five stocks, with Apple being 50\% of the total \citep{Best2023-fz}? Such stock selection could improve Berkshire’s returns and its ability to generate positive alphas if their price increase more than the S\&P500 average. Yet, what will happen if something goes wrong with one of the five stocks, particularly Apple? Is there too much reliance on the ability to accurately forecast future relative returns and weight these against the associated increase in risk? For an excellent, comprehensive analysis of 'Buffett's Alpha', the interested reader is referred to \cite{Frazzini1}.

One of the primary objectives of the M6 competition was to investigate the value of accurate forecasting in investment decision making. In our data, we found no clear connection between the two. What was even more interesting was the fact that on average, the top performing teams in the forecasting task constructed relatively inefficient portfolios, while the top performing teams in the investment decisions submitted forecasts of lower accuracy levels. At the same time, the top teams performed relatively better, with 23\% beating the forecasting benchmark and an achieving an R-Squared of 0.099 between the forecasting and the investment decision ranks. We believe that even small improvements in forecasting accuracy can provide significant value for improving investment decisions and we plan to explore such improvements in the remaining of this paper by further investigating the value of more accurate forecasts in improving investment decisions.

\section{Design and execution}

The M6 competition was a financial forecasting competition where participating teams\footnote{Each team could consist of up to 5 participants: a team leader who would be responsible for making the submissions and 4 members. None of the teams members could be part of another team.} were asked to submit their forecasts regarding the relative performance of a number of tradable assets as well as their investment positions for these assets. The target was to achieve the best performance in two challenges, forecasting and investment, and also the overall best performance taking in to account both criteria. In this section, we will present in detail the design of the M6 competition. We will discuss the design innovations compared to previous forecasting competitions, present the selected assets and discuss the selection process and present the process of the competition, the submission points, and the submission requirements. We will then define how performance was measured for each part of the duathlon but also how we measured the overall performance; and, finally, we will present how prizes for the top performing submissions were awarded.

\subsection{Design innovations}
The M6 forecasting competition offers three major design innovations compared to the previous M competitions \citep{Makridakis2021-xa}. These design innovations are as follows:

\subsubsection*{Live evaluation on real data and over multiple periods}
Out of the previous five M forecasting competitions, four of them (M1 \citep{MakridakisM1}, M3 \citep{MAKRIDAKIS2000451}, M4 \citep{MAKRIDAKIS202054}, and M5 \citep{MAKRIDAKIS20221325}) focused on measuring the performance of the submitted forecasts on a single evaluation window; a technique also known as ``fixed origin evaluation'' \citep[see also][]{TASHMAN2000437}. The organizers of the respective competitions split the available data for each time series into two sets, in-sample and out-of-sample. The in-sample data was disclosed to the participants, without any other information that would allow them to recover the nature or identity of the target time series and, thus, the values of the out-of-sample observations. The participants were asked to submit forecasts for the out-of-sample period, and their forecasts were then compared to the withheld data. This design is relatively straightforward and allows the organizers to analyze the performance of the participating methods on data that are not by default available to the participants (such as Walmart’s sales data in the M5 forecasting competition).

The only exception to the above was the M2 competition \citep{Makridakis1993-bw}, where a three-phase approach was adopted. The organizers first offered to the participants the first batch of data; which allowed the participants to produce their first set of forecasts. The organizers then sent to the participants an updated version of the data based on which the participants submitted their updated forecasts. This process was repeated one more time, when the organizers offered not only an updated version of the data, but also an analysis of the participants’ forecasts. This iterative process allowed the participants to improve their forecasts via the provision of feedback on their previous forecasts. However, as the competition made use of proprietary data which were masked via the use of multipliers, it is not clear if the participants had a fair chance of using personal judgment to improve their forecasts.

The M6 forecasting competition builds on real, live data through a series of twelve rolling submission points that cover almost an entire calendar year. More importantly, the data used in the M6 competition are fully identified, open, and publicly available. In fact, the M6 forecasting competition is the first competition where data was not provided directly to participants; instead, a set of 100 publicly tradable assets (50 stocks and 50 ETFs) were chosen and their unique identifiers were shared with the participants. The organizing team did offer an interface to collect data associated with the adjusted closing values of the selected assets, but the participants had the option to use their own data sources (either from freely available sources such as Yahoo Finance, Google Finance, etc or perhaps alternatively from subscription based services) to collect the historical prices of these assets, and decide on the amount of history they wish to collect as well as its frequency (tick data, hourly prices, daily/closing prices or even weekly prices). As the names and identifiers of the assets were fully transparent, the participants also had the opportunity to collect additional supporting data (again from freely available or subsription based sources) that would allow them to augment the performance of their forecasts. Such supporting data could include contextual information from news and (social) media, fundamental, accounting, economic data etc.

For each of the twelve submission points, participating teams were asked to submit their forecasts (and investment positions) in the light of new information (most recent data). The competition featured a live evaluation system that was updated every day, and the participants’ performance of the current month and all previous months and quarters was public information. This system offered the participants appropriate feedback on their submissions, allowing them to monitor and compare their performance relatively to that of other participating teams, and make adjustments to their approaches over time.

\subsubsection*{The nature of the forecasting task}
Previous M forecasting competitions were primarily time series forecasting competitions. Participants were given a set of historical values for the target series (sometimes along with other exogenous variables) and the target was to forecast the future values in terms of point forecasts for these time series. In the last two competitions, M4 and M5, additionally to point forecasts, one or many pairs of prediction intervals (for predefined confidence levels) around these point forecasts also formed part of the submissions. To address the challenge, some participants in the M4 and M5 forecasting competitions employed approaches that rely on cross-learning. While this was straightforward and, to a degree, implicitly suggested from the design of M5 and the use of hierarchical structured data, this was not the case for M4 where the observations across the series were not even aligned in terms of time. However, this did not discourage the winning submissions of the M4 competition \citep{SMYL202075,MONTEROMANSO202086} in successfully applying global models.

The M6 forecasting competition moves away from a pure time series task and focuses on a cross-sectional forecasting exercise. The participating teams were not asked to submit forecasts of the future values of each of the assets nor the associated prediction intervals. Instead, we asked the participants to estimate the probability that each of the assets would be ranked within the first, second, third, fourth or fifth quintile with regards to the relative percentage returns across all 100 assets. In other words, forecasts in the M6 competition are not values associated with a single asset but are relative predictions that take into account the predicted performance of one asset relative to that of all other assets.

\subsubsection*{Evaluating forecast utility}
The previous five M forecasting competitions focused on the accuracy of the point forecasts, while the last two additionally focused on the evaluation of forecast uncertainty via (multiple in the case of M5 competition) prediction internals. Additionally, the latter competitions offered insights with regards to the trade-offs of forecasting performance and computational complexity. However, none of the previous forecasting competitions explicitly focused on how the forecasting performance is (implicitly or explicitly) linked to the forecast utility, i.e. how better forecasts could allow us to make better decisions and to what are the measurable (utility) benefits of such decisions. For example, in the context of the M5 forecasting competition and retail forecasting, forecast utility could be measuring the inventory performance of the various forecasts and how improvements in forecast accuracy/uncertainty translate to decreases in inventory performance (holding and backlog costs as well as achieved service levels).

In the M6 forecasting competition we explicitly evaluate utility (and provide  an incentive to maximize a specific definition thereof). The participants were asked to submit not only forecasts regarding the various assets but also their investment positions for each of these assets. In other words, we want to see how participants utilize (or not) the forecasts they produce to make informed investment decisions and how the performance of such decisions translates in to returns (direct monetary gains) adjusted for the risk taken. The M6 competition is the first competition in the M series that goes beyond the narrow definition of ``forecasting performance'' \citep[see also][]{YardleyP2021} and extends to how forecasts (explicit or implicit) are used in practice within the particular context of financial forecasting.

\subsection{Data}\label{sec:data}
The investment universe of the M6 competition consisted of two classes of assets:
\begin{itemize}[noitemsep]
\item 50 stocks from the Standard and Poor's (S\&P) 500 index, and 
\item 50 international exchange-traded funds (ETFs). 
\end{itemize}

The 50 stocks and 50 ETFs were selected such that they are broadly representative of the market. In particular to the selection of stocks, the sampling was done so that an appropriate proportion of stocks is selected per sector (communication services, energy, financial, healthcare, materials, etc). Then, we computed (in November 2021) the features for each of the stocks:
\begin{itemize}[noitemsep]
\item Average stock price (over the last 250 days)
\item Coefficient of variation for the stock price (over the last 250 days)
\item Coefficient of variation for the stock price (since the beginning of 2018)
\item Average daily returns (since the beginning of 2018)
\item Standard deviation of daily returns (since the beginning of 2018)
\item Total returns (over the last 250 days)
\item Total returns (since the beginning of 2018)
\item Average volume (over the last 250 days)
\item Coefficient of variation for the volume (over the last 250 days)
\end{itemize}

Given these features, we performed K-means clustering for each sector to get some insights on the diversity of the stocks within a sector. Then, we randomly sampled the desired number of stocks from each sector (so that the total stocks selected from each sector reflects to the size of the sector), making sure however that the population of each cluster is taken into consideration (so larger clusters will contribute more series, but still some series will be sampled from the rest). For more details about process used for constructing the M6 universe of assets, please refer to the supplementary material.

Note that the organizing team provided an easy way for the participants to access the asset prices data, updated daily, via a customized submission website dedicated to the M6 forecasting competition (https://m6competition.com/). However, this was not considered to be the sole available source of historical data. Given the open and public nature of the data and the assets, the participants were able to collect historical asset prices data from alternative sources/providers, while also selecting the length of the history of the collected data as well as their frequency. Additionally, participants were free to collect data from other data sources (such as news articles) or even historical prices for other assets that may allow them to produce better forecasts and make better investment decisions. Finally, the organizers have provided a forum (https://mofc.unic.ac.cy/forum/list/) to the participants where they could discuss on topics they considered important, ask questions and clarifications, and exchange ideas, code, or information\footnote{Exchanging information privately was not allowed as this would be equivalent to participating with multiple teams. As a result, the organizers disqualified teams for which there was enough evidence of private information exchange, as well as teams whose members were part of multiple teams.}. The forum was also used by the M6 organizers for making important announcements and posting potential clarifications.

\subsection{Process and timeline}
The M6 competition was a live forecasting competition, lasting for twelve months, starting in February 2022, and ending a year later in 2023. It consisted of a single month trial run and 12 rolling origins where participants were asked to provide their submissions and were evaluated once the actual data became available. For each submission point, participants were asked to provide their forecasts and investment decisions over the next four weeks (usually, the next 20 trading days). The submission deadline for each point was on 18:00 GMT the Sunday before the start of the corresponding investment period. All submissions were made through the customized submission website dedicated to the M6 forecasting competition.

The deadline for submissions for the trial practice run, which was not taken into account against the overall rankings and results of the M6 competition, was February 6, 2023. Four weeks after the deadline for submissions for the trial run, the first actual submission point took place, followed by another eleven submission points. The interval between two consecutive submission points was four weeks (i.e. equal to the forecasting horizon). In other words, the rolling origin evaluation process of the competition involved 12 non-overlapping four-week periods. The 12 submission points were divided into four quarters, as presented in Table \ref{tab:timeline}.

\begin{table}[h]
\small
\centering
\caption{The timeline of the M6 forecasting competition.}
\begin{tabular}{cccc}
\hline
\textbf{Quarter}	& \textbf{Month 1}& \textbf{Month 2}& \textbf{Month 3}\\
\hline
1 & March 6, 2022 & April 3, 2022 & May 1, 2022 \\
2 & May 29, 2022 & June 26, 2022 & July 24, 2022 \\
3 & August 21, 2022 & September 18, 2022 & October 16, 2022 \\
4 & November 13, 2022 & December 11, 2022 & January 8, 2023 \\
\hline
\end{tabular}
\label{tab:timeline}
\end{table}

\subsection{Submission requirements}

The objective was to submit (i) forecasts and (ii) investment decisions for all 100 assets that specify the participants’ forecasts and investment strategy over the next four-week period. At each submission point, a participating team was asked to submit a single file consisting of 100 rows (one row for each asset) and seven values per row. Below, we specify the seven values that the participants needed to provide for each of the 100 assets.

\begin{itemize}[noitemsep]
\item The first value of each row should indicate the asset for which the forecasts and the investment decisions of the respective row refer to. The acronym of each asset served as an identifier (e.g. ``GOOGL'' or ``ATVI'').
\item The second to sixth values should be values summing to unity that refer to the probabilities with which the asset will have percentage returns that are within the first, second, third, fourth, or fifth quintile across all assets (stocks and ETFs). In other words, we asked the participants to provide probabilistic forecasts for the performance (percentage returns) of each asset relative to the other assets. We automatically checked for invalid submissions (i.e. submissions where the sum of the five probabilities does not sum to one; or submissions with negative probabilities values) and returned such submissions to the participants.
\item The seventh value for each asset should be a numerical value corresponding to the weight for investing on that asset. Such weights should be positive for long positions, negative for short positions, or zero for no position. For instance, if three assets were assigned weights 0.5, 0.3, and -0.2, respectively, and all other assets a weights of 0, this would mean that the participant invested in only three assets with positions long, long, and short and with a budget allocation of 50\%, 30\% and 20\% respectively. If the sum of the absolute weights exceeded 1 (or 100\%), then the submission was automatically considered to be invalid and returned to the participants. If the sum of the absolute weights was less than 1 (less than 100\%), then the remainder was assumed to be assigned to an asset with zero return and zero risk (i.e. no investment). However, if the sum of the absolute weights was below 0.25 (25\%), then a warning message would be given and the submission would be considered to be invalid, being returned to the participants. In other words, the competition expected that the participants should make at least some investment and take some risk. 
\end{itemize}

Table \ref{tab:examplesubmission} presents the first 8 rows of a fictitious submission file. In this case, the participant decided to invest in three assets (``GOOG'', ``OGN'', ``DRE'') with weights 50\%, 30\% and 20\% (or 0.5, 0.3, and 0.2) and positions long, long and short respectively. Additionally, the participant predicted that there was a probability of 0.1, 0.2, 0.5, and 0.2 that the first asset (``ABBV'') would have percentage return within the 2\textsuperscript{nd}, 3\textsuperscript{rd}, 4\textsuperscript{th}, and 5\textsuperscript{th} quintile respectively. Equally, the participant was confident that the expected percentage returns for the second asset (``CNC'') will be within the 3\textsuperscript{rd} quintile across all assets (where a probability of 1 was assigned).

\begin{table}[h]
\small
\centering
\caption{An example of the submission format for the M6 forecasting competition.}
\begin{tabular}{lllllll}
\hline
\textbf{ID}	& \textbf{Rank 1}& \textbf{Rank 2}& \textbf{Rank 3}	& \textbf{Rank 4}& \textbf{Rank 5}& \textbf{Decision}\\
\hline
ABBV & 0 & 0.1 & 0.2 & 0.5 & 0.2 & 0 \\
CNC & 0 & 0 & 1 & 0 & 0 & 0 \\
GOOG & 0.1 & 0.1 & 0.1 & 0.1 & 0.6 & 0.5 \\
EWG & 0.5 & 0.4 & 0.05 & 0.05 & 0 & 0 \\
BMY & 0.2 & 0.2  & 0.2 & 0.2  & 0.2 & 0 \\
OGN & 0 & 0 & 0.1 & 0.4 & 0.5 & 0.3 \\
DRE & 0.7 & 0.3 & 0  & 0  & 0  & -0.2 \\
UNH & 0 & 0 & 1 & 0 & 0 & 0 \\
$\dots$ & $\dots$ & $\dots$ & $\dots$ & $\dots$ & $\dots$ & $\dots$ \\
\hline
\end{tabular}
\label{tab:examplesubmission}
\end{table}

Note that on occasions participants decided not to submit forecasts and investment decisions at particular submission points, we assumed that their previous (latest) submission carried over. In other words, their forecasts and investment decisions did not change. In this regard, if for instance a team made a single submission at the first month of the competition, this same submission would be used to evaluate its performance across all 12 submission points of M6.

Note also that although participating teams were allowed to change their submissions for each point up to 5 times per day till the submission deadline, the last submission made was the only to be considered for evaluation. 

\subsection{Measuring performance}\label{sec:perfromance}
\subsubsection*{Measuring the performance of the forecasts}
The forecasting performance for a particular submission point was measured by the Ranked Probability Score ($RPS$). The realized percentage total returns of all assets (stocks and ETFs) over the period were divided into quintiles, ranking from 1 (worst performing) to 5 (best performing). Given 100 assets, 20 of these would receive a rank of 5, 20 a rank of 4, and so forth. In cases involving a tie on the margins of the classes, the tied assets would all be assigned the respective average rank. For instance, if four assets were tied at the 18\textsuperscript{th} place, then they would all get a rank of $(5+5+5+4)/4=4.75$, with the three ``5’s'' in this expression being the rank of the 3 assets in the first quintile, and the ``4'' being the rank of the asset in the second quintile.

The actual return ranking of each asset, $i$, and each time, $T$, is described by a vector $q_{i,T}$ of order 5. 
\begin{itemize}[noitemsep]
\item In the case of no ties on the margins of the classes, the elements in this vector, $q_{i,T,k}$ with $k \in 1, \dots, 5$, are set equal to one if the asset is ranked in quintile $k$ and zero otherwise. For instance, if asset $i$ is ranked in the third quintile at time $T$, then $q_{i,T}$ has values 0, 0, 1, 0, and 0.
\item In the case of ties on the margins of the classes, then the values assigned to the elements of the vector $q_{i,T}$ are calculated such that the tied classes are assigned non-zero weights, with the respective weighted average being equal to the actual rank. For instance, following the above example of a 4.75 rank, the values of $q_{i,T}$ would be 0, 0, 0, 0.25, and 0.75, such that $0\times1+0\times2+0\times3+0.25\times4+0.75\times5 = 4.75$.
\end{itemize}

Similarly, we construct a vector denoting the probabilities of each rank for a particular asset, $f_{i,T}$, as submitted by a participating team.

The $RPS$ for asset $i$ in period $T$ is then calculated as
\begin{eqnarray}
RPS_{i,T} = \frac{1}{5}\sum_{j=1}^{5}{\left( \sum_{k=1}^{j}{q_{i,T,k}} - \sum_{k=1}^{j}{f_{i,T,k}} \right)^2}.
\end{eqnarray}

The $RPS$ for a given competitor for period $T$ is a simple average of the $RPS$ values across all assets:
\begin{eqnarray}
RPS_{i,T} = \frac{1}{100}\sum_{i=1}^{100}{RPS_{i,T}}.
\end{eqnarray}

The overall $RPS$ for multiple submission points $T_1$ to $T_2$ is
\begin{eqnarray}
RPS_{T_1:T_2} = \frac{1}{100(T_2-T_1+1)}\sum_{T=T_1}^{T_2}{\sum_{i=1}^{100}{RPS_{i,T}}}.
\end{eqnarray}

The $RPS$ is zero for a perfect score, and positive otherwise. The $RPS$ for a naive method for which the probabilities of each quintile being realized are equal to 0.2 for all assets (i.e. each asset is equally likely to have a performance in the 1\textsuperscript{st}, 2\textsuperscript{nd}, 3\textsuperscript{rd}, 4\textsuperscript{th}, or 5\textsuperscript{th} quintile) is equal to 0.16. From now on, this naive method will be referred as the forecasting benchmark.

\subsubsection*{Measuring the performance of the investment decisions}
The performance of the investment decisions is measured by means of a variant of the Information Ratio, $IR$, defined as the ratio of the portfolio return, $ret$, to the standard deviation of the portfolio return, $sdp$. Namely, risk adjusted returns are defined as
\begin{eqnarray}
IR = \frac{ret}{sdp},
\end{eqnarray}

\noindent where $ret$ denotes continuously compounded portfolio returns, and $sdp$ denotes the standard deviation of these returns, measured at a daily frequency. 

Note that all reported $IR$ values are annualized. Additionally, $IR$ is a variant of the typical Information Ratio, but with the benchmark return set equal to 0; and is also a variant of the Sharpe Ratio, but with the risk free rate set to 0. All return calculations begin with the daily portfolio holding period return, calculated as
\begin{eqnarray}
RET_t = \sum_{i=1}^{N}{w_i\left( \frac{S_{i,t}}{S_{i,t-1}} - 1 \right)},
\end{eqnarray}

\noindent where $N$ denotes the number of assets, $w_i$ is a portfolio weight, and $S_{i,t}$ denotes the price of asset $i$ at the end of trading day $t$, with $t-1$ referring to the previous trading day. In all return calculations, prices are adjusted closing prices. Continuously compounded portfolio returns are then calculated as $ret_t = \text{ln}(1+RET_t)$.

In the above expressions, $RET_t$ is measured for a single day, $t$, and is the percentage return (gain/loss) associated with each asset selected for investment, averaged by the corresponding investment decision weight for each asset. Returns for a holding period longer than one day are calculated as the sum of daily returns. In particular, the return for the holding period from $t_1$ to $t_2$ is calculated as
\begin{eqnarray}
ret_{t_1:t_2} =\sum_{t=t_1}^{t_2}{ret_t}.
\end{eqnarray}

The standard deviation, $sdp_{t_1:t_2}$, is calculated using the same $t_2-t_1+1$ values of $ret_t$ as those used in the calculation of $ret_{t_1:t_2}$. In particular, $varp_{t_1:t_2} = \frac{1}{T-1}\sum_{t=t_1}^{t_2}{\left( ret_t - T^{-1}ret_{t_1:t_2} \right)^2}$ and $sdp_{t_1:t_2} = \sqrt{varp_{t_1:t_2}}$ with $T=t_2-t_1+1$.

Higher IR values suggest better investment performance generated per unit risk taken. In order to benchmark the performance of the participating teams, an approach where equal long positions are taken for all 100 assets (investment weights of 0.1) was assumed.

\subsubsection*{Measuring the combined performance of the forecasts and the investment decisions}
The combined performance is measured by means of the arithmetic mean of the ranks of the ranked probability score, $RPS$, and performance of the investment decision, $IR$, which assumes equal importance between the two tasks/challenges. As such, the overall rank for a submission, $OR$, is calculated as
\begin{eqnarray}
OR = \frac{\text{rank}(RPS) + \text{rank}(IR)}{2},
\end{eqnarray}

\noindent where $\text{rank}(\cdot)$ returns the rank of a participant relative to all other participants for that measure ($RPS$ or $IR$). To calculate the overall forecasting rank, across all 12 submission points, we take the arithmetic mean of the $RPS$ as calculated in each month.

Note that the M6 GitHub repository (https://github.com/Mcompetitions/M6-methods) provided sample code (R and Python) for evaluating the submissions in terms of RPS and IR with the objective to clarify the details of the evaluation process and facilitate the replicability of the competition's results. A MS excel file with similar computations was also available for teams with limited programming background.

\subsection{Prizes}
In the M6 forecasting competition, we offered a total of \$300,000 in prizes that were awarded based on the participants’ performance on each of the competition’s challenges (forecasting and investment) but also their overall performance.

We offered \$42,000 in performance prizes for each of the four quarters of the competition. The (nine) quarterly prizes were awarded to the participants with the first, second, and third
\begin{itemize}[noitemsep]
\item best performance in terms of forecasting (evaluated by the ranked probability score);
\item best performance in terms of investment decisions (evaluated by the information ratio);
\item best overall performance of the above two challenges, winning the quarterly duathlon prize.
\end{itemize}

In addition, we offered \$124,000 in global prizes where we considered the performance across all four quarters. The (fifteen) global prizes were awarded to the participants with the first to fifth
\begin{itemize}[noitemsep]
\item best performance in terms of forecasting.
\item best performance in terms of investment decisions.
\item best overall performance of the above two challenges, winning the global duathlon prize.
\end{itemize}

Finally, we offered \$8,000 (\$2,000 per quarter) to the best performing submissions by teams with students\footnote{In order for a team to be eligible for the student prize, it had to consist solely of students, with the exception of one participant who could serve as supervisor} as members.

Note that if a participant did not submit forecasts and investment decisions in the first month of a particular quarter, and if there were no submissions to be carried over from the previous quarter, then they were automatically not eligible for the prize for that particular quarter. Equally, if a participant was not eligible for a prize for a single quarter, then they were automatically not eligible for the global prizes (awards based on the performance across all 12 submission points). In other words, in order for a participating team to be eligible for a global prize, they had to submit forecasts and investment decisions from the very first month of the competition (after the trial run).

In order to be able to offer this significant prize pool, we relied on sponsorships by multiple organizations: Google (Platinum Sponsor), Meta (Gold Sponsor), JP Morgan (Diamond Sponsor), SAS, International Institute of Forecasters, Kinaxis, Intech, ForecastPro, causaLens, Rutgers University, Erasmus Business School, University of Nicosia, and Makridakis Open Forecasting Center.

\section{Participating teams and overview of submissions}
\label{sec:teams}

The M6 competition involved 318 participants on 226 teams. About 80\% of the teams consisted of a single participant, 10\% involved two members, while the remaining 10\% more than two members. Almost half of the participants originated from the United States of America, India, France, China, Turkey, Germany, and Greece, while the remaining from 43 different countries. Although the affiliation and the background of the participants was not always clear, we concluded based on the answers collected through the questionnaire that around 60\% of the participants were independent researchers, consultants or data scientists, 25\% worked in the industry, and 15\% were academics. Moreover, 13 of the participants were students.

From the 226 teams that have entered the M6, 163 teams (about 72\%) did that on the beginning of the competition, thus being eligible for the ``Global'' prizes. The remaining 63 teams joined the competition mostly within the first three months of the competition and the beginning of the second, third, and fourth quarters, probably in order to be eligible for the corresponding quarterly prizes. Moreover, 59 teams participated in the trial run to familiarize themselves with the submission system and the evaluation setup of the competition. Table \ref{tab:stats_sub} summarizes the number of active teams and new submissions made per submission period.

\begin{table}[h]
\small
\centering
\caption{Number of active teams and new submissions made per submission period. For each period, the proportion of previous submissions used for evaluation purposes is also reported.}
\begin{tabular}{lrrrrrrrrrrrrrr}
\hline
\multirow{2}{*}{\textbf{Period}} & \textbf{Active} & \textbf{New} & \multicolumn{12}{c}{\textbf{Submissions used for evaluation (\%)}}\\
& \textbf{Teams} & \textbf{Submissions}  & \textbf{1\textsuperscript{st}} & \textbf{2\textsuperscript{nd}} & \textbf{3\textsuperscript{rd}} & \textbf{4\textsuperscript{th}} & \textbf{5\textsuperscript{th}} & \textbf{6\textsuperscript{th}} & \textbf{7\textsuperscript{th}} & \textbf{8\textsuperscript{th}} & \textbf{9\textsuperscript{th}}& \textbf{10\textsuperscript{th}} & \textbf{11\textsuperscript{th}} & \textbf{12\textsuperscript{th}}\\
\hline
1\textsuperscript{st} & 163 & 163 & 100.0	\\
2\textsuperscript{nd} & 176 & 112 & 36.4 & 63.6	\\
3\textsuperscript{rd} & 185 & 104 & 28.1 & 15.7 & 56.2	\\
4\textsuperscript{th} & 197 & 114 & 20.8 & 10.7 & 10.7 & 57.9	\\
5\textsuperscript{th} & 197 & 86 & 20.3 & 10.2 & 8.1 & 17.8 & 43.7	\\
6\textsuperscript{th} & 200 & 85 & 20.0 & 10.0 & 7.5 & 14.0 & 6.0 & 42.5	\\
7\textsuperscript{th} & 208 & 88 & 19.2 & 9.6 & 7.2 & 12.5 & 3.8 & 5.3 & 42.3	\\
8\textsuperscript{th} & 208 & 66 & 18.8 & 9.6 & 7.2 & 12.0 & 3.4 & 3.8 & 13.5 & 31.7	\\
9\textsuperscript{th} & 214 & 72 & 18.2 & 9.3 & 7.0 & 11.7 & 2.8 & 3.7 & 9.3 & 4.2 & 33.6	\\
10\textsuperscript{th} & 223 & 82 & 17.5 & 9.0 & 6.7 & 7.6 & 2.7 & 3.1 & 8.1 & 3.1 & 5.4 & 36.8\\
11\textsuperscript{th} & 226 & 71 & 17.3 & 8.8 & 6.6 & 7.5 & 2.7 & 3.1 & 7.5 & 3.1 & 4.0 & 8.0 & 31.4	\\
12\textsuperscript{th} & 226 & 63 & 17.3 & 8.8 & 6.6 & 7.1 & 2.7 & 3.1 & 6.2 & 2.7 & 3.5 & 6.6 & 7.5 & 27.9\\
\hline
\end{tabular}
\label{tab:stats_sub}
\end{table}

\begin{figure}[!h]
    \centering
    \includegraphics[width=0.55\textwidth]{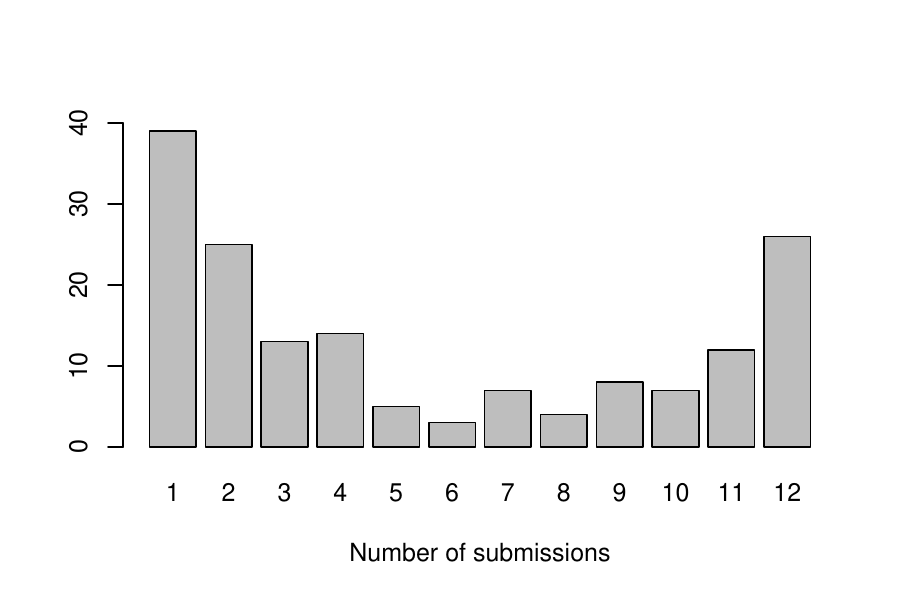}
    \caption{Number of teams per submission count.}
    \label{fig:submission_frequency}
\end{figure}

As explained in Subsection \ref{sec:perfromance}, the participating teams were not obliged to update their submissions at every single submission point. Instead, a team could retain a previous submission for one or multiple evaluation rounds. In this regard, teams that opted to stick with their previous forecasts and investment decisions did not have to re-submit the same submission file, while teams that did not have the time to work on their submission for a certain submission point were not disqualified. In this context, Table \ref{tab:stats_sub} summarizes the proportion of previous submissions used for evaluation purposes at each round. It is evident that as the competition proceeded, less and less teams were capable of regularly updating their submissions. Indicatively, in the second submission point, 36.4\% of the teams were evaluated based on their initial submission, while in the third submission point 28.1\% and 15.7\% of the teams were evaluated based on the submissions they made at the first and second submission points, respectively. Ultimately, at the last submission point of the competition, only 27.9\% of the teams were evaluated based on a completely updated submission, with the remaining 17.6\%, 12.0\%, and 42.5\% utilizing submissions created more than one, three, and six months, respectively. In addition, from the 163 teams included in the ``Global'' leaderboard, only 26 made an original submission at all 12 submission points, 64 made more than 6 submissions, and 39 made a single submission at the very beginning of the competition, as displayed in Figure \ref{fig:submission_frequency}.

The observations above highlight the difficulties present when organizing live competitions that involve multiple evaluation rounds and cover a long period of time. Despite the incentives we tried to provide through the quarterly prizes, few participants were actually attracted after the launch of the competition, probably due to their exclusion from the major M6 prizes. Moreover, they demonstrate how challenging it is in practice for a team to remain dedicated to such a demanding duathlon. Fortunately, said dedication seemed to pay off in many cases as all five winners in the forecasting track and four of the five winners in the investment track updated their submission at every single round, while the same was true for three of the duathlon winners\footnote{The two other teams had made 2 and 3 submissions throughout the competition}. Nevertheless, the overall correlation between the number of submissions made and the performance of the team was minor, suggesting that active participation is a sufficient but not necessary condition for winning a forecasting competition like M6.

\begin{figure}[!h]
    \centering
    \includegraphics[width=0.9\textwidth]{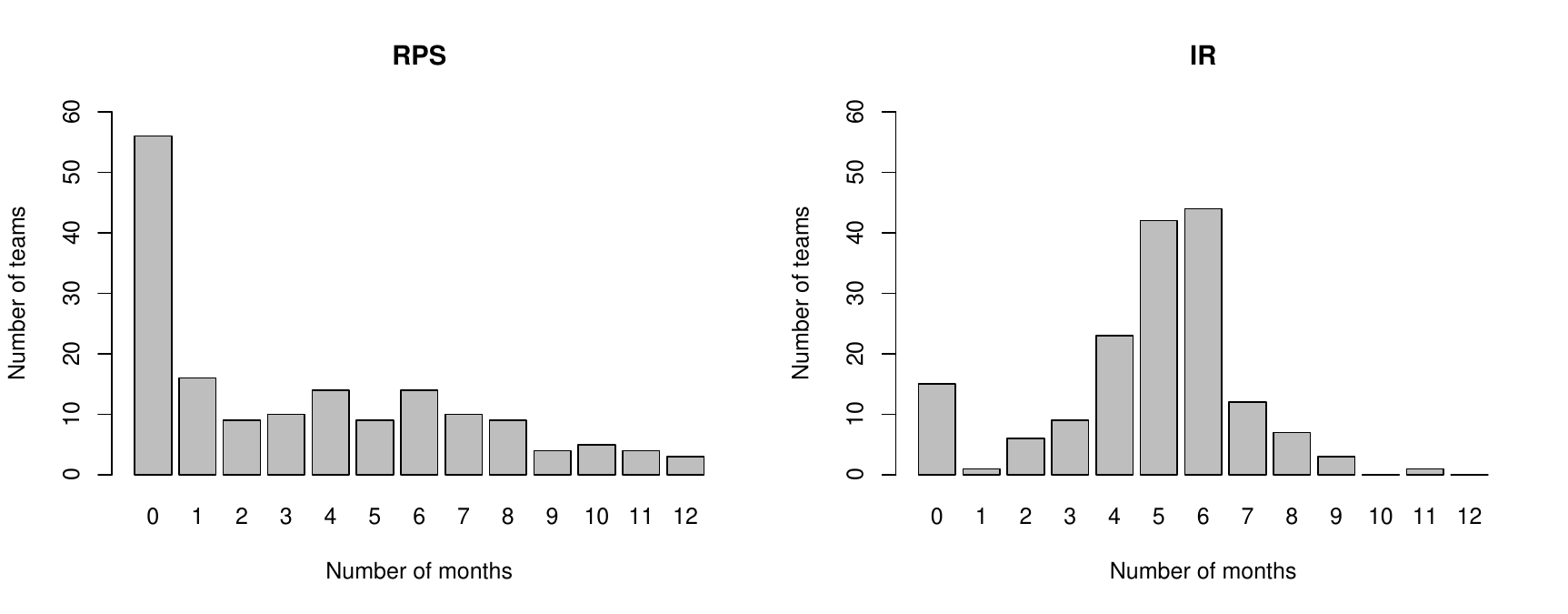}
    \caption{Number of teams included in the ``Global'' leaderboard that managed to outperform the benchmark in terms of RPS or IR in $N$ months out of the 12 months the competition covered.}
    \label{fig:outperform_month}
\end{figure}

In terms of performance, from the 163 teams included in the global leaderboard, 38 (23.3\%) managed to provide more accurate forecasts than the benchmark, 47 (28.8\%) to construct better portfolios, and 11 (6.7\%) to achieve both higher IR and RPS scores. It is also interesting that, as shown in Figure \ref{fig:outperform_month}, only 3 teams outperformed the benchmark's forecasts in all 12 months of the competition, while none the benchmark's investment decisions (a single team achieved higher IR than the benchmark in 11 months and 3 teams in 9 months). 

\begin{figure}[!h]
    \centering
    \includegraphics[width=0.9\textwidth]{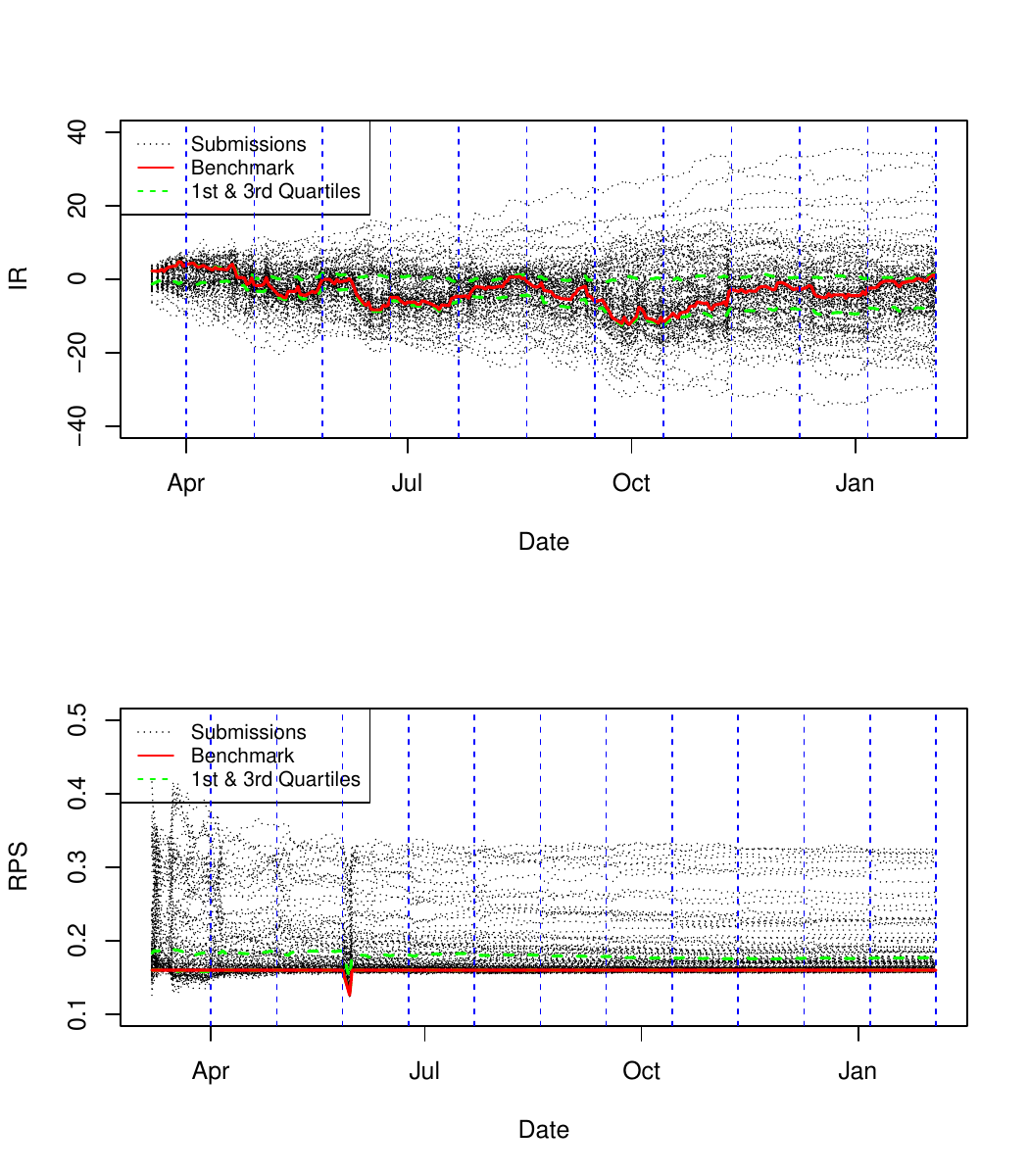}
    \caption{Daily evolution of the RPS and IR scores of the 163 teams included in the ``Global'' leaderboard. The performance of the benchmark as well as the 1\textsuperscript{st} and 3\textsuperscript{rd} quartiles of the submission scores are also reported. Blue vertical lines indicate the end points of the 12 evaluation rounds.}
    \label{fig:score_evolution}
\end{figure}

Figure \ref{fig:score_evolution} visualizes the daily evolution of the RPS and IR scores of the 163 teams included in the ``Global'' leaderboard. When it come to RPS, we observe that teams perform either similarly well or significantly worse than the benchmark throughout the competition. On the contrary, there is a notable number of teams that perform either significantly better or significantly worse than the benchmark, with the majority however of the teams reporting lower IR scores in most of the periods. Drawing from the above statistics and Figure \ref{fig:score_evolution}, it becomes evident that beating the benchmark consistently or consistently was particularly difficult in practice, despite the simplistic forecasting and investing approaches the benchmark employed.  

\section{The ten hypotheses and their evaluation}

\noindent\textbf{Hypothesis No.1: }\textit{The efficient market hypothesis will hold for the great majority of teams but this will not be the case for the top performing ones.}

Clearly we can not (and do not) claim to accept or reject the efficient market hypothesis (EMH) based on our data. Our teams competed for a period of one year only, limiting the amount of data we collected, and the portfolios submitted were not particularly representative of those typically found in the institutional settings. Moreover, the investment decisions would have to be tested for a much longer period of time so that any biases were removed from the results and ``skill'' was effectively separated from ``luck''. Nevertheless, given the practical difficulties present in running a competition for several consecutive years and maintaining a sufficiently large number of active teams, the M6 competition offers a realistic testing ground for the EMH. 

To evaluate the EMH hypothesis, we focus on the 148 teams included in the ``Global'' leaderboard whose investment submissions were not identical to the benchmark. We did so in order to exclude from the analysis the teams that did not submit original investment decisions and draw conclusions based on those whose performance can be assessed in longer-term fashion, i.e. across all 12 submission points. 

Table \ref{tab:sum_hyp1} summarizes the performance of the benchmark and the teams in terms of returns, risk, and IR across the 12 submission points individually and in total. As seen, although the vast majority of the teams (75\%) have managed to construct less risky portfolios than the benchmark (this is not particularly challenging, as participants were able to hedge away market risk via short positions), only 31\% have realized higher returns and IR. Moreover, we observe that the percentage of teams that outperformed the benchmark was usually higher when the benchmark return was positive, meaning that many teams adopted a directional bias. Therefore, it is unsurprising that, overall, the benchmark did better than the "average" team. But, we also find that some teams managed to beat the market to a significant extent. Focusing on ``Global'' scores, where the benchmark realized an IR of 0.453, the teams reported a score of $-3.421\pm9.832$. In other words, assuming a normal distribution, about 16\%\footnote{In fact, 16 teams (11\%) reported an IR higher than 6.411.} of the teams (one standard deviation higher than the mean) have managed to score an IR higher than 6.411, which should be regarded as a notable improvement (the returns increase from 0.5\% to more than 5.6\%, respectively). 

\begin{table*}[h]
\centering
\caption{Statistics summarizing the performance of the benchmark and the teams (mean and standard deviation) in terms of returns, risk, and IR across the 12 submission points and in total. The percentage of teams that outperformed the benchmark for each measure is also reported. The results focus on the 148 teams included in the ``Global'' leaderboard whose investment submissions were not identical to the benchmark.}
\resizebox{\textwidth}{!}{ \begin{tabular}{lcccrrrrrr}
\hline
\multirow{2}{*}{\textbf{Period}} & \multicolumn{3}{c}{\textbf{Better than the Benchmark (\%)}} & \multicolumn{3}{c}{\textbf{Benchmark}} & \multicolumn{3}{c}{\textbf{Teams - Mean(St. Deviation)}} \\
& \textbf{Returns} & \textbf{Risk} & \textbf{IR} & \textbf{Returns} & \textbf{Risk} & \textbf{IR} & \textbf{Returns} & \textbf{Risk} & \textbf{IR}\\
\hline
1\textsuperscript{st} Submission & 59.46 & 72.97 & 59.46 & 0.044 & 0.011 & 3.990 & 0.015(0.032) & 0.010(0.006) & 1.285(3.467)\\
2\textsuperscript{nd} Submission & 18.92 & 77.70 & 24.32 & -0.063 & 0.010 & -5.972 & -0.028(0.043) & 0.008(0.005) & -2.957(4.433)\\
3\textsuperscript{rd} Submission & 55.41 & 56.76 & 58.11 & 0.018 & 0.015 & 1.215 & 0.006(0.036) & 0.011(0.007) & 0.649(3.319)\\
4\textsuperscript{th} Submission & 20.95 & 56.08 & 22.30 & -0.063 & 0.015 & -4.139 & -0.029(0.049) & 0.011(0.007) & -2.186(3.609)\\
5\textsuperscript{th} Submission & 27.03 & 89.86 & 35.14 & 0.005 & 0.009 & 0.577 & -0.003(0.016) & 0.007(0.005) & -0.361(2.342)\\
6\textsuperscript{th} Submission & 64.19 &  91.22 & 66.89 & 0.051 & 0.008 & 6.060 & 0.019(0.036) & 0.007(0.005) & 2.658(4.526)\\
7\textsuperscript{th} Submission & 25.00 & 73.65 & 29.73 & -0.064 & 0.012 & -5.273 & -0.022(0.037) & 0.008(0.005) & -1.891(4.858)\\
8\textsuperscript{th} Submission & 30.41 & 58.11 & 33.78 & -0.073 & 0.015 & -4.834 & -0.019(0.048) & 0.010(0.006) & -1.020(4.679)\\
9\textsuperscript{th} Submission & 63.51 & 61.49 & 66.22 & 0.110 & 0.014 & 7.839 & 0.028(0.067) & 0.010(0.006) & 2.223(5.968)\\
10\textsuperscript{th} Submission & 27.03 & 95.95 & 38.51 & 0.000 & 0.008 & -0.017 & -0.004(0.028) & 0.006(0.003) & -0.529(4.007)\\
11\textsuperscript{th} Submission & 35.14 & 84.46 & 47.30 & 0.006 & 0.011 & 0.570 & 0.001(0.020) & 0.008(0.005) & -0.015(2.245)\\
12\textsuperscript{th} Submission & 50.00 & 91.22 & 54.73 & 0.034 &0.007 & 5.122 & 0.005(0.049) & 0.006(0.005) & 0.021(5.754)\\
\hline
Global & 31.08 & 75.00 & 31.76 & 0.005  & 0.012 & 0.453 & -0.031(0.087) & 0.009(0.004) & -3.421(9.832)\\
\hline
\end{tabular}}
\label{tab:sum_hyp1}
\end{table*}

\begin{figure}[!h]
    \centering
    \includegraphics[width=0.9\textwidth]{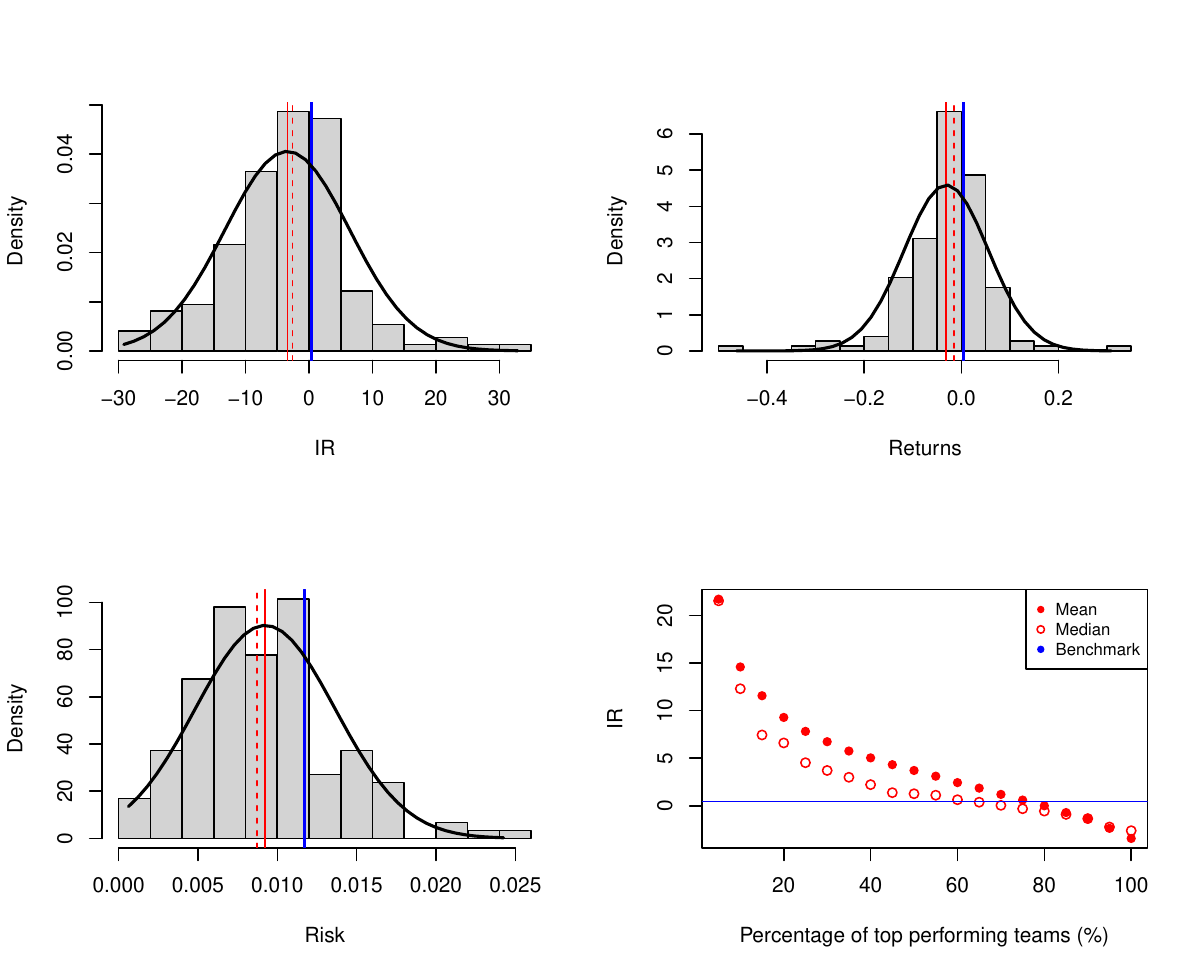}
    \caption{Distribution of IR, returns, and risk of the 148 teams included in the ``Global'' leaderboard whose investment submissions were not identical to the benchmark. A normal distribution is fitted over the histograms to facilitate comparisons. On the bottom right plot, the mean and the median IR of the top performing teams according to the IR is also presented over the benchmark for various percentages.}
    \label{fig:hyp1}
\end{figure}

The finding that the EMH holds true for the great majority of teams but not for the top performing ones is shown clearly in Figure \ref{fig:hyp1}. In alignment with Table \ref{tab:sum_hyp1} we observe that although the mean and median performance of the teams is worse than the benchmark, a small group of teams achieved strongly positive IR values, which correspond to an impressive rate of return of about 30\%. Furthermore, it is evident that the improvements in terms of IR grow exponentially as we move from the worse to the top performing teams. At the same time, the performance of the teams is rather symmetric around the mean in the sense that more than one fourth of the teams realized losses that exceeded 7\%, reaching up to 46\%.\\

\noindent\textbf{Hypothesis No.2: }\textit{There will be a small group of participants that clearly outperform the average both in terms of forecast accuracy and portfolio returns.}

\begin{figure}[!h]
    \centering
    \includegraphics[width=0.9\textwidth]{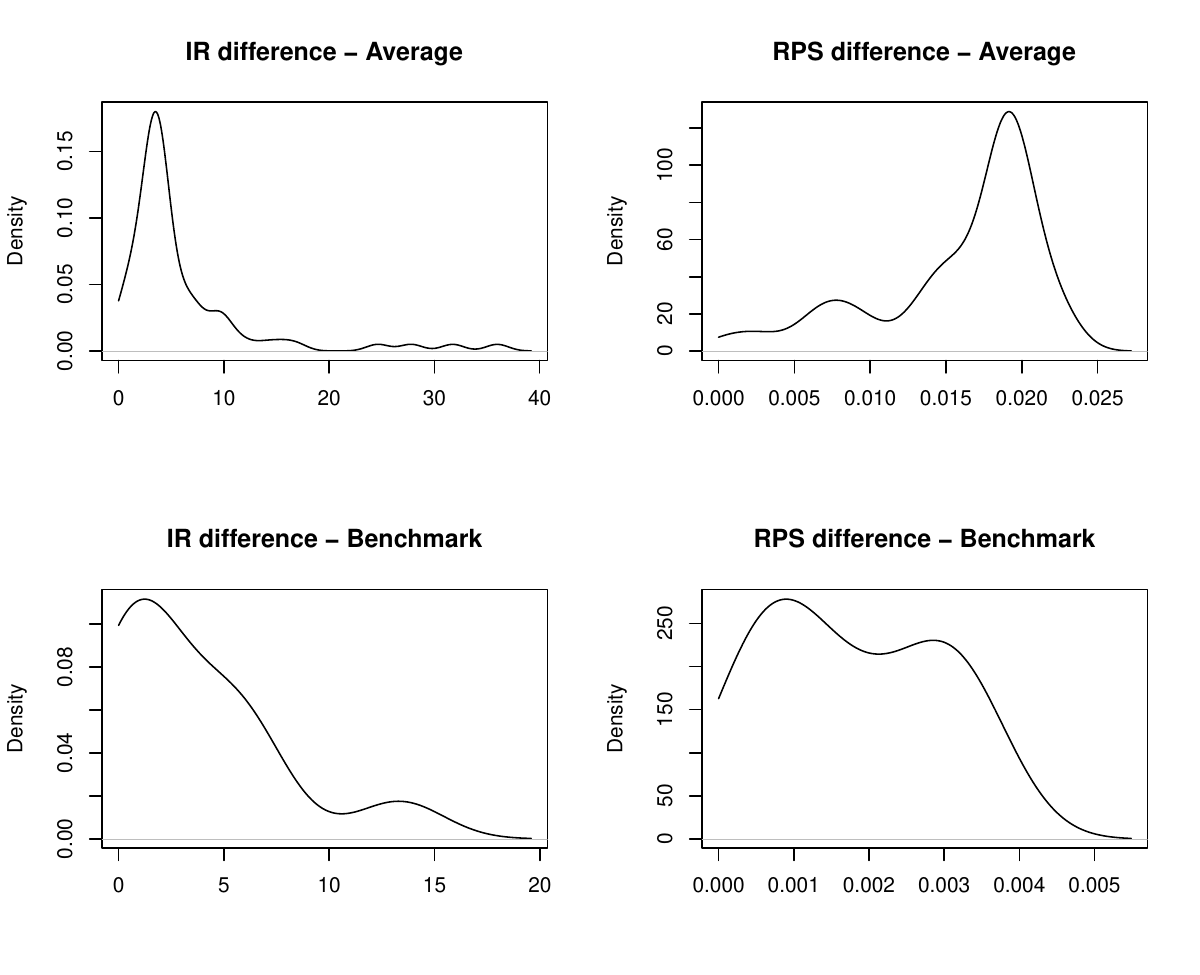}
    \caption{Performance difference in terms of IR and RPS of the teams that outperformed the average submission (top) and the benchmark (bottom).}
    \label{fig:hyp6}
\end{figure}

In order to assess this hypothesis, we focus on the 162 teams included in the ``Global'' leaderboard, whose average IR and RPS scores were -3.087 and 0.179, respectively, while the medians of said scores were -1.473 and 0.162. From these teams, 75 (46.3\%) managed to outperform the average submission, both in terms of forecast accuracy and portfolio returns, and 41 (25.3\%) to outperform the median submission. The performance differences of the former teams are shown in Figure \ref{fig:hyp6}. 

We observe that in accordance to the EMH (see Figure \ref{fig:hyp1}), only a small group of participants clearly outperform the average and the median submissions. It is indicative that 19 of the 75 teams (and 5 of the 41 for the median) report negative returns, while the average forecast accuracy improvement is less than 10\% (and less than 4\% for the median). At the same time, the maximum accuracy improvement is 12.7\% for the average submission and 3.7\% for the median submission. The number of out-performing teams is even smaller when the benchmark is used as a point of reference. Specifically, just 11 teams report better IR and RPS scores than the benchmark and although notable improvements can be identified in the investment challenge, forecast accuracy improvements do not surpass 2.2\%. In this context, the hypothesis is accepted.\\ 

\noindent\textbf{Hypothesis No.3: }\textit{There will be a weak link between the ability of teams to accurately forecast individual rankings of assets and risk adjusted returns on investment. The magnitude of this link will increase in tandem with team rankings, on average. Additionally, team portfolios will in general be more concentrated and risky than can be theoretically justified given the accuracy of their forecasts.}

In order to evaluate this hypothesis, we first compute the correlation coefficient, $r$, between IR and RPS. We focus on the 138 teams included in the ``Global'' leaderboard whose forecast submissions were not identical to the benchmark so that any unoriginal forecasts are excluded and the results become more representative.

\begin{figure}[!h]
    \centering
    \includegraphics[width=0.9\textwidth]{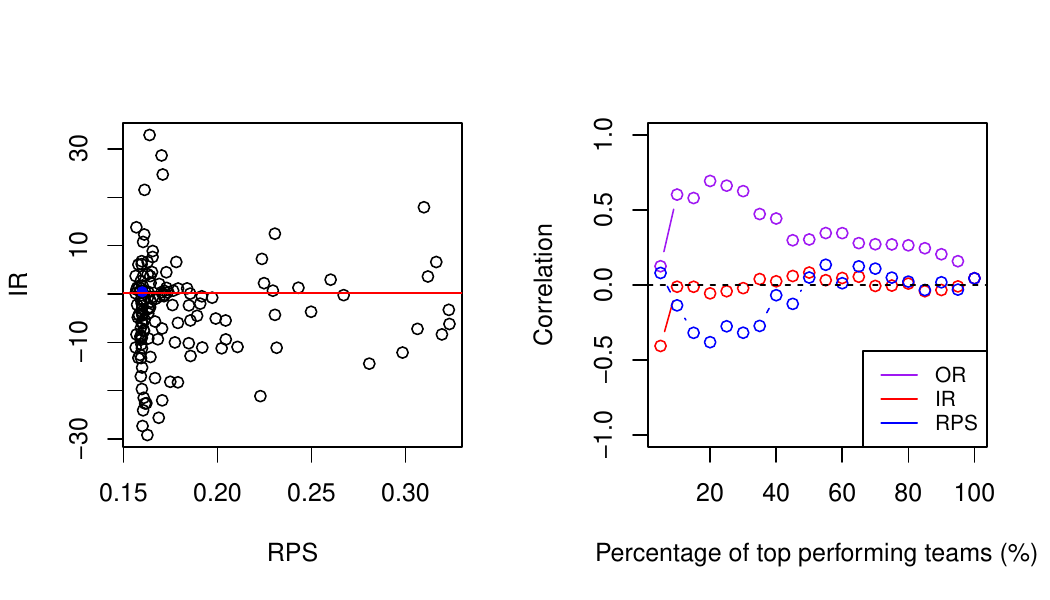}
    \caption{Left panel: Correlation between IR and RPS of the 138 teams included in the ``Global'' leaderboard whose forecast submissions were not identical to the benchmark. Right panel: Correlation between IR and RPS, reported for various percentages of the top performing teams and ranked based on OR, IR, and RPS.}
    \label{fig:hyp3}
\end{figure}

When the complete set of teams is considered, no connection is identified between the two variables ($r=0.04$), as shown in Figure \ref{fig:hyp3}. Nevertheless, since the M6 was a duathlon competition, it should be the case that at least the top performing teams according to the OR have managed to achieve relatively high scores, both in terms of IR and RPS. Figure \ref{fig:hyp3} confirms this hypothesis to some extent, suggesting that teams of higher OR tend to construct more efficient portfolios and at the same time produce more accurate forecasts. However, this link is weak, being maximized ($r=0.7$) for the top 20\% of the teams and vanishing when more than 40\% of the teams are considered. Moreover, there seems to be no association ($r=0.12$) between the two measures for the top 5\% of the teams, meaning that the teams that submitted the best forecasting submissions did not perform similarly well in terms of investment decisions and vice versa. 

The latter finding is confirmed when the same correlation analysis is conducted, but this time the teams are ranked according to their IR and RPS scores instead of OR. As shown in Figure \ref{fig:hyp3}, the top performing teams in the forecasting challenge constructed relatively inefficient portfolios on average (negative or close to zero coefficients), while the top performing teams in the investment decisions challenge have submitted forecasts of various accuracy levels (zero or even negative coefficients). 

To validate the second part of our hypothesis, we introduce two concentration proxy variables, namely the average number of invested assets and the average absolute investment weight per asset. As their descriptions imply, the first variable measures concentration in terms of number of assets involved in the constructed portfolios (more assets, lower concentration and risk), while the second in terms of capital invested per asset (larger investment weights, higher concentration and risk). Our analysis, measuring the correlation between the two concentration proxy variables and RPS, identified small negative correlations ($r=-0.05$), thus confirming that, in general, the risks taken by the teams cannot be justified by the accuracy of their forecasts.\\ 

\noindent\textbf{Hypothesis No.4: }\textit{Top performing teams in the investment challenge will build their portfolios using assets that they can forecast more accurately.}

In order to evaluate this hypothesis, we focus on the 138 teams included in the ``Global'' leaderboard whose forecast submissions were not identical to the benchmark, i.e. put some effort in the forecasting challenge of the competition. For each team and submission point we compute the RPS score of each asset separately, as well as the corresponding proportion of invested capital (investment weight). In order for the hypothesis to be true, assets that are assigned with higher investment weight by the top performing teams in the investment challenge should also display relatively lower RPS values.

To simplify the comparisons, we group the forecasts into three classes based on the realized accuracy, namely ``high'', ``moderate'', and ``low'' accuracy. The first class involves forecasts where RPS is lower than 0.1, the second class forecasts where RPS ranges from 0.1 to 0.22, while the third class forecasts where RPS is higher than 0.22\footnote{The thresholds were selected so that they are close to the first (0.10) and third (0.24) quantile of the RPS scores achieved by the teams and symmetric around the RPS score of the benchmark (0.16).}. Then, the average weight assigned to each class of assets is computed for each team across the twelve submission points, as well as the average correlation between the investment weights and RPS scores.

\begin{figure}[!h]
    \centering
    \includegraphics[width=0.9\textwidth]{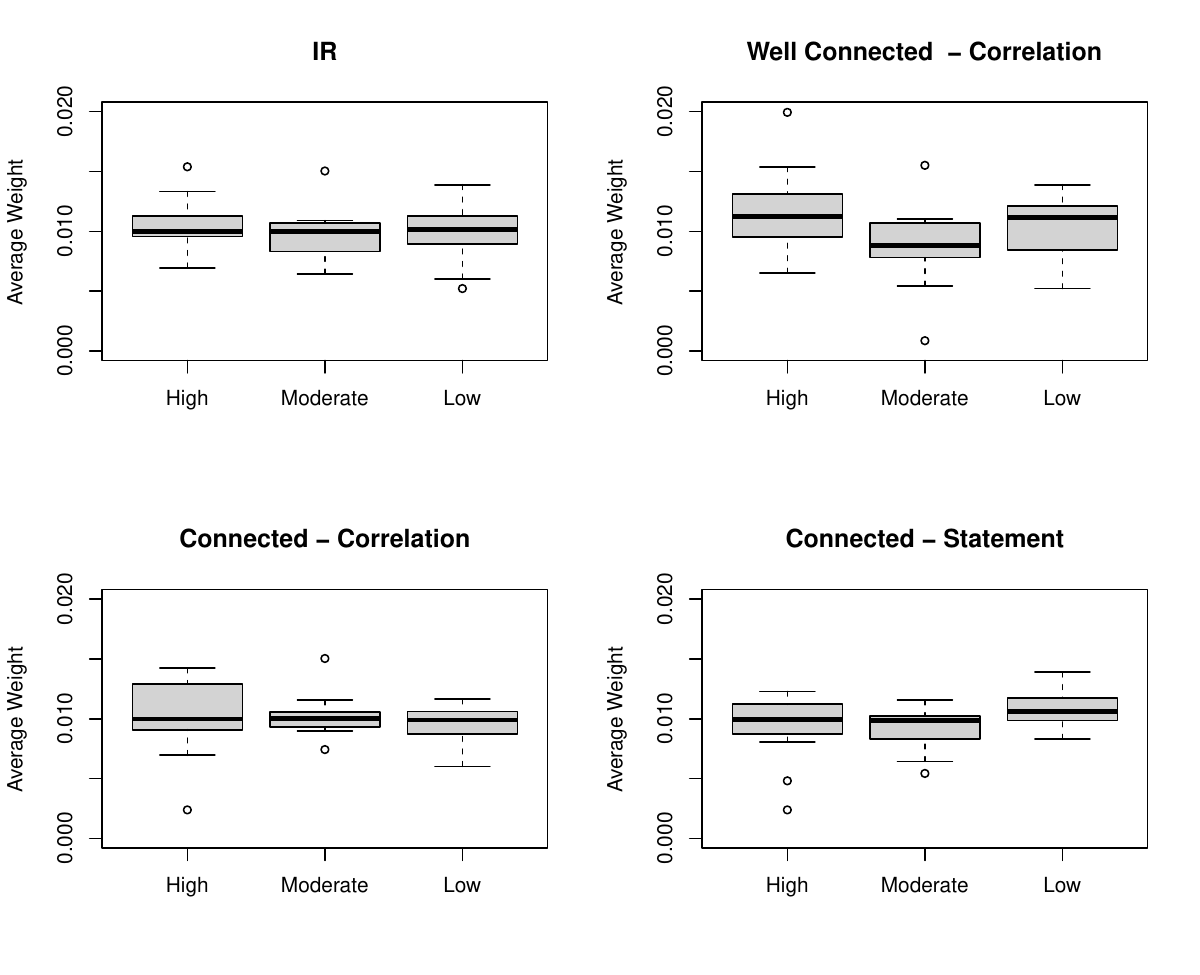}
    \caption{Average Investment weights used by the 138 teams included in the ``Global'' leaderboard whose forecast submissions were not identical to the benchmark in assets that were forecast with high (RPS<0.10), low (RPS>0.22) or moderate accuracy. The relationship between the accuracy reported per asset and the corresponding investment weight is examined for the top 15 teams of the competition in terms of IR (top left) as well as for the top 15 teams whose forecasts are ``well connected'' with the investments (top right), whose forecasts are ``connected'' with the investments (bottom left), and who claimed in the questionnaire that they were going to connect their forecasts with their investment decisions (bottom right).}
    \label{fig:hyp4}
\end{figure}

Figure \ref{fig:hyp4} presents the distribution of the investment weights per class of assets for the top 15 performing teams according to the IR. As seen, there is no evidence that the teams we examined built their portfolios using assets that they could forecast more accurately, a finding that is also supported by the insignificant correlation between the investment weights and the RPS scores ($r=0.06$).

Since we showed (see Figure \ref{fig:hyp3}) that the top performing teams in the investment challenge did not perform similarly well in the forecast challenge, we could argue that the lack of correlation between the investment weights and the accuracy of the forecasts could be attributed to the general lack of connection between the forecasts and the investment decisions. In order to take this disconnect into account, we computed for each team the average correlation between the predicted ranks and the investment weights. In this regard, submissions where higher amounts of capital were invested in assets of higher predicted ranks were classified as ``well connected'' or ``connected'', while the rest as ``weakly connected'', ``disconnected'', or of ``opposite connection'' (for more details about this classification, please refer to the supplementary material in the appendix). In this context, Figure \ref{fig:hyp4} also presents the distribution of the investment weights per class of assets for the top 15 performing teams (according to the IR) whose forecasts were either ``well connected'' or ``connected'' with the investment decisions. Once again, the distributions of weights largely overlap across the three classes, indicating that forecast accuracy did not affect the investment decisions of the teams. Similar conclusions can be made if we focus on the top 15 performing teams (according to the IR) that claimed in the questionnaire that their forecasts would be linked with their investment decisions. Based on the above, this hypothesis is rejected.\\

\noindent\textbf{Hypothesis No.5: }\textit{Teams that employ consistent strategies throughout the competition will perform better than those that change their strategies significantly from one submission point to another.}

In order to measure the impact of strategy consistency on the investment performance of the teams, one has to define first the structural elements of an investment strategy. For reasons of brevity, we decided upon the following four elements.

\begin{itemize}[noitemsep]
\item Exposure: We measure the total amount of capital invested at a certain submission point. According to the rules of the competition, exposure could range between 0.25 and 1. Therefore, the submissions were classified as ``lowly'' [0.25,0.50),  ``moderately'' [0.50,0.80) or ``highly'' [0.80,1.00] exposed.
\item Diversification: We measure the concentration of the constructed portfolios in terms of number of invested assets. The submissions were classified as ``lowly'' [1,10),  ``moderately'' [10,80) or ``highly'' [80,100] diversified.
\item Investment weight range: We measure the range of the investment weights considered within a portfolio, normalized by the exposure. In practice, this measure provides information about whether the portfolio considered similar investment weights for all assets or focused on a particular set of assets. The weight range could be either ``small'' (0,0.1) or ``large'' [0.1,1].
\item Investment direction: Although the position of each asset could be either long or short, the teams could also place all their investments in a single direction. Accordingly, the submissions could be classified as ``directional'' or ``non-directional''.  
\end{itemize}

Note that the thresholds of the aforementioned clarifications were based on the distributions of the defined elements and defined so that any class change would also signify a significant strategy change.

\begin{figure}[!h]
    \centering
    \includegraphics[width=0.9\textwidth]{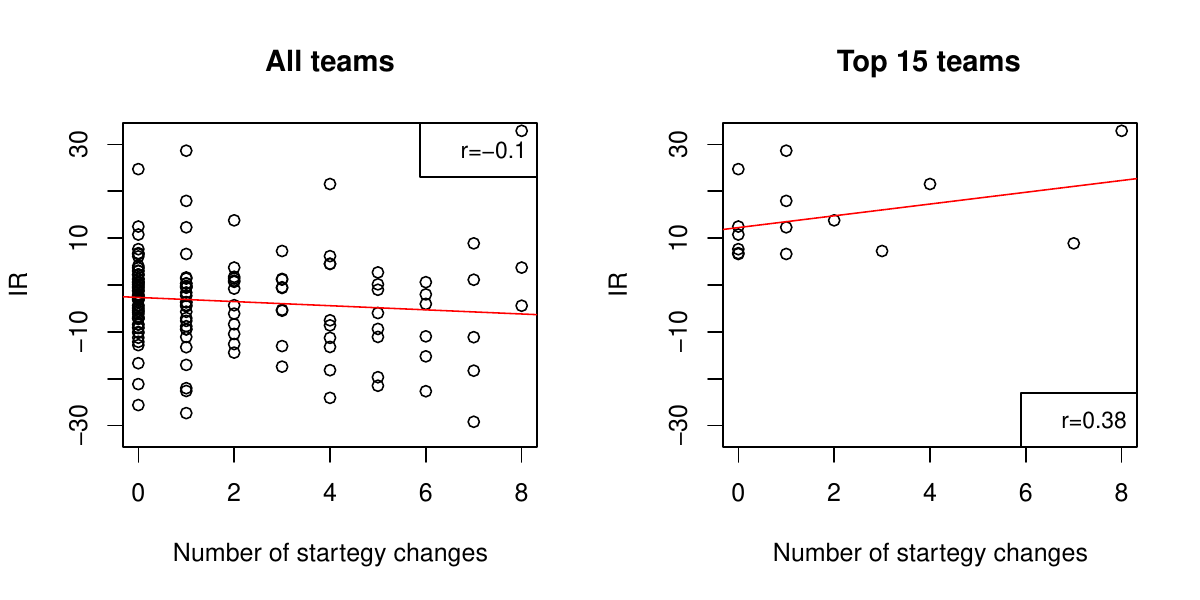}
    \caption{Correlation between IR and number of strategy changes. The sample involves the 148 teams included in the ``Global'' leaderboard whose investment submissions were not identical to the benchmark (left panel) and the top 15 performing teams of the investment challenge (right panel).}
    \label{fig:hyp8_1}
\end{figure}

Having measured the exposure, diversification, investment weight range, and investment directionality for all the submissions made by each team, we computed the number of strategy changes and the corresponding investment performance (IR) for the complete duration of the competition. The correlation of the latter two measures is visualized in Figure \ref{fig:hyp8_1}, both for the complete sample of teams and the top 15 performing ones in the investment challenge.

By observing Figure \ref{fig:hyp8_1} we find that, overall, there is a weak negative connection (r=-0.1) between strategy changing and investment performance. However, it is evident that teams of the same number of strategy changes can report significantly different IR scores, ranging e.g. from -24 to  22 when four strategy changes have occurred. Moreover, the distributions of the IR for different number of strategy changes largely overlap. The latter is confirmed if we focus on the top 15 performing teams: Although 6 teams never changed their strategy, 4 teams changed their strategy only once, and the remaining 6 up to eight times, they all achieved comparable IR scores. Interestingly, the winning team of the investment challenge was found to have changed its strategy several times (8 in particular). In addition, when the focus is on the top performing teams, the correlation of the measures we examine becomes slightly positive ($r=0.38$). In light of that evidence, this hypothesis is rejected.

\begin{figure}[!h]
    \centering
    \includegraphics[width=0.9\textwidth]{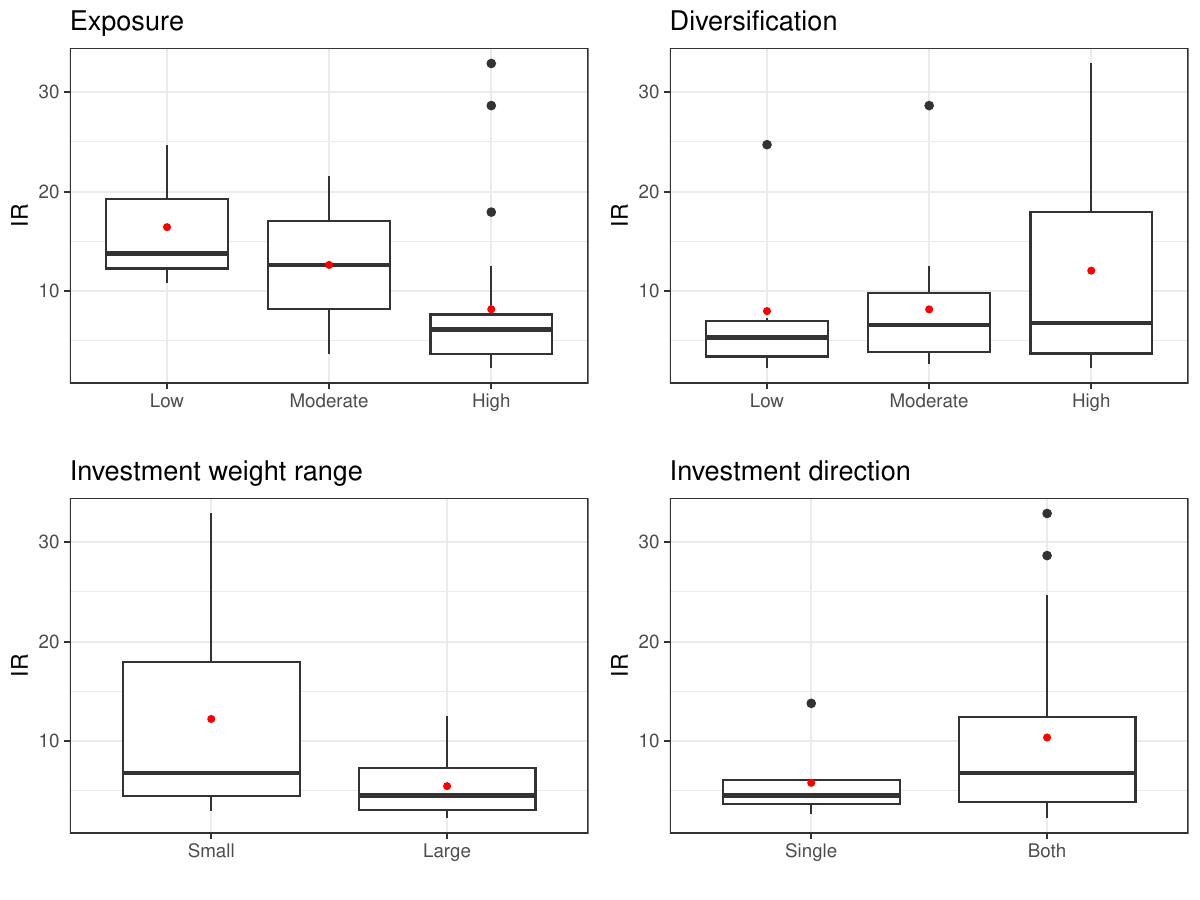}
    \caption{IR of the top 30 performing teams in the investment challenge, distinguished based on structural strategic elements, namely exposure, diversification, investment weight range, and investment direction. The classification of the teams is performed based on the average strategy they followed.}
    \label{fig:hyp8_2}
\end{figure}

To provide further insights on the strategic elements that defined the winners in the investment challenge, we complete this analysis by grouping the top 30 performing teams (according to the IR) based on the exposure, diversification, investment weight range, and investment direction classes they fell into, on average, across the complete competition. Figure \ref{fig:hyp8_2} summarizes the results. We find that lower levels of exposure were particularly beneficial in achieving higher IR scores. Moreover, we observe that more diversified portfolios of comparable investment weights typically performed better. Finally, it is evident that most of the top performing teams invested in both short and long positions, thus being able to effectively adjust to the directional changes of the market.

\noindent\textbf{Hypothesis No.6: }\textit{Team rankings based on information ratios will be different from rankings based on portfolio returns or rankings based on the volatility of portfolio returns.}

IR is optimized when returns are realized with the minimum possible risk or, equivalently, when risk is realized with the maximum possible returns. In this regard, although it is generally expected that teams ranked higher in the leaderboard constructed portfolios that simultaneously reported higher returns and lower risk, this may not always have been be the case. For instance, two teams with similarly risky portfolios may ultimately realize significantly different returns and, therefore, IR scores. This is the basis of the present hypothesis - we investigate whether low risk or high returns contribute in particular to superior information ratios.

\begin{figure}[!h]
    \centering
    \includegraphics[width=0.9\textwidth]{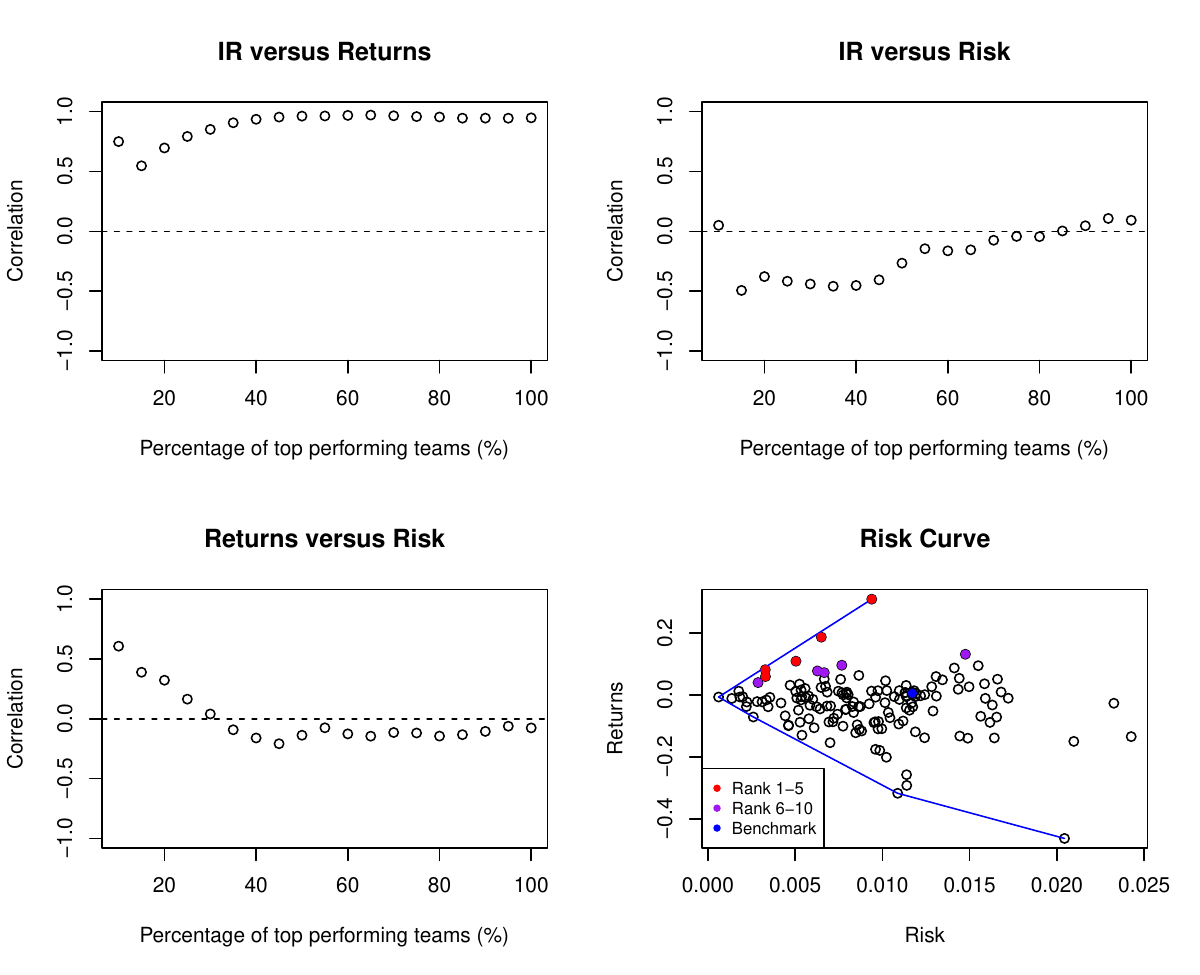}
    \caption{Spearman correlation between IR and returns, IR and risk, and returns and risk of the 148 teams included in the ``Global'' leaderboard whose investment submissions were not identical to the benchmark. The correlations are presented for various percentages of the top performing teams, ranked based on IR. On the bottom right plot, an estimate of the empirical risk curve is also provided using convex hull, i.e. straight lines that optimally enclose every point of the data set.}
    \label{fig:hyp2}
\end{figure}

Figure \ref{fig:hyp2} presents the correlation coefficients, $r$, between IR, returns, and risk in a pairwise fashion, computed for various percentages of the top performing teams (ranked according to the IR). There are several notable observations to make. First, when all teams are considered, IR and returns are highly correlated ($r=0.94$), in contrast to IR and risk that are barely associated ($r=0.09$). Second, the correlation between IR and returns decreases significantly for the top ranked teams, reaching a minimum of 0.55 for the top 20\%. Third, IR and risk are negatively correlated for up to the top 45\% of the teams, while returns and risk positively associated for up to the top 20\% of the teams. 

In general we find that successful teams inhabited the 'goldilocks zone' - risk was 'not too hot' - well controlled, but sufficient - 'not too cold' to allow a reasonable excess return to be generated.

To further elaborate on the last point, Figure \ref{fig:hyp2} presents the empirical risk curve estimated based on the portfolio returns and risk measured for each of the participating teams. As seen, with the exception of one team, all the top 10 performing teams have constructed portfolios that are significantly less risky than the benchmark. Moreover, the top 5 teams have effectively managed to maximize their returns given a certain amount of risk, which was either particularly small (around 0.003) or moderate (around 0.008). Based on the above we conclude that team rankings based on information ratios would be different from rankings based on portfolio returns and particularly on rankings based on portfolio risk.\\

\noindent\textbf{Hypothesis No.7: }\textit{Teams will be measurably overconfident in the accuracy of their forecasts, on average. Namely, forecasts will be less dispersed and have smaller variance than observed in the data.}

In order to evaluate this hypothesis, we look at overprecision in the assessed probabilities, in terms of unwarranted certainty for the outcomes.  In other words, we would expect that for very low assessed probabilities (near zero) the relative frequency of the outcomes would be higher than the assessed probabilities, and for very high assessed probabilities (near one) the relative frequency of the outcomes would be lower than the assessed probabilities 
\citep[see, for example, ][]{Lichtenstein1982-fb}.

For each asset, the teams provided probabilities (summing to one) that percentage return will be within the first, second, third, fourth, of fifth quintile across all assets. These probabilistic forecasts were evaluated by RPS.  We focus on the 38 teams with RPS less than 0.16, i.e., the teams that did better than the benchmark in their probabilistic forecasts.

Overconfidence in assessed probabilities can be explored through a calibration curve which plots relative frequency of outcomes against assessed probabilities of those outcomes. Figure \ref{fig:overconfidence} below shows the relative frequency of the outcomes corresponding to the average assessed probabilities within intervals of size 0.05 of assessed probabilities from 0 to 1 across the 38 teams and across all quintiles for all assets in all the submissions.  The dotted diagonal line represents perfect calibration and the solid line shows the actual performance.  Note that for assessed probabilities higher than 0.3 (somewhat higher than 0.2 in the benchmark), the relative frequency is less than the assessed probability, substantially so as assessed probability increases.  Similarly, for very low assessed probabilities, the relative frequency is higher than that assessed. In other words, extreme probabilities (in the direction of 0 and 1) show unwarranted certainty for the outcomes, i.e., overconfidence, and we accept the hypothesis.\\

\begin{figure}%[!ht]
    \centering
    \includegraphics[width=3.8in]{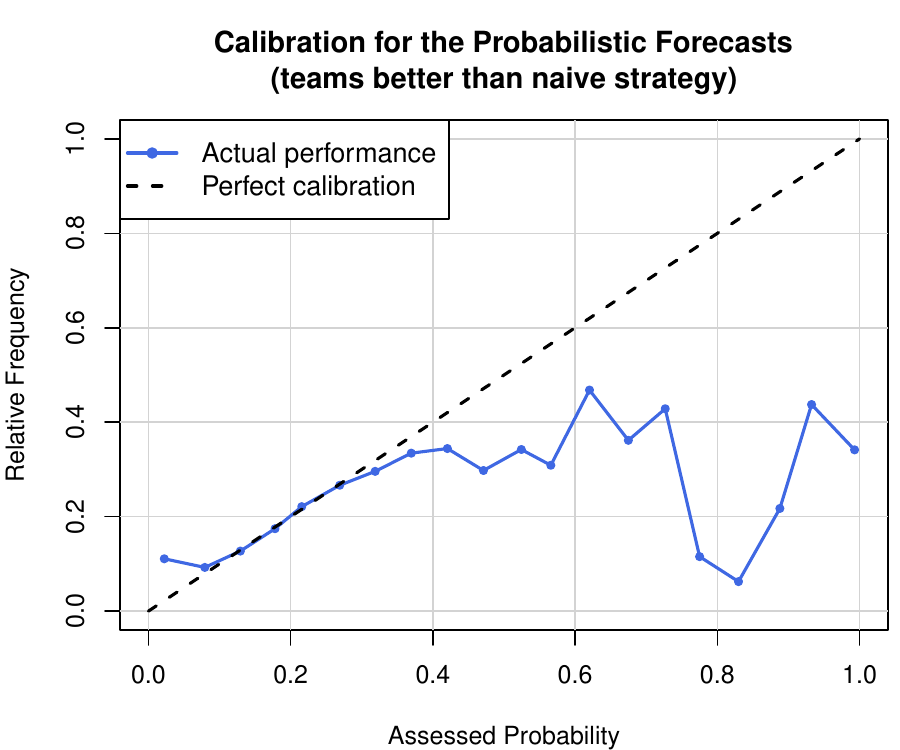}
    \caption{Assessing overconfidence of forecasts.}
    \label{fig:overconfidence}
\end{figure}

\noindent\textbf{Hypothesis No.8: }\textit{Averaging forecast rankings (investment weights) across all teams for each asset will yield rankings (weights) that outperform those of the majority of the teams, except in cases where the very worst teams are removed from the average.}

The ``wisdom of crowds'' is a popular concept according to which the aggregation of information in groups typically results in better decisions than those made by any individual member of the group \citep{Surowiecki2005}. The benefits of combining forecasts have been confirmed in all the previous M competitions, as well as multiple other forecasting studies \citep{PETROPOULOS2022705}, while similar encouraging conclusions have been reached in several financial applications \citep{GOTTSCHLICH201452, CHAU2020100741, DAI2021561}. In this context, this hypothesis aims to validate the value of combining, both in the forecasting and investment domains.

In order to reach more representative conclusions, our analysis focuses on the 138 teams included in the ``Global'' leaderboard whose forecast submissions were not identical to the benchmark. We proceed by ranking said teams based on their OR, IR, and RPS scores and averaging their submissions for different proportions of the sample, namely the top $5, 10, \dots, 95, 100$ percent, consecutively computing the corresponding IR and RPS scores of the averages. Note that within this process the teams were ranked on an ex-post basis, i.e. according to the ``Global'' scores they realized and not their expected performance.

\begin{figure}[!h]
    \centering
    \includegraphics[width=0.9\textwidth]{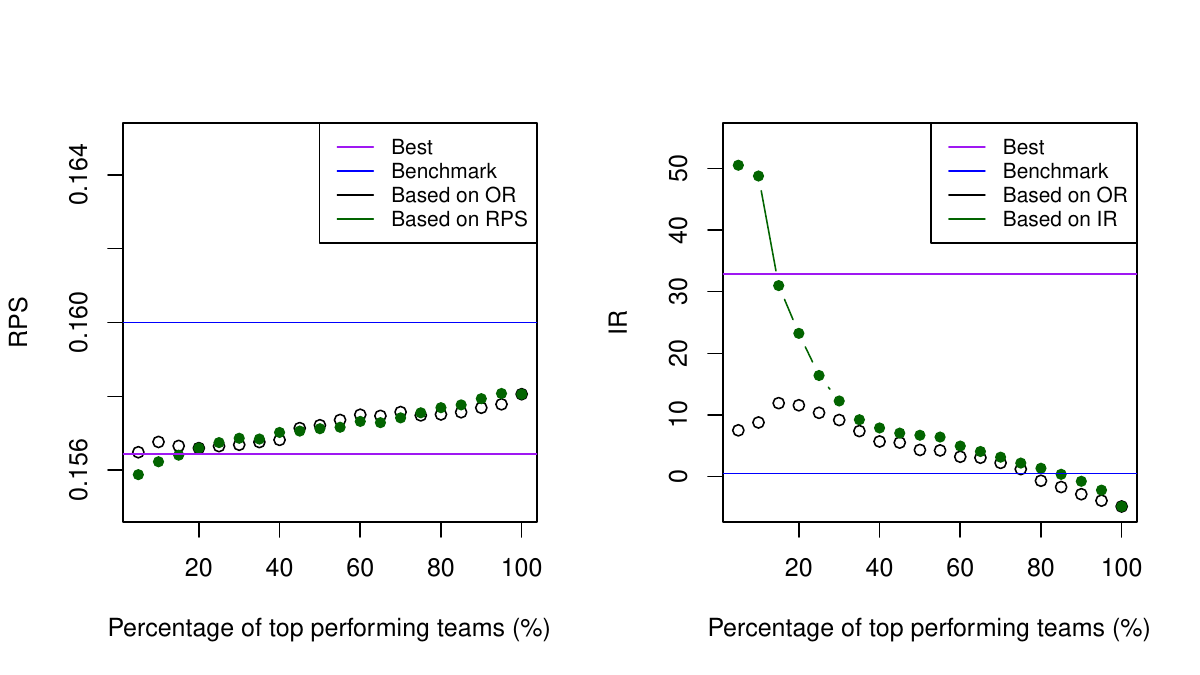}
    \caption{RPS and IR of the average forecast and investment submission when various percentages of the top performing teams are considered either per track (ranked based on RPS and IR, respectively) or overall (ranked based on OR). The sample of submissions being averaged involves the 138 teams included in the ``Global'' leaderboard whose forecast submissions were not identical to the benchmark.}
    \label{fig:hyp7}
\end{figure}

Figure \ref{fig:hyp7} summarizes the results for the RPS and IR measures separately. In the first case, the top $N$ percentage of the teams is ranked based on the OR and RPS measures, while in the latter based on the OR and IR measures. This is done because averaging the submissions of the top performing teams in the forecasting (investment decisions) challenge is expected to yield better forecasts (investment decisions) compared to the scenario where the ranking is performed according to the OR.

Focusing on forecast accuracy, Figure \ref{fig:hyp7} suggests that averaging the forecasts of the top 5, 10 or 15\% of the teams according to the RPS results to superior performance than the best performing team. Moreover, when the teams are ranked based on the OR, the accuracy of the combined forecasts is similar to that of the top performing team, even when the top 40\% of the teams in included in the aggregation process. More importantly, when the forecasts are averaged, the forecast error is always lower than that of the benchmark. 

The results are similarly encouraging in terms of investment decisions. Averaging the investment weights of the top 5 or 10\% of the teams according to the IR results to significantly better performance than the best performing team. In addition, the averages of the investment decisions outperform the benchmark, at least when the worse 20\% of the teams is excluded from the aggregation process. Finally, similarly to the RPS case, using IR instead of OR to decide which submissions should be included in the aggregation process, always yields better results. In light of the above, this hypothesis is approved.\\

\noindent\textbf{Hypothesis No.9: }\textit{Submissions based on pure judgment or that rely heavily on judgment will perform worse than those based on data-driven methods, on average.}

In order to investigate this hypothesis, we classify participating submissions into four categories. This classification was made qualitatively based on the participants' responses to the questionnaire accompanying their submissions, and in particular to their descriptions of the forecasting methods implemented. Given that in most there were multiple submissions per team, and between submissions teams had the option to change their forecasting method, in this analysis we classified the descriptions by considering the most ``encompassing'' approach. For instance, if in a particular submission point the forecasting approach described by a particular team was simply data-driven, whereas later on it was also informed by judgment, this would fall under the category ``judgment-informed''. The categories that we decided upon were:
\begin{itemize}[noitemsep]
\item Data-driven approaches (time series, ML, combinations);
\item Judgment-informed (data-driven approaches informed by judgment);
\item Pure judgment;
\item Not specified (where participants' descriptions did not allow us to categorize them in one of the above three groups).
\end{itemize}

The absolute and ranked values of $RPS$ and $IR$ for each of the above categories of forecasting methods (mean as well as 90\textsuperscript{th} percentile) are presented in tables \ref{tab:h9t1} and \ref{tab:h9t2}. We also show, in the second and third columns, the counts and percentages of the submissions associated with each category.

\begin{table}[ht]
\small
\centering
\caption{Participants' performance per category of forecasting method.}
\begin{tabular}{ccccccc}
\hline
\textbf{Category} & \textbf{N} & \textbf{\%} & 
\multicolumn{2}{c}{\textbf{RPS}} & \multicolumn{2}{c}{\textbf{IR}}\\
&&& Mean & 90\textsuperscript{th} Perc. & Mean & 90\textsuperscript{th} Perc. \\
\hline
Data-driven & 171 & 68.4 & 0.182	& 0.159	& -3.374	& 6.562\\
Judgment-informed & 8	& 3.2	& 0.181	& 0.158	& -0.193	& 7.044\\
Pure judgment & 14	& 5.6	& 0.175	& 0.160	& -6.832	& 0.036\\
Not specified & 57	& 22.8	& 0.169	& 0.160	& -1.493	& 4.555\\
\hline
\end{tabular}
\label{tab:h9t1}
\end{table}

\begin{table}[h]
\small
\centering
\caption{Ranked participants' performance per category of forecasting method.}
\begin{tabular}{ccccccc}
\hline
\textbf{Category} & \textbf{N} & \textbf{\%} & 
\multicolumn{2}{c}{\textbf{RPS Rank}} & \multicolumn{2}{c}{\textbf{IR Rank}}\\
&&& Mean & 90\textsuperscript{th} Perc. & Mean & 90\textsuperscript{th} Perc. \\
\hline
Data-driven & 171 & 68.4 & 84.0	& 15.1	& 83.8	& 16.1 \\
Judgment-informed & 8	& 3.2	& 79.3	& 14.5	& 76.2	& 35.5 \\
Pure judgment & 14	& 5.6	& 89.7	& 49.3	& 101.8	& 66.5\\
Not specified & 57	& 22.8	& 73.7	& 27.6	& 71.5	& 22.4\\
\hline
\end{tabular}
\label{tab:h9t2}
\end{table}

It is clear that approaches that were heavily based on judgment (pure judgment) were in general inferior to those based on data-driven approaches. For example, the 90\textsuperscript{th} percentile of $IR$ for data-driven approaches is 6.562 compared to 0.036 for pure judgmental approaches. However, it seems that there is some merit in introducing judgment to data-driven forecasting approaches. Judgment-informed forecasting approaches (albeit, very few) perform on par (if not better) compared to pure data-driven approaches. So, judgment utilized correctly (and in conjunction with a data-driven approach) can offer good performance. Based on the above, there is empirical support for accepting this hypothesis.\\

\noindent\textbf{Hypothesis No.10: }\textit{The top performing teams in the forecasting challenge will employ more sophisticated methods compared to the top performing teams in the investment challenge.}

To address this hypothesis, we also focused on the qualitative responses provided by the participating teams regarding their description of the methods. In this case, though, our classification focused on separating time series (TS) based methods from ML-based methods. Also, we pooled all approaches related to the used of judgment together. In balance, we considered the following four categories:
\begin{itemize}[noitemsep]
\item Judgment-based (either pure judgmental or judgment-informed);
\item TS-based (time series approaches, but also their combinations);
\item ML-based (including ML approaches integrated with TS and the respective combinations);
\item Not specified (where participants' descriptions did not allow us to categorize them in one of the above three groups).
\end{itemize}

We present the counts of the top 5, 10, 15 and 20\% of the teams that described their forecasting approaches as above and we present these in tables \ref{tab:h10t1} (for $RPS$) and \ref{tab:h10t2} (for $IR$). We observe that there is little evidence that teams in the forecasting challenge employed more sophisticated (i.e. ML-based) approaches than the top teams in the investment challenge. So, this hypothesis is inconclusive.

\begin{table}[h]
\small
\centering
\caption{Frequency of forecasting methods categories employed by the top performing teams in terms of $RPS$.}
\begin{tabular}{ccccccc}
\hline
\textbf{Category} & \textbf{Top 5\%} & \textbf{Top 10\%} & \textbf{Top 15\%} & \textbf{Top 20\%}\\
\hline
Judgment-based & 1 & 1 & 2 & 3\\
TS-based & 4 & 7 & 9 & 12\\
ML-based & 3 & 7 & 10 & 14\\
Not specified & 1 & 2 & 4 & 5\\
\hline
\end{tabular}
\label{tab:h10t1}
\end{table}

\begin{table}[h]
\small
\centering
\caption{Frequency of forecasting methods categories employed by the top performing teams in terms of $IR$.}
\begin{tabular}{ccccccc}
\hline
\textbf{Category} & \textbf{Top 5\%} & \textbf{Top 10\%} & \textbf{Top 15\%} & \textbf{Top 20\%}\\
\hline
Judgment-based & 1 & 2 & 2 & 3\\
TS-based & 4 & 8 & 10 & 14\\
ML-based & 3 & 6 & 9 & 10\\
Not specified & 1 & 1 & 4 & 6\\
\hline
\end{tabular}
\label{tab:h10t2}
\end{table}

%\section{Winning submissions}
%\label{sec:wmethods}

%The forecasting methods of the winning teams can be summarized as follows:

%\textcolor{red}{To be completed by Evangelos once we receive the drafts of the invited methodological papers.}

%\begin{itemize}
    
%    \item \textbf{First place (\textit{XXX}; XXX):} 
    
%    \item \textbf{Second place (\textit{XXX}; XXX):}
    
%    \item \textbf{Third place (\textit{XXX}; XXX):} 
    
%\end{itemize}

\section{An investment risk model for the M6 competition}\label{sec:riskmodel}

\emph{`Quantitative active management is the poor relation of modern portfolio theory. It has the power and structure of modern portfolio theory without the legitimacy'}\footnote{\cite{Grinold1999-it}.}.
This quote is from a foundational reference book for quantitatively orientated portfolio managers who were learning their trade in the 1990s and 2000s, a period which saw a significant increase in practitioner interest in more systematic approaches to investment decision making. Via this statement, the authors acknowledge that while their approach is academically rigorous, the entire premise of active investment decision making is called in to question by modern portfolio theory. The M6 competition was in part motivated by this debate and as a means to contribute, in this section we describe an investment risk model constructed to analyze the \emph{investment} submissions made by participants in the M6 competition. Our model is designed to be relatively simple to fit and to use data which was readily available to competition participants, while taking advantage of recent advances in the understanding of the structure of multivariate volatility across asset classes. Applying our model to the M6 competition submissions, we find that most participants were measurably overconfident - they assumed much more investment risk than was justified by the accuracy of the submissions made in the \emph{forecasting} leg of the competition. A number of additional findings related to this one are discussed in the sequel.

In the remainder of this section, we first summarize key features of the M6 associated with the investment part of the competition. We then dive more deeply into the measurement of investment risk. Thereafter, we outline our investment risk model. Finally, we summarize a number of findings based on implementation of the said model.

\subsection{Key features of the investment part of the competition}

Recall that participants in the M6 were asked to submit their entries in two parts. The first part of each submission comprised a set of \emph{forecasts} summarizing their expected probability distributions for the returns on the universe of investment assets specified in the competition rules. The second part of each submission comprised a set of \emph{investment decisions}, which were expected to be made on the basis of these forecasts. Participants were given a clear mandate -  to maximize the Sharpe ratio of the resulting portfolio.

In order to explain why we use the Sharpe ratio as our metric for assessing investment performance, recall that economic theory posits that investors choose portfolios (represented by a set of weights allocating capital to positions in some subset of the universe of potential investments) by maximizing a function representing their `utility'. The procedure to do this was set out in Nobel winning work of \cite{Markowitz1959-gy}. The idea is that investors make capital allocation decisions by forecasting returns and variance/covariance on some universe of investment assets. For each investor, the set of `optimal' investment weights are then chosen by maximizing a utility function, given a parameter summarizing the investor's level of risk tolerance. In technical terms, and assuming a quadratic utility function, the investor chooses a weight vector, $w$, that maximizes their expected utility function:

\begin{equation}
    U_w = \boldsymbol{\alpha}\mathbf{w} - \lambda \mathbf{w} \boldsymbol{\Sigma} \mathbf{w}',
\end{equation}
where the risk aversion parameter $\lambda$ is a scalar value chosen by the investor. 

While this is a simple and elegant theory, there are significant hurdles to overcome in practical application of the Markowitz approach. Aside from various conceptual (and well documented behavioral) difficulties that non-technical investors may experience when specifying a utility function, choosing an investment time horizon and calibrating their risk aversion, implementing a Markowitz style approach require sensible \emph{estimates} of $\alpha$ and $\Sigma$. In general, producing accurate return ($\alpha$) and covariance ($\Sigma$) forecasts can be challenging. For example, while estimating univariate variance terms (i.e., the diagonal component of $\Sigma$) is relatively easy, since volatility is somewhat persistent, deriving reliable estimates of the off-diagonal elements of this covariance matrix is somewhat more challenging, given that investors often select from a large number of potential investments ($n$), and given that covariance change over time, so that the effective time window ($t$) for constructing reasonably accurate estimates tends to be rather small. Moreover, experience has shown that using noisy and/or badly calibrated estimates of risk and return leads to poorly structured and poorly performing investment portfolios - optimization can easily become an estimation error, rather than utility maximizing tool.

Notwithstanding the forecasting challenges discussed above, the structure of the M6 competition enabled participants to abstract away from the issues discussed above regarding the choice of utility function, time horizon and calibration of risk aversion. This was done by giving a clear investment mandate to participants; we asked them to produce portfolios designed to maximize the Sharpe Ratio measured over the 4 trading weeks commencing on the Monday following the weekend during which they submitted their portfolios. That is, participants were asked to submit a weight vector $w$ which would maximize the Sharpe ratio:

\begin{equation}
    SR_w = \frac{\boldsymbol{\alpha}\mathbf{w}}{\mathbf{w} \boldsymbol{\Sigma} \mathbf{w}'}
\end{equation}
This ratio of portfolio return to portfolio risk is commonly used by practitioners and academics as a measure of historic risk adjusted investment performance. Notice that in the above expression, portfolio return is simply the weighted sum of the returns on the individual assets in the portfolio. In this sense, return forecasts made in the first part of the competition are directly relevant. In addition, in order to maximize the Sharpe Ratio, participants needed to model or at least make assumptions regarding portfolio risk, given that $\Sigma$ appears in $SR_w$. The advantage of posing the M6 investment objective in this way is that no risk aversion parameter is required. The disadvantages were that submitted portfolios varied significantly in the level of risk assumed (although this was also interesting academically) and that the optimization objective is somewhat non standard (although doing so is well within the capabilities of many freely available software packages). For a full discussion of Sharpe Ratio optimization and its relationship to the traditional Markowitz approach, the reader is referred to \cite{Lassance1}.

\subsection{Measuring and managing investment risk}

In this subsection we discuss our approach to measuring investment risk. We begin with a brief discussion of the stylized facts of stock market volatility, illustrating these with examples from the competition and elsewhere. We then briefly summarize some of the models commonly used to capture these features of the data.

\begin{figure}[h]
\centering
\includegraphics[scale=.65]{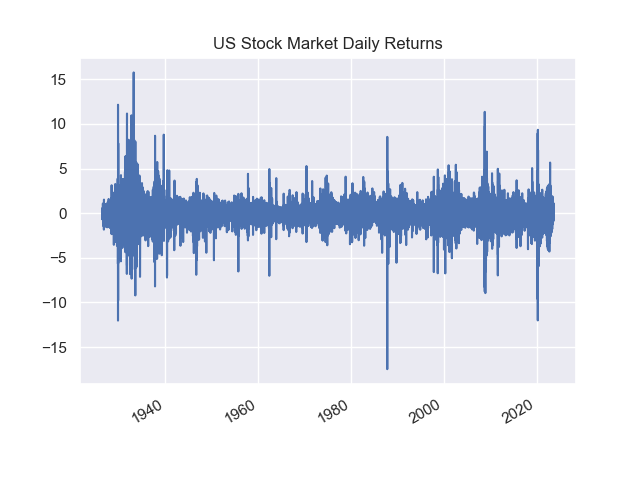}
\caption{Daily US Stock Market Returns from 1926. The chart shows daily returns for the US Market from 1926. (Sourced from the data library of Kenneth R. French\cite{French_undated-pg})}
\label{fig:USMkt_retns}
\end{figure}

Perhaps the most important stylized fact is that return variance is not constant over time. Figure \ref{fig:USMkt_retns} plots daily returns of U.S. equities over time 1926. Note the clear evidence of clusters of volatility in the 1930s and 1940s associated with the great depression and the Second World War. Note also the short period of extreme volatility associated with the 1987 stock market crash and more recent episodes of volatility associated with the Asian Debt / Long Term Capital Management crisis, the Great Financial Crisis of 2007-8 and the COVID pandemic. In between these periods, volatility tends to return to lower levels.

\begin{figure}[h]
\centering
\includegraphics[scale=.65]{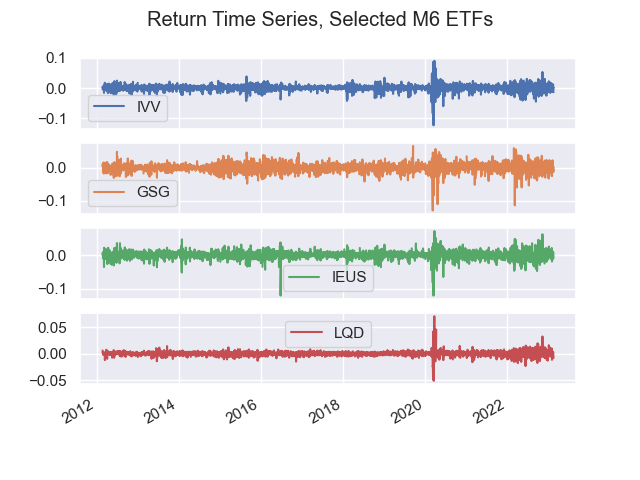}
\caption{Daily return time series from 2012 for selected M6 ETF securities - (Ishares Core S\&P500 ETF (IVV), Ishares Commodity Indexed ETF (GSG), Ishares MSCI Europe Small Cap ETF (IEUS) and Ishares Ibox \$ Investment Grade ETF (LQD))}
\label{fig:4_etfs_retns}
\end{figure}

Another stylized fact is that markets are characterized by volatility spillover across assets and asset classes. To illustrate this feature, we plot, in Figure \ref{fig:4_etfs_retns}, the daily returns for several ETFs selected from the M6 investment universe (these include the Ishares Core S\&P500 ETF (IVV), the Ishares Commodity Indexed ETF (GSG), the Ishares MSCI Europe Small Cap ETF (IEUS) and the Ishares Ibox \$ Investment Grade ETF (LQD)). Note that the time series of changing volatility are similar across assets. This is particularly evident for the spike in volatility associated with the COVID pandemic, and increased volatility during the M6 competition period associated with the Russian invasion of Ukraine.

\begin{figure}[h]
\centering
\includegraphics[scale=.65]{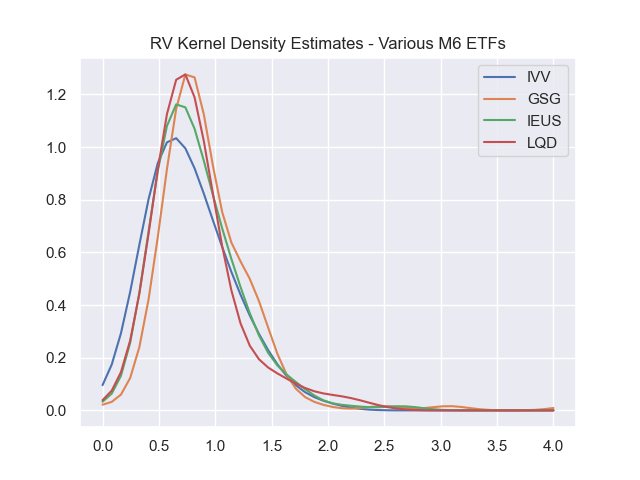}
\caption{Kernel Density Estimates for the 4 ETF securities displayed in \ref{fig:4_etfs_retns} above. The chart plots Kernel Density Estimates of the distribution of daily realized variance scaled by its long run average for each asset.}
\label{fig:ETF_KDEs}
\end{figure}

Another stylized fact concerns the common practice of fitting volatility models on an asset by asset basis. This is done despite empirical evidence suggesting that fitted model parameters often cluster in a reasonably tight range. The models described in \cite{Bollerslev2018-ee}, for example, take advantage of this feature. They justify the use of a common volatility model for many assets / asset classes by showing how similar the distribution of realized volatility become when scaled by each asset's long term or `expected' volatility. We repeat their analysis for the same subset of four M6 ETF assets discussed above, and plot our results in Figure \ref{fig:ETF_KDEs}. Note that the resulting realized volatility distributions closely mirror those examined in \cite{Bollerslev2018-ee}.

Turning now to the specification of volatility models, we first consider the case of univariate models. These are models used to describe the risk patterns associated with an individual asset or or market aggregate. A simple (perhaps simplistic) approach to this is to use a short term (for example 20 day or 100 day) moving average of squared returns. A related but more sophisticated approach is to use an exponential smoothing type model, again on squared daily returns. Here, variance is modeled as an exponentially weighted moving average, so that for instance, if the location of a time series $y_t$ is modeled as a function of some vector of predictors $x_t$:

\begin{align}
    y_t|F_{t-1} &\sim [x_t'\beta,\sigma^2_t] & \sigma^2_t &=  (1-\gamma) \sigma^2_{t-1} + \gamma e^2_{t-1}
\end{align}
where $F_{t-1}$ denotes a conditioning information set that includes data up to time period $t-1$, $[x_t'\beta,\sigma^2_t]$ denotes a distribution with mean $\mu$ and standard deviation of $\sigma_t$, $\beta$ and $\gamma$ are fixed parameters that must be estimated, and $e_t$ is a stochastic disturbance term (error). Such models are often fitted assuming that the errors are non-Gaussian (see \cite{Jondeau2007-eh}) for a comprehensive treatment of these models). A major (and Nobel prize winning) advance in univariate modeling of volatility was made with the introduction of the Auto-regressive Conditional Heteroskedasticity (ARCH) model introduced in \cite{Engle1982-si}. ARCH (and more generalized ARCH (i.e., GARCH) models are time varying, capture volatility clustering, and are widely used in industry. The ARCH(p) model is specified as above, but with:
\begin{equation}
    \sigma^2_t = \alpha_0 + \sum_{i=1}^p \alpha_i e_{t-i}^2
\end{equation}

The GARCH(p,q) generalization of this model is due to \cite{Bollerslev1986-dj} who noted that current conditional volatility ($\sigma^2_t$) is likely to depend not only only lagged squared errors ($e_t^2$) but also on lagged conditional volatility, leading to the following formulation for conditional volatility: 
\begin{equation}
    \sigma^2_t = \alpha_0 + \sum_{i=1}^p \alpha_i e_{t-i}^2+ \sum_{i=1}^q \beta_j \sigma_{t-i}^2
\end{equation}

Empirically, perhaps due to its relative parsimony the GARCH (1,1) model has been to be very competitive with more variants of the above models (\cite{hansen-lunde}). Note also that Bollersev's 1986 paper led to a voluminous literature on time series volatility estimation and an alphabet soup of ARCH/GARCH derivative models for both univariate and multivariate time series (see \cite{Bollerslev2009-nh}) . 

Other univariate volatility models are also used by some practitioners. A key example is the stochastic volatility model in which a system of stochastic differential equations are used to describe the return and volatility of an asset. These models are specified in continuous time, and require high frequency intra-daily data for estimation. They are discussed in detail in the Risk and Volatility survey paper appearing in this special issue (see also  \cite{ANDERSEN200143}). For a discussion of stochastic volatility modeling using Bayesian methods, see for example \cite{TriantafyllopoulosKostas2021BIoS}, \cite{Prado2021-xl}. For further discussion that focuses on the granularity of data used to estimate volatility models, see \cite{GarmanMarkB.1980OtEo}, \cite{ParkinsonMichael1980TEVM}, \cite{RogersL.C.G.1991EVFH} and \cite{YangDennis2000DVEB}.

Needless to say, multivariate volatility models are also crucial to portfolio management. In particular, it is important to consider cases where $n$ is large and $t$ may be small. 

Possibly the simplest approach to multivariate modeling is to use exponential smoothing on the cross products of daily returns for the universe of assets (see \cite{BrockwellPeterJ1991Ts:t}). This approach has the attraction of needing only one exponential smoothing parameter, which can be estimated from the data, or chosen by the researcher. Conceptually similar approaches extend the univariate  GARCH type models discussed above to the multivariate setting. For example, see the Baba, Engle, Kraft and Kroner (BEKK) model developed in \cite{Engle1995-td} and the Dynamic Conditional Correlation (DCC) model of \cite{Engle2002-if} and \cite{Engle2009-jk}. Both of these models can be parameterized heavily or simplified so that very few fitted parameters are required. 

A particular modeling approach which considerably simplifies model fitting, and leads to parsimony in the underlying model is to use a strategy known as `variance targeting' (univariate models) or `covariance targeting' (multivariate models) as described in \cite{Engle_undated-qv}. To give a univariate example, consider the GARCH(1,1) model:

\begin{equation}
   \sigma^2_t = (1-\alpha-\beta) \sigma^2_0  + \alpha e^2_{t-1} + \beta \sigma^2_{t-1}
\end{equation}
where $\sigma^2_0$ is a long run estimate of the variance to which the process tends to revert. We utilize a variant of this model in our below analysis.

When $n$ is large, components of modern multivariate model building involve key (Bayesian) ideas - variable selection, dimension reduction and parameter shrinkage, also common in the Machine Learning literature. Indeed, modern portfolio optimization approaches which use the full covariance matrix without application of some or all of these ideas are virtually non-existent, to the best of our knowledge (largely for good Darwinian reasons). (See \cite{Black1992-nw} for an early discussion of this topic). Arguably the earliest and perhaps currently the most popular and successful approach to achieve dimension reduction involves specifying and estimating a so-called factor model. 

In the context of factor models, common factors are assumed to underlie the co-movements of a set of variables, such as asset volatility, where the number of factors is $k$, with $k<<n$. The idea is to achieve dimension reduction by modeling the covariance matrix of $n$ assets as a function of $k$ of factors. When carrying out dimension reduction for the purpose of volatility estimation using returns, factor analysis yields a set of `factor returns' (sometimes orthogonal), and can be used to easily estimate a covariance matrix. For example, one might consider a model such as the following:

\begin{equation}
    {y_t} = {F_t} {\Delta} + {e_t},
\end{equation}
where ${y_t}$ is an $n$x$1$ vector of asset returns, ${F_t}$ is an $n$ x $r$ matrix in which each row collects the $r$ values of each of the $r$ common factors that are associated with each of the $n$ returns, ${\Delta}$ vector of factor loading, and ${e_t}$ is a vector of stochastic disturbance such that ${e_t} \sim [0,\Omega ]$, estimates of which can be constructed after jointly estimating the factors and loading coefficients. This framework is convenient. For example, when $\Omega$ is diagonal it is easy to estimate 
$\Sigma_y = \mathbf{F}\Sigma_F\mathbf{F}' +  \Omega$. A simple one factor model (i.e.  set $r=1$), where the single factor is a latent `market' portfolio and a security's loading is referred to as its `beta' underpins most of Modern Portfolio Theory. In practice, more useful models tend to have more factors, and complicated dynamics can be assumed to characterize the stochastic disturbance term. For further discussion of estimation of these models as well as more sophisticated variants thereof, refer to \cite{Swanson1}, \cite{Swanson2}, and \cite{Liao1}.  In addition, \cite{Connor2019-kf} describe several other conceptual approaches to building such models. 

A parallel approach to calibrating return forecasts is outlined in \cite{Grinold1999-it}. The idea is to use a Linear Bayes approach (\cite{Goldstein2007-it} to pre-process forecasts into a form compatible with a given risk model. This approach borrows from the analysis of expert opinion (see \cite{West1992-ml} and \cite{West1992-mg}). Here, relative return forecasts are shrunk towards those of a benchmark (in practice the benchmark is usually set up to have zero expected return). We adopt this approach in the sequel. 

For complete technical details describing the implementation of our investment risk model, refer to the appendix.   

\subsection{Findings based on application of our investment risk model to the M6 competition}

\begin{figure}[h]
\centering
\includegraphics[scale=.65]{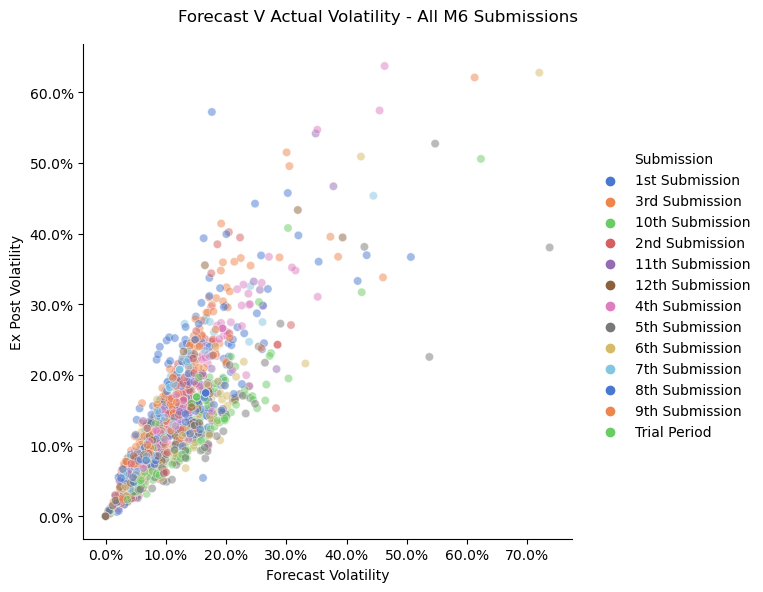}
\caption{Forecast M6 Portfolio Volatility compared to realized 20 day volatility for all M6 Submissions. We estimate M6 Portfolio ex-ante using our risk model, and compare this to ex post volatility 20 day portfolio returns.}
\label{fig:FcvAc}
\end{figure}

Bringing it all together, we list a set of key features of the M6 competition based on the application of forecasts associated with our investment risk model. First, however, we explore how well the risk model captures the realized returns of the submitted investment portfolios. We do this by calculating the realized variance of each submitted portfolio over each 20 day test period, and then comparing this with our (model based) forecast volatility as at submission date. Figure \ref{fig:FcvAc} plots outcome volatility vs. actual volatility for each portfolio at all submission points. We note that there is a relatively stable relationship between risk forecast and risk outcome, with a margin of error increasing steadily with the level of risk assumed. It is also clear from this figure that participants submitted portfolios with risk profiles substantially higher than those typically associated with institutionally managed portfolios.

Now, consider the distribution of forecast portfolio volatility forecast errors (i.e, forecast volatility - actual volatility), set out in Table \ref{tab:vol_fcast_errs}. We do this separately for all submitted portfolios, and for portfolios with a more conventional risk profile (here we choose a realized volatility of less than 10\% to denote `conventional risk').

\begin{table}[h]
\centering
\caption{Distribution of volatility forecast errors for M6 Portfolios.} 
\begin{tabular}{ c c c c c c }
         \hline
 Portfolios & Mean & SD & 25\% & 50\% & 75\% \\
 \hline
 All & 1.97 & 5.36 & -1.3 & 1.37 & 4.89\\
 Conventional risk & -0.05 & 2.35 & -1.36 &0.07 & 1.39\\
  \hline
\end{tabular}
\label{tab:vol_fcast_errs}
\end{table}

These results demonstrate a negative bias (portfolio risk is on average under-forecast by 1.95\%) across all portfolios, albeit with a substantial margin of error. Focusing on the more conventional portfolios, we note that this bias all but disappears.

\begin{figure}[h]
\centering
\includegraphics[scale=.65]{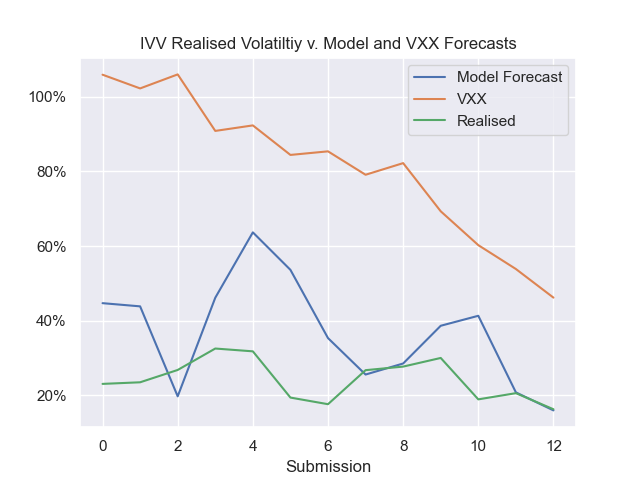}
\caption{Volatility forecast for the IVV S\&P500 ETF as at each submission point from our risk model, 'Market' forecast of S\&P500 Volatility based n the price of the VXX ETF, along with realized ex-post daily volatility (measured over each separate 20 day evaluation period) }
\label{fig:IVVvVXX}
\end{figure}

We now explore how well our model forecasts volatility for individual assets. An interesting exercise in this context involves examining (annualized) volatility forecasts for the S\&P500 ETF (IVV), as these can be compared to a `market forecast' (i.e., the Short Term VIX Futures ETF (VXX) provides a market price for S\&P500 annualized volatility approximately one month ahead)\footnote{https://www.spglobal.com/spdji/en/indices/indicators/sp-500-vix-short-term-index-mcap/$\#$overview.}. Consider the volatility plots in Figure \ref{fig:IVVvVXX}. Evidently, the volatility forecasts embedded in the VXX index were somewhat too high for the period, as were, to a lesser extent, forecasts from our risk model. 

In the following subsections we summarize some of the key features of the M6 competition that arise when comparing the results from our model with participants competition entries.

\subsubsection*{Risk profile of the M6 portfolio submissions}

As shown above, participants in general assumed substantial levels of investment risk in their portfolio submissions. It is of interest to examine the levels of risk assumed in more detail. Throughout this discussion, and unless otherwise noted, we focus on active M6 \emph{submissions} made as at a particular point (as opposed to portfolio entries carried forward from previous submission points).

\begin{figure}[h]
\centering
\includegraphics[scale=.65]{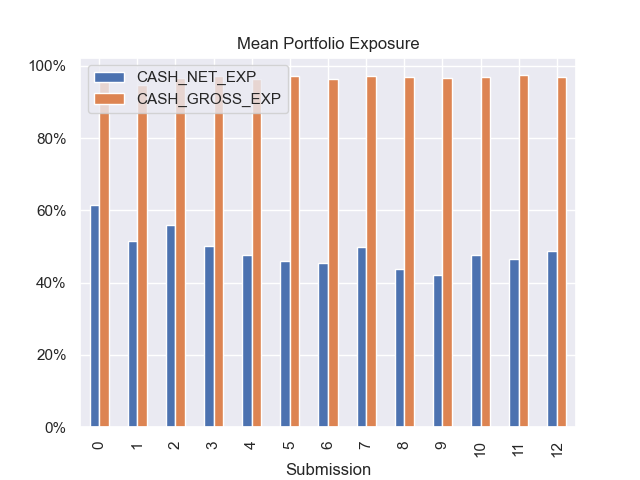}
\caption{Net and Gross cash exposure of M6 Portfolios by submission date. Net exposure is the sum of all portfolio weights. Gross exposure is the sum of all absolute portfolio weights. Submission 0 corresponds to the Trial Period.}
\label{fig:exposure}
\end{figure}

First, we examine the levels of market exposure. Note that competition guidelines were that to be eligible for prizes, participants needed to have an absolute level of between 25\% and 100\% of their notional assets invested at each submission point. Without use of a risk model, we can examine the degree of net market exposure (the sum of all positions) and `gross' market exposure (the sum of the absolute value of all positions) as a first approximation to the level of aggressiveness of a portfolio. Managing gross portfolio exposure is generally the simplest tool to control risk in a hedge fund style portfolio. Exposure figures are contained in Figure \ref{fig:exposure}.

\begin{figure}[h]
\centering
\includegraphics[scale=.65]{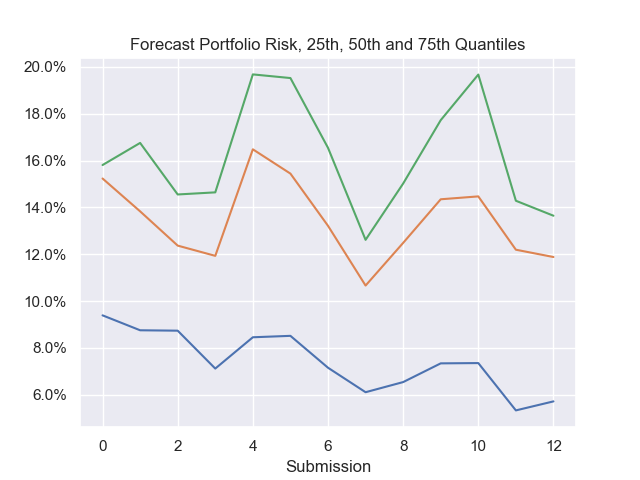}
\caption{Evolution of forecast volatility for M6 portfolio submissions. At each submission point we produce a risk forecast for each M6 portfolio. The chart illustrates the 25\textsuperscript{th}, 50\textsuperscript{th} (median) and 75\textsuperscript{th} percentile of this cross-sectional distribution at each submission point. Submission 0 corresponds to the Trial Period.}
\label{fig:risk ev}
\end{figure}

Inspection of this bar chart indicates that participants generally maintained a significant gross portfolio exposure throughout the competition, although total market exposure tended to decline throughout over time. Recall that participants were not required to assume any directional market exposure. Focusing on cash exposure takes no account of predictable variation in the level of volatility of each asset, nor of correlation between assets. We can sharpen this analysis by using the risk model to make a direct forecast of portfolio risk using our estimates of these quantiles. Figure \ref{fig:risk ev} shows the evolution of the level of risk in the submissions across the course of the competition. Here we plot the level of risk assumed for the 25\textsuperscript{th}, 50\textsuperscript{th}, and 75\textsuperscript{th} quantiles of the distribution of all portfolios (as opposed to fresh submissions made at each point). It becomes evident that the risk taken by the participants is constantly much higher than the predicted one.

\begin{figure}[h]
\centering
\includegraphics[scale=.65]{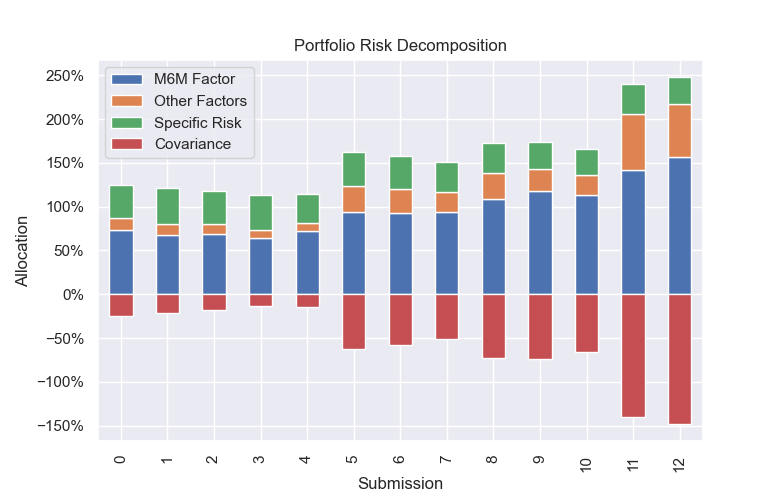}
\caption{Alongside each risk forecast used in Figure \ref{fig:risk ev}, we decompose the corresponding portfolio \emph{variance} in to 4 components as described in the main text. The chart shows the time series evolution of this decomposition of risk for the distribution of M6 portfolios. Submission 0 corresponds to the Trial Period.}
\label{fig:risk qtiles}
\end{figure}

Further insight can be gleaned from analyzing portfolio risk decomposition using the model. We are able to decompose forecast portfolio variance by source of risk. In the following, we do so by splitting \emph{variance} (which is additive, whereas of course standard deviation is not) into 4 components; exposure to the M6M factor, exposure to other systematic factors, specific risk (uncorrelated with systematic factors), and a covariance effect. These results are summarized in Figure \ref{fig:risk qtiles}. On initial inspection of these results, we suspected that the change in risk profile during the competition may have been driven by 'rolling forward' of previous submissions where these were not being actively updated by participants (see Table \ref{tab:stats_sub}). However, although we repeated the analysis for explicitly updated submissions, we discovered a similar patter to that shown in Figure \ref{fig:risk qtiles}.

We see from this chart that risk contribution from diversification tended to increase across the course of the competition, whereas the proportion of specific risk assumed by participants remained broadly constant. This finding leads to three questions about how participants chose to structure their portfolios: How much investment risk did participants assume, where did this risk come from, and how did this change over time? We address these questions in the sequel.

\subsubsection*{Portfolio optimization based on forecast submissions}

Here we examine whether participants could have improved the performance / Sharpe ratios of the portfolios they submitted using the risk information from our model. We construct notional portfolios using only the forecast information submitted by participants and the risk estimates from our model. We compare the characteristics and out-of-sample performance of these notional portfolios to the portfolio submissions made by competition participants. To do so, we proceed as follows, adopting the procedure largely as set out in \citet{Grinold1999-it}.

First, we score each asset at each submission point, by taking the product of the vector $[1,2,3,4,5]$ with the vectors containing the forecast submissions of each participating team. We then standardize these scores across assets for each participant. In several cases, participants submitted identical scores for each asset, meaning that the cross sectional standard deviation of the scores becomes equal to zero. In such cases, we replaced the standard deviations of the scores by the standard deviation of the set $[1,2,3,4,5] \approx 1.41$.

Second, we assume an investment coefficient (IC). This is a number representing the expected correlation between scores and outcomes. This  is a measure of the `edge' of an investor - the extent to which their forecasts add value relative to the consensus - and plays a key role in the mechanics set out in \citet{Grinold1999-it}. Because competition among investors makes active management difficult, ICs tend to be close to zero in practice, although efficient implementation and careful risk control can extract value from even a small positive IC. The interested reader is referred to \citet{Grinold1999-it} for extensive further discussion. For the analysis below, we assume that IC takes three values 0.05 (Moderate), 0.1 (Good) and 0.15 (Exceptional)  and produce portfolios based on each value.

Third, we take the (annualized) expected 20 day volatility $\hat{\sigma}_{20}$ for each asset. The linear Bayes refined return estimate for each asset $E(r_i|g_i)$ is now a function of the prior expectation $r_i$ (which we take to be zero) and its forecast $g_i$  for each asset / participant is given by:
\begin{equation}
\label{LB eqn}
    \alpha_i = E(r_i|g_i) = E(r_i) + Cov(r_i,g_i) Var^{-1}(g_i) (g_i - E(g_i)).
\end{equation}

\noindent We have $E(r_i) = 0$ and the second component on the right hand side of equation (\ref{LB eqn}) can be written as $IC \times \sigma_{20} \times Score_i$, where $Score_i$ is the standardized forecast ($(g_i-E(g_i))/Std(g_i)$) and $IC$ is the assumed correlation between $r_i$ and $g_i$, as described above. This procedure has the effect of shrinking the return forecasts for the assets towards zero. The degree of shrinkage is governed by the value of $IC$, and appropriate choices of this hyper parameter guard against overly risky portfolios. We then choose a set of weights, $w$, to optimize the portfolio:
\begin{align}
    argmax_w(\mathbf{w}\boldsymbol{\alpha} / \mathbf{w}\boldsymbol{\Sigma}\mathbf{w}'),
\end{align}

\noindent where all quantities are annualized. We use the Sequential Least Squares algorithm in the the \textit{Scipy} Python library, with first derivative of the objective function calculated via the JAX Python library to perform our calculations. We illustrate the results of this exercises by displaying statistics for portfolios aggregated according to the realized IC for each set of submitted forecasts. We estimate the IC for each submission by calculating the rank correlation of the scores with the subsequent returns across each evaluation period for all competition assets. We then group all submissions into quintiles based on realized IC, for each submission period. In the table below, we report median risk, returns, and information ratios for each group, for all cases where our optimization software returned a flag reporting that the procedure had terminated successfully.

\begin{table}[h]
\small
\centering
\caption{Sharpe ratio optimized portfolios. The table sets out the (median) realized returns and volatility of portfolios constructed using our risk model and optimization procedure, compared to those submitted by competition participants. For full details see main text.}
\begin{tabular}{ c c c c c c c c}
\hline
IC  & Realized  & Submission  & Optimal  & Submission  & Optimal  & Submission  & Optimal  \\
Quintile &  IC &  Ex-ante Risk (\%) & Ex-ante Risk (\%)&  Return (\%)&  Return (\%)&  IR &  IR \\
\hline
1 & -0.24 & 10.8 & 0.8 & -12.4 & -3.9 & -1.4 & -3.2\\
2 & -0.08 & 11.7 & 0.9 & -5.2 & -1.6 & -0.8 & -1.4\\
3 & 0.00 & 11.9 & 0.9 & 2.5 & 0.0 & 0.5 & 0.5\\
4 & 0.08 & 11.5 & 0.9 & 4.7 & 1.4 & 0.7 & 1.4\\
5 & 0.22 & 10.3 & 0.9 & 16.5 & 3.7 & 2.5 & 3.3\\
\hline
\end{tabular}
\normalsize
\label{tab:sharpe_opt}
\end{table}

It is clear from a cursory inspection of these results (Table \ref{tab:sharpe_opt}) that the optimization routines produced portfolios with substantially lower levels of risk than those assumed by the competitors, and consequently returns of lower magnitude. In cases where the original forecasts were of low quality, the portfolio optimization routine faithfully mapped these forecasts into under-performing portfolios, with substantially lower information ratios than the original portfolios. It appears that if underlying forecasts are poor, then sub-optimal implementation is, in-fact, optimal. Nevertheless, Table \ref{tab:sharpe_opt} shows that given a set of reasonably accurate forecasts (IC$\geq$.05), the IR can be substantially improved. 

\subsubsection*{Portfolio optimization based on forecast and portfolio submissions}

As noted, the original set of optimal portfolios described above differed materially in risk profile to the submissions made by M6 participants, and although the optimal portfolios delivered superior risk adjusted returns, their absolute returns were relatively low in magnitude, reflecting the level of risk assumed. In an institutional context, portfolios with such low levels of risk are unlikely to generate returns justifying an economically relevant management fee. We therefore undertook a second optimization based analysis. For this exercise (Table \ref{tab:risk_tgt_opt_sharpe}) we again took as inputs the return forecasts submitted by the participants, but supplemented this information with the estimated investment risk of each corresponding submitted portfolio. We then added an additional constraint to the optimization routine, such that the resulting optimal portfolio had a risk level at least equal to that of the submission made by each participant. Note that the optimization problem as set up for this exercise is more challenging than that conducted above, and submissions corresponding to more extreme portfolios are removed from this sample. Moreover, note that the results of Table \ref{tab:risk_tgt_opt_sharpe} involve optimal portfolios that were constructed with returns calibrated using a `Good' IC value of $0.1$ (results for `Moderate' and `Exceptional' IC differ very little from those reported below).

\begin{table}[h]
\small
\centering
\caption{Sharpe ratio optimized portfolios, with risk level targeting. The table sets out (median) realized returns and volatility of portfolios constructed using our risk model and optimization procedure, compared to those submitted by competition participants, grouped by the quintile ranking of the submission IC. In this instance, the optimizer was configured to target the levels of portfolio volatility assumed by participants. For full details see main text.}
\label{tab:risk_tgt_opt_sharpe}
\begin{tabular}{ c c c c c c c c}
\hline
IC  & Realized  & Submission  & Optimal  & Submission  & Optimal  & Submission  & Optimal  \\
Quintile &  IC &  Ex-ante Risk  (\%) &  Ex-ante Risk  (\%)  &  Return  (\%) &  Return  (\%) &  IR &  IR \\
\hline
1 & -0.24 & 10.8 & 6.2 & -12.4 & -20.9 & -1.4 & -2.8\\
2 & -0.08 & 11.7 & 5.9 & -5.2 & -7.9 & -0.8 & -1.2\\
3 & 0.00 & 11.9 & 5.9 & 2.5 & -0.1 & 0.5 & -0.1\\
4 & 0.08 & 11.5 & 5.8 & 4.7 & 8.3 & 0.7 & 1.6\\
5 & 0.22 & 10.3 & 6.1 & 16.5 & 25.4 & 2.5 & 4.3\\
\hline
\end{tabular}
\normalsize
\end{table}

Evidently, the optimizer failed to obtain the objective of matching the risk profile of the submitted portfolios in many cases. Investigation reveals that this was the case mostly for instances when submitted portfolios had high or very high ($>10\%$) risk forecasts. Despite risk profiles which remained substantially lower than those of the portfolio submissions, where underlying forecasts were of good quality, the optimal portfolios outperformed the submissions in terms of return, and consequently exhibit higher information ratios. In this sense, the optimizer displays the same efficiency as noted above in turning poor quality forecasts into wealth-damaging portfolios.

\subsubsection*{Reverse optimization}

Finally, we attempt to deduce the set of asset return forecasts that would, given the risk model, make each set of portfolio submissions made by M6 participants `optimal'. We then compare these sets of implied asset return forecasts to the forecast submissions. In order to do this, we set sensible upper and lower bounds on implied alphas for each asset. We calculate these using an implied IC of 0.3, a `score' of 3 standard deviations, and a volatility estimate from the risk model. This ensures that for a given score, more volatile assets have higher expected returns. We then optimize to choose a set of $\alpha$ based on the following functional optimization:
\begin{align}
    argmax_{\alpha}(\mathbf{w}\boldsymbol{\alpha} / \mathbf{w}\boldsymbol{\Sigma}\mathbf{w}')
\end{align}

Subsequently, we take the values returned from this optimization, rank them from 1 to 5 and set our forecast submission equal to 1 for rank of each asset. We first examine the cross sectional correlation of the reverse optimized scores with those actually submitted by participants. In particular, for each submission we calculate the correlation coefficient between the submitted and reverse optimized rankings. The results are set out in Table \ref{tab:rev_opt}.

\begin{table}[h]
\centering
\caption{Reverse optimized portfolios. Implied portfolio $\alpha$ correlation with actual submitted forecasts. See main text for full details.}
\begin{tabular}{ c c c c c c }
 \hline
Mean & SD & 25\% & 50\% & 75\% \\
\hline
0.32 & 0.38 & -0.02 & 0.29 & 0.64 \\
\hline
\end{tabular}
\label{tab:rev_opt}
\end{table}

These findings indicate a relatively weak, but positive relationship between the two sets of rankings, suggesting that participants' portfolios reflect, on average, their forecasts. This is confirmed by a highly significant t-statistic of 117 on the reverse optimized $\alpha$ obtained by regressing the submitted forecasts against the this (plus a constant) across all submissions. 

\subsection{Analysis of winning submissions}

In this section we briefly analyze the performance of the winning submissions. We select and analyze separately the top ten competitors by RPS and IR.  We find that the top ranked forecasters (best RPS) were in general well calibrated in terms of investment decision making; they submitted portfolios with lower than average levels of investment risk. Our optimized portfolios were still able to demonstrate significant improvements in IR however, both by virtue of reduced risk and increased returns (Table \ref{tab:RPS_top}).

\begin{table}[h]
\centering
\caption{Statistics for the top 10 competitors, ranked by RPS across the entire competition. The table shows the median Ex-ante \& Ex-ante volatility, Ex-post return and IR for the top 10 competitors raked by RPS (across the entire submission history), compared to re-optimised portfolios constructed using the risk model and the submitted set of forecasts.}
\begin{tabular}{ c c c c c c }
\hline
& Ex-ante Volatility (\%) & Ex-post Volatility  (\%) & Ex-post Return  (\%) & Ex-post IR \\
\hline
Submission (Mean) & 7.2 &8.4 & 5.3 &0.23\\
Re-optimized (Mean) & 5.9 &6.5 & 5.6 & 0.87\\
Submission (Median) & 4.5 &5.7 & 0.9 &0.37\\
Re-optimized (Median) & 4.9 &5.8 & 2.6 & 0.50\\
\hline
\end{tabular}
\label{tab:RPS_top}
\end{table}

We now turn to the winning IR portfolios. We compare the submitted portfolios to the re-optimized portfolios constructed in two different ways. Firstly, we re-optimize based on submitted return forecasts for each submission. Secondly, we re-optimize based on the `reverse optimization $\alpha$' as described above. Once more, winning teams assumed lower levels of portfolio risk than the average participant. We also find that the performance of these portfolios was in general not well explained either by the explicit return forecasts submitted by participants, or by the forecasts implied by the starting portfolio weights. To an extent these participants may have benefited from `luck' in their investment decision making - there is little evidence to support the hypothesis that their positive IRs were generated by superior investment insights.

\begin{table}[h]
\centering
\caption{Statistics for the top 10 competitors, ranked by IR across the entire competition. The table shows the median Ex-ante \& Ex-ante volatility, Ex-post return and IR for the top 10 competitors raked by IR (across the entire submission history), compared to re-optimized portfolios constructed using the risk model and the submitted set of forecasts.}
\begin{tabular}{ c c c c c c }
\hline
& Ex-ante Volatility (\%) & Ex-post Volatility (\%)& Ex-post Return (\%)& Ex-post IR \\
\hline
Submission (Mean) & 9.4 & 10.4 & 32.0 & 2.05\\
Re-optimized (Mean) & 5.8 & 6.7 & 5.2 & 0.65\\
Reverse $\alpha$ (Mean) & 6.3 & 6.1 & 6.7 & 0.92\\
Submission (Median) & 6.5 & 8.2 & 6.0 & 1.15\\
Re-optimized (Median) & 5.3 & 5.7 & -0.0 & -0.12\\
Reverse $\alpha$ (Mean) & 5.5 & 5.3 & 1.0 & 0.18\\
\hline
\end{tabular} 
\label{tab:IR_top}
\end{table}

\subsection{Final remarks based on the findings of our investment risk model}

The M6 competition guidelines asked competitors explicitly for return forecasts on a selection of assets, and for investment portfolio decisions. However, the link between the two parts of the competition was not specified, and competitors were left to make their own choices as to how to tie forecasts to investment decisions. Moreover, some incentive was given to participants to properly `link' the two parts of their submission, given that fiduciary prizes were allotted to the `best' forecasts, investment decisions, and overall performance. In this section we discussed a risk model that allowed us to assess the strength of the connection between participants forecasts and investment decisions. Our findings can be summarized as follows. Namely, we find that participants assumed significantly greater risk in their portfolios than was justified by the accuracy of their forecasts. For the most accurate sets of return forecasts, portfolio optimization along with our risk model was able to produce portfolios with substantially better returns and much lower levels of risk. On the other hand, when return forecasts were poor, our model based optimization efficiently translated these in to wealth destroying portfolios. This is because assigning appropriate levels of risk to assets that are `poorly' forecast leads to substantially sub-optimal investment portfolio weight selection.

Our overall characterization of these results is that sub-optimal portfolio construction added a (large) additional random component to achieved portfolio returns. Some 'lucky' participants made relatively poor forecasts, but benefited when this random component turned out positive ex-post. Other competitors made relatively accurate forecasts, but poor implementation and 'bad luck' (negative ex-post return realizations) led to poorly performing portfolios.

\section{Major findings and insights}

Below is a summary of the findings related to the analysis of the submission data, the evaluation of the ten hypotheses made, the performance of the top ranked methods, and the results obtained through the developed risk model:

\textbf{Finding 1: The challenging task of forecasting the relative performance of assets.}

Due to the volatile nature of tradable assets, rendering the prediction of their exact prices a hard task, the M6 competition focused on forecasts that resemble the relative performance of the assets. Unfortunately, our results clearly demonstrate that producing such forecasts is far from straightforward either. 

The benchmark selected for the forecasting track of the competition assigned equal probabilities to all five quintiles, assuming that all assets are equally probable to realize relatively higher or lower percentage returns. Although simplistic, less than 25\% of the teams managed to estimate the requested probabilities more precisely than the benchmark. Moreover, those that did so overall, improved the accuracy by less than 2.5\% and found it difficult to outperform the benchmark consistently. It is indicative that only 3 teams performed better than the benchmark in all 12 months of the competition and less than 15 teams in more than 9 months. Our analysis also showed that participants had unwarranted certainty for the outcomes of their forecasts, thus being overconfident in general. 

Note that this finding does not suggest that the accuracy improvements achieved by the top performing teams were minor. On the contrary, it highlights the inherit uncertainty of the requested forecasts that makes any improvement challenging to achieve, especially in the long run. Therefore, it is encouraging that, according to our experiments, even minor accuracy gains can result in significant improvements in terms of IR. For instance, our analysis shows that re-optimizing the portfolios of the top ten performing teams according to RPS based on the developed risk model, improved on average the IR from 0.23 to 0.87. Similarly, our results indicate that re-optimizing portfolios of high IC values can result to better IR values, which can be realized either by taking more reasonable risks or by optimally translating forecasts into investment weights given a certain level of risk. Interestingly, in the latter case, both the IR and the returns can be improved. 

\textbf{Finding 2: The difficulty of consistently outperforming the market.} 

Although testing the EMH based on a certain sample of teams and a set of submissions that cover a period of one year is challenging, our results confirm that beating the market is particularly difficult. 

Focusing on the ``Global'' investment performance of the teams, about 60\% of the participants realized negative returns that usually exceeded 7\% and reached up to 46\%. In addition, less than one third of the teams have managed to outperform the benchmark. Moreover, a very small group of participants have managed to 
consistently outperform the market. It is impressive that none of the teams has achieved higher IR than the benchmark in all 12 months of the competition and just 4 have beat the market in more than 8 months. 

Our analysis, apart from confirming the EMH, also demonstrates that the risks taken by the participants could rarely be justified by the returns of the constructed portfolios and that, when there were any positive returns, these were often marginal, meaning that in the long run they could be diminished. The fact that the majority of the teams tended to perform better than the benchmark when the market was bearish is also indicative, suggesting that most of the participants have failed either to precisely short assets when losing value or to construct portfolios that are robust is such periods. 

On the positive side, some teams have managed to beat the market to a significant extent, realizing impressive returns that surpassed 30\%. Although these were pleasant, rare exceptions and, according to our analysis, they may also be subject to ``luck'', we hope that we can offer valuable lessons. 

\textbf{Finding 3: The limited connection of the submitted forecasts and investment decisions as well as the potential benefits of their association.} 

One of the primary objectives of the M6 competition was to investigate the value that accurate forecasts can add to investment decisions. In this regard, participants were asked to predict the relative performance of some tradable assets and, based on said predictions, construct portfolios.

The analysis conducted to test the 3\textsuperscript{rd} of our hypotheses demonstrates that a limited number of teams actually chose to exploit their forecasts to define the weights of their investments. When analyzing the complete set of submissions, we found no connection between the RPS and IR scores. Moreover, although we found some association between the two scores for the top 20\% of the teams (according to OR), this connection diminished when more than 40\% of the teams were considered or when our focus turned on the duathlon winners. What was even more interesting, was the fact that the top performing teams in the forecasting challenge have constructed relatively inefficient portfolios on average, while the top performing teams in the investment decisions challenge have submitted forecasts of various accuracy levels. 

The analysis on our 4\textsuperscript{th} hypothesis is also relevant to this topic. By investigating the investment weights assigned to assets predicted more accurately, we found that there is no evidence that the top performing teams (according to the IR) built their portfolios by primarily using assets that they could forecast more accurately. In fact, the weights assigned to all assets were similar on average, regardless how accurately the participants could forecast their relative performance. Interestingly, this was true even for the teams that, overall, tended to put higher/lower weights to assets that were predicted to have higher/lower returns.  

These findings indicate that most of the teams probably decided to work on the forecasting and investment tracks of the competition individually, using approaches that are barely connected. Nevertheless, they do not imply that better forecasts cannot assist towards more profitable investments. In fact, based on the analysis conducted using our risk model, it becomes evident that investment decisions can be improved, provided of course that the risk is properly measured and the forecasts are of adequate quality (since the portfolio optimization routine is aligned with the forecasts, inaccuracies will inevitably result to under-performing portfolios).

\textbf{Finding 4: The value added by information exchange and the ``wisdom of crowds''.} 

Aggregating information in groups has been shown to result to better decisions than those made by individuals, while combining forecasts of different methods has long been regarded as one of the most successful strategies for improving accuracy. The analysis conducted for testing our 8\textsuperscript{th} hypothesis is clearly aligned with these expectations. 

Our results show that the averages of the forecasts submitted by the top performing teams (according to the RPS) consistently outperform the forecasts made by the best performing participant. In addition, the accuracy of such combinations remains similar to that of the most accurate submission, even in extreme cases where about half of the teams are included in the aggregation process. Moreover, said combinations always outperform the benchmark. 

In a similar manner, averaging the investment decisions of the top performing teams (according to the IR) results to significantly better performance than the best performing team, while averaging the investment weights of multiple teams can also outperform the benchmark, at least when the worst performing teams are excluded from the combination.

These findings highlight the robustness of forecast combinations and decisions aggregation, confirming the beneficial effects that the ``wisdom of crowds'' can offer. 

%\textbf{Finding 5: The dominance of relatively simple and standard methods in the investment challenge.} 
%\textcolor{red}{To be drafted once we receive the methodological papers and read the posts made by the winners.}

\textbf{Finding 5: The positive effect of adapting to changes.}

Before the competition started, we hypothesized that teams that would employ consistent strategies throughout the competition would perform better than those that changed their strategies significantly from one submission point to another. However, our results suggest otherwise. By focusing on the top performing teams, we found that changing investment strategies proved to be beneficial overall. Specifically, the winning team of the investment challenge was found to have changed its strategy 8 times, while most the top performing ones, more than once. 

In a similar fashion, we found that teams that updated their submissions on a regular basis tended to perform better than those that did not. In particular, all five winners in the forecasting track and four of the five winners in the investment track updated their submission at every single round, while the same was true for most of the duathlon winners.

Adapting based on external information and judgment was also proved to have a positive effect on forecasting and investment performance. Although few in number, judgment-informed forecasting approaches were found to perform better or similarly well to pure data-driven approaches, suggesting that when judgment is utilized correctly, it can offer good performance. 

Although these findings cannot be generalized for the complete set of teams and submissions, they do provide sufficient empirical evidence that adapting to changes can have a positive effect, both to forecasting and investment performance. What is critical, however, is to carefully balance the risk that said changes imply with the expected improvements.  

\section{Discussion, limitations, and directions for future research}

Following the rich and successful tradition of the past Makridakis (M) competitions, M6 was overall a success. Following the M5 forecasting competition, we set up to empirically investigate an interesting and specific problem: how well forecasters and investors can forecast various tradable assets and set their investment weights. Makridakis’ forecasting competitions have always been in the forefront of forecasting research, pushing boundaries, setting research agendas, and shaping the field. The M6 competition was no different - its impact to the community is already evident.

\subsection{Discussion}
Compared to past forecasting competitions \citep{Makridakis2021-xa}, we implemented three key changes in our competition design. First, the M6 competition was a live competition, meaning that participants collected the most up-to-date data and submitted forecasts in real time. Their forecasts were evaluated against the actuals as time progressed, and then they submitted their forecasts for the next period. This cycle was repeated twelve times (over almost one calendar year) to achieve certain levels of robustness. Second, the nature of the forecasting task differed compared to past competitions. In the past, the objective was to submit forecasts for the future values of each target variable. In the M6 competition, the forecasting task involved forecasting the relative performance of different tradable assets. Third, this competition did not stop in evaluating the performance of the forecasting task, but also directly evaluated the implications of the forecasts by measuring the performance of the investment decisions.

One of the key results of the competition was the inability of participants to significantly outperform the forecasting benchmark (that assumed equal probabilities) in predicting the probabilities for the returns quintiles of the various assets. Despite that, participants were able to form efficient portfolios that balanced returns with risk. We observed a considerable disconnect between forecasting performance and investment performance, where there was almost zero correlation between the two if all participants were taken into account. Our empirical results confirmed the EMH, with only one team outperforming the investment benchmark (which referred to an equal-weighted portfolio across all assets with long positions) across all 4 quarters, while none of the teams managed to outperform the investment benchmark across all 12 months of the competition. Another key finding that reiterates the results of past forecasting competitions is that the use of forecast combinations (or aggregation, in this case through ``wisdom of crowds'') is a simple but efficient approach to improve performance. Finally, again in-line with insights from past M competitions, simple and standard methods can yield good performance overall.

The disconnect of the performance between the two tasks, the forecasting and the investment challenge, was an interesting finding that was discussed during the presentation of the results in the 2023 International Symposium on Forecasting. Some attendees argued that it is likely that participants simply focused on making informed investments and simply avoided producing any forecasts. We disagree with this argument. We strongly believe that any decision, in this case the investment weights for the different assets, requires a forecast. This forecast is sometimes implicit and, thus, not properly quantified. However, it is still a forecast: one opts for a high weight and long position in a particular asset only because they implicitly forecast that asset will perform better compared to other available assets. In fact, by construction, M6 enabled the submission of either qualitative (i.e, pure judgmental) or quantitative forecasts, however seemingly some participants bypassed the forecasting task completely, failing to efficiently record what a forecast, which implicitly or explicitly informed their investment decisions, could look like. Our results in section \ref{sec:riskmodel} clearly show that if participants have used a realistic risk model (as the one we present in this paper) then there would be a clear link between forecasting and investment performance. Submissions with forecasting performance in the best or second-best quintile could have yielded a higher investment performance (as measured by the IR) and higher return.

\subsection{Limitations}
One of the limitations of this competition was its duration. The competition involved twelve non-overlapping submission points, each 28 days after the previous one. As a result, the competition ran for almost one year. This had a negative impact on both the total number of participating teams as well as engagement. In fact, we anticipated this negative impact and we attempted to mitigate it by offering rich monetary awards for the top-performing teams (both overall and per quarter). However, still, only a small percentage of the teams updated their forecasts regularly during the duration of the competition, while many teams simply submitted once or twice at the very beginning of the competition. Could we have designed the competition such that it lasted less than a year to increase engagement and decrease drop-outs? Probably not. On the contrary, we believe that, in evaluating the EMH, it would be even better to have the competition running for a longer time period, possibly 2-5 years, to be able to include significant cycles in the economy. As a result, we believe that while neither the competition’s duration nor the achieved participation were perfect, we managed to achieve a good balance between these two aspects.

The second limitation of the M6 competition design has to do with the selected one hundred assets to form the investment universe. As with any empirical exercise that uses a finite and finely defined set of data, one can argue that the results of the competition are limited to its data. In an ideal scenario, we could have included all possible investable assets available worldwide. This, though, would render the exercise impractical from the participants’ point of view, as the forecasting and investment task would be too complex. Another consideration would be to just allow participants to focus (and provide forecasts of) only specific subsets of the one hundred assets (for instance, stock only or ETFs only). We decided against doing so, as this would inevitably have led us to have more distinct evaluation categories (resulting in a further split on the prizes) as the forecasting performance of one subset (such as stocks) would not be comparable to that of the other subset (such as ETFs), not allowing us to rank all participants together.

The third limitation of the M6 competition was that, on top of the “duathlon” prizes, we offered prizes for each of the tasks/challenges separately. This meant that participants could focus exclusively on one of the challenges (forecasting or investment), receive significant awards, and ignore the other challenge. For instance, the top-performing teams in the forecasting task did not perform very well in the investment task. As a consequence, this limited our learning opportunities with regards to the actual connectedness between forecasting and investment performance. Our post-competition analysis suggested that if the top-performers in one challenge made use of state-of-the-art risk models, then they could have performed much better in the other challenge compared to their actual performance.

The final limitation of the M6 competition has to do with the limited opportunities it offers for reproducing or even replicating the forecasts and investment decisions of the participating teams. Given the nature of the competition (financial forecasting), and in order to attract participation from expert financial analysts, we decided to only ask for a brief, bird's eye view, description of the participants’ forecasting and investment approach. The main argument was that an expert analyst would not participate in such a competition if they had to fully “expose” their approach as this would render them “obsolete”. Making the submission of a detailed description of the teams’ approaches, or even the submission of the respective code, a prerequisite for participation would allow us to replicate the submissions but, at the same time, would decrease participation from experts and would potentially render submissions based on pure judgment not viable.

\subsection{Directions for future research}
We believe that the live aspect of the M6 forecasting competition offered a unique take on forecasting competitions. We, as the organizers, did not need to conceal the data or their source and did not need to define a suitable hold-out sample. At the same time, the participants had access to all the information related to the data and were able to also apply, apart from quantitative approaches, judgmental approaches either exclusively or in conjunction with formal models. An additional advantage is that live competitions will allow us to be in a position to evaluate the domain expertise of the participants on top of the technical expertise. We would like to see more forecasting competitions moving away from concealed data and hold-out samples towards live set-ups using real-time data. Associated challenges include data availability and the imminence of the task (forecasts need to be produced very shortly after the actual values have been published). 

In this and the past M forecasting competitions, participants had to submit the outputs of their approaches (forecasts and decisions). An alternative way forward would be to ask the participants to submit their solutions directly (i.e., code or executable). While this would limit the scope of their approaches to strictly quantitative ones, it would offer a considerable advantage. Competition organizers would be able to run the participants’ solutions directly, being in a position to evaluate their effectiveness over longer time periods, in a rolling origin manner. Also, the submission of solutions instead of outputs would guarantee the reproducibility of the submitted forecasts and decisions.

Three of the most cited M forecasting competitions, M1, M3 and M4, offered a diverse forecasting challenge, encompassing data from various domains (micro, macro, industry, finance, demographic, etc.) and a variety of frequencies (yearly, quarterly, monthly, weekly, daily, hourly). Such competitions allow us to identify methods and approaches that are robust and efficient over a variety of settings and contexts. The latter two competitions, M5 and M6, focused on specific contexts, that of retail and financial forecasting. We believe that focusing on particular industries will allow us to better identify the “horses for the course”. In that sense, we can see future competitions focusing on specific industries such as energy forecasting, pharmaceutical forecasting, and medical forecasting. Equally, more “generalist” competitions push the boundaries of innovation towards the development of generic robust forecasting approaches.

Finally, we would like to reiterate the value of the M forecasting competitions in bridging theory and practice via robust and wide-scale empirical evaluations. As we move forward, we invite companies and industries that face challenges with their forecasting tasks to consider M forecasting competitions as a platform of innovation that will offer them a diverse pool of efficient forecasting solutions that is unique to their data and challenges. 

\bibliographystyle{model5-names}  
\bibliography{main}

\begin{thebibliography}{59}
\expandafter\ifx\csname natexlab\endcsname\relax\def\natexlab#1{#1}\fi
\providecommand{\bibinfo}[2]{#2}
\ifx\xfnm\relax \def\xfnm[#1]{\unskip,\space#1}\fi
%Type = Article
\bibitem[{Alexander(2002)}]{AlexanderC.2002PCMf}
\bibinfo{author}{Alexander, C.} (\bibinfo{year}{2002}).
\newblock \bibinfo{title}{Principal component models for generating large garch
  covariance matrices}.
\newblock {\it \bibinfo{journal}{Economic Notes}\/},  {\it
  \bibinfo{volume}{31}\/}, \bibinfo{pages}{337--360}.
%Type = Article
\bibitem[{Andersen et~al.(2001)Andersen, Bollerslev, Diebold \&
  Ebens}]{ANDERSEN200143}
\bibinfo{author}{Andersen, T.~G.}, \bibinfo{author}{Bollerslev, T.},
  \bibinfo{author}{Diebold, F.~X.}, \& \bibinfo{author}{Ebens, H.}
  (\bibinfo{year}{2001}).
\newblock \bibinfo{title}{The distribution of realized stock return
  volatility}.
\newblock {\it \bibinfo{journal}{Journal of Financial Economics}\/},  {\it
  \bibinfo{volume}{61}\/}, \bibinfo{pages}{43--76}.
%Type = Misc
\bibitem[{Armour(2023)}]{Armour2023-aq}
\bibinfo{author}{Armour, B.} (\bibinfo{year}{2023}).
\newblock \bibinfo{title}{Active funds continue to fall short of their passive
  peers}.
\newblock
  \bibinfo{howpublished}{https://www.morningstar.com/etfs/active-funds-continue-fall-short-their-passive-peers}.
\newblock \bibinfo{note}{Accessed: 2023-09-01}.
%Type = Misc
\bibitem[{Best(2023)}]{Best2023-fz}
\bibinfo{author}{Best, R.} (\bibinfo{year}{2023}).
\newblock \bibinfo{title}{Top 5 positions in warren buffett's portfolio}.
\newblock
  \bibinfo{howpublished}{https://www.investopedia.com/articles/investing/022816/top-5-positions-warren-buffetts-portfolio.asp}.
\newblock \bibinfo{note}{Accessed: 2023-09-01}.
%Type = Article
\bibitem[{Black(1992)}]{Black1992-nw}
\bibinfo{author}{Black, F.} (\bibinfo{year}{1992}).
\newblock \bibinfo{title}{Global portfolio optimization}.
\newblock {\it \bibinfo{journal}{Financial Analysts Journal}\/},  {\it
  \bibinfo{volume}{48}\/}, \bibinfo{pages}{28--44}.
%Type = Article
\bibitem[{Bollerslev(1986)}]{Bollerslev1986-dj}
\bibinfo{author}{Bollerslev, T.} (\bibinfo{year}{1986}).
\newblock \bibinfo{title}{Generalized autoregressive conditional
  heteroskedasticity}.
\newblock {\it \bibinfo{journal}{Journal of Econometrics}\/},  {\it
  \bibinfo{volume}{31}\/}, \bibinfo{pages}{307--327}.
%Type = Misc
\bibitem[{Bollerslev(2009)}]{Bollerslev2009-nh}
\bibinfo{author}{Bollerslev, T.} (\bibinfo{year}{2009}).
\newblock \bibinfo{title}{{{Glossary}} to {{ARCH} ({GARCH)}}, in {{T.
  Bollerslev, T., Russel, J., Watson, M., Volatility and Time Series
  Econometrics: Essays in Honor of Robert F. Engle. Oxford University Press,
  London.}}}
%Type = Article
\bibitem[{Bollerslev et~al.(2018)Bollerslev, Hood, Huss \&
  Pedersen}]{Bollerslev2018-ee}
\bibinfo{author}{Bollerslev, T.}, \bibinfo{author}{Hood, B.},
  \bibinfo{author}{Huss, J.}, \& \bibinfo{author}{Pedersen, L.~H.}
  (\bibinfo{year}{2018}).
\newblock \bibinfo{title}{{Risk Everywhere: Modeling and Managing Volatility}}.
\newblock {\it \bibinfo{journal}{Review of Financial Studies}\/},  {\it
  \bibinfo{volume}{31}\/}, \bibinfo{pages}{2729--2773}.
%Type = Book
\bibitem[{Brockwell(1991)}]{BrockwellPeterJ1991Ts:t}
\bibinfo{author}{Brockwell, P.~J.} (\bibinfo{year}{1991}).
\newblock {\it \bibinfo{title}{Time series : theory and methods}\/}.
\newblock Springer series in statistics (\bibinfo{edition}{2nd} ed.).
\newblock \bibinfo{address}{New York}: \bibinfo{publisher}{Springer-Verlag}.
%Type = Misc
\bibitem[{Buffett(2023)}]{Buffett2023-dr}
\bibinfo{author}{Buffett, W.~E.} (\bibinfo{year}{2023}).
\newblock \bibinfo{title}{Letter 2022 to berkshire shareholders}.
%Type = Article
\bibitem[{Chau et~al.(2020)Chau, Lin \& Lin}]{CHAU2020100741}
\bibinfo{author}{Chau, M.}, \bibinfo{author}{Lin, C.-Y.}, \&
  \bibinfo{author}{Lin, T.-C.} (\bibinfo{year}{2020}).
\newblock \bibinfo{title}{Wisdom of crowds before the 2007–2009 global
  financial crisis}.
\newblock {\it \bibinfo{journal}{Journal of Financial Stability}\/},  {\it
  \bibinfo{volume}{48}\/}, \bibinfo{pages}{100741}.
%Type = Article
\bibitem[{Cheng et~al.(2021)Cheng, Swanson \& Yang}]{Swanson2}
\bibinfo{author}{Cheng, M.}, \bibinfo{author}{Swanson, N.~R.}, \&
  \bibinfo{author}{Yang, X.} (\bibinfo{year}{2021}).
\newblock \bibinfo{title}{Forecasting volatility using double shrinkage
  methods}.
\newblock {\it \bibinfo{journal}{Journal of Empirical Finance}\/},  {\it
  \bibinfo{volume}{62}\/}, \bibinfo{pages}{46--61}.
%Type = Article
\bibitem[{Connor(2019)}]{Connor2019-kf}
\bibinfo{author}{Connor, G.} (\bibinfo{year}{2019}).
\newblock \bibinfo{title}{The three types of factor models: A comparison of
  their explanatory power}.
\newblock {\it \bibinfo{journal}{Financial Analysts Journal}\/},  {\it
  \bibinfo{volume}{51}\/}, \bibinfo{pages}{42--46}.
%Type = Article
\bibitem[{Dai et~al.(2021)Dai, Jia \& Kou}]{DAI2021561}
\bibinfo{author}{Dai, M.}, \bibinfo{author}{Jia, Y.}, \& \bibinfo{author}{Kou,
  S.} (\bibinfo{year}{2021}).
\newblock \bibinfo{title}{The wisdom of the crowd and prediction markets}.
\newblock {\it \bibinfo{journal}{Journal of Econometrics}\/},  {\it
  \bibinfo{volume}{222}\/}, \bibinfo{pages}{561--578}.
%Type = Article
\bibitem[{Doung \& Swanson(2015)}]{Diep1}
\bibinfo{author}{Doung, D.}, \& \bibinfo{author}{Swanson, N.~R.}
  (\bibinfo{year}{2015}).
\newblock \bibinfo{title}{Empirical evidence on the importance of aggregation,
  asymmetry, and jumps for volatility prediction}.
\newblock {\it \bibinfo{journal}{Journal of Econometrics}\/},  {\it
  \bibinfo{volume}{187}\/}, \bibinfo{pages}{606--621}.
%Type = Article
\bibitem[{Engle(2002)}]{Engle2002-if}
\bibinfo{author}{Engle, R.} (\bibinfo{year}{2002}).
\newblock \bibinfo{title}{Dynamic conditional correlation - a simple class of
  multivariate garch models}.
\newblock {\it \bibinfo{journal}{Journal of Business and Economic
  Statistics}\/},  {\it \bibinfo{volume}{20}\/}, \bibinfo{pages}{339--350}.
%Type = Book
\bibitem[{Engle(2009)}]{Engle2009-jk}
\bibinfo{author}{Engle, R.} (\bibinfo{year}{2009}).
\newblock {\it \bibinfo{title}{Anticipating Correlations: A New Paradigm for
  Risk Management}\/}.
\newblock \bibinfo{publisher}{Princeton University Press, New Jersey}.
%Type = Article
\bibitem[{Engle \& Mezrich(1996)}]{Engle_undated-qv}
\bibinfo{author}{Engle, R.}, \& \bibinfo{author}{Mezrich, J.}
  (\bibinfo{year}{1996}).
\newblock \bibinfo{title}{Garch for groups}.
\newblock {\it \bibinfo{journal}{Risk}\/},  {\it \bibinfo{volume}{9}\/},
  \bibinfo{pages}{36--40}.
%Type = Article
\bibitem[{Engle(1982)}]{Engle1982-si}
\bibinfo{author}{Engle, R.~F.} (\bibinfo{year}{1982}).
\newblock \bibinfo{title}{Autoregressive conditional heteroscedasticity with
  estimates of the variance of {U}nited {K}ingdom inflation}.
\newblock {\it \bibinfo{journal}{Econometrica}\/},  {\it
  \bibinfo{volume}{50}\/}, \bibinfo{pages}{987}.
%Type = Article
\bibitem[{Engle \& Kroner(1995)}]{Engle1995-td}
\bibinfo{author}{Engle, R.~F.}, \& \bibinfo{author}{Kroner, K.~F.}
  (\bibinfo{year}{1995}).
\newblock \bibinfo{title}{Multivariate simultaneous generalized arch}.
\newblock {\it \bibinfo{journal}{Econometric Theory}\/},  {\it
  \bibinfo{volume}{11}\/}, \bibinfo{pages}{122--150}.
%Type = Book
\bibitem[{Fama(1969)}]{Fama1969-qj}
\bibinfo{author}{Fama, E.~F.} (\bibinfo{year}{1969}).
\newblock {\it \bibinfo{title}{Papers and Proceedings of the twenty-eight
  annual meeting of the American Finance Association}\/}.
\newblock \bibinfo{address}{New York}.
%Type = Article
\bibitem[{Frazzini et~al.(2018)Frazzini, Kabiller \& Pedersen}]{Frazzini1}
\bibinfo{author}{Frazzini, A.}, \bibinfo{author}{Kabiller, D.}, \&
  \bibinfo{author}{Pedersen, L.~H.} (\bibinfo{year}{2018}).
\newblock \bibinfo{title}{Buffett’s alpha}.
\newblock {\it \bibinfo{journal}{Financial Analysts Journal}\/},  {\it
  \bibinfo{volume}{74}\/}, \bibinfo{pages}{35--55}.
%Type = Misc
\bibitem[{French(2023)}]{French_undated-pg}
\bibinfo{author}{French, K.~R.} (\bibinfo{year}{2023}).
\newblock \bibinfo{title}{{Kenneth R. French Data Library}}.
\newblock
  \bibinfo{howpublished}{https://mba.tuck.dartmouth.edu/pages/faculty/ken.french/data\_library.html}.
%Type = Article
\bibitem[{Garman \& Klass(1980)}]{GarmanMarkB.1980OtEo}
\bibinfo{author}{Garman, M.~B.}, \& \bibinfo{author}{Klass, M.~J.}
  (\bibinfo{year}{1980}).
\newblock \bibinfo{title}{On the estimation of security price volatilities from
  historical data}.
\newblock {\it \bibinfo{journal}{The Journal of Business}\/},  {\it
  \bibinfo{volume}{53}\/}, \bibinfo{pages}{67--78}.
%Type = Article
\bibitem[{Ghysels et~al.(2007)Ghysels, Sinko \& Valkanov}]{Ghysels1}
\bibinfo{author}{Ghysels, E.}, \bibinfo{author}{Sinko, A.}, \&
  \bibinfo{author}{Valkanov, R.} (\bibinfo{year}{2007}).
\newblock \bibinfo{title}{Midas regressions: further results and new
  directions}.
\newblock {\it \bibinfo{journal}{Econometric Reviews}\/},  {\it
  \bibinfo{volume}{26}\/}, \bibinfo{pages}{53--90}.
%Type = Book
\bibitem[{Goldstein \& Wooff(2007)}]{Goldstein2007-it}
\bibinfo{author}{Goldstein, M.}, \& \bibinfo{author}{Wooff, D.}
  (\bibinfo{year}{2007}).
\newblock {\it \bibinfo{title}{Bayes Linear Statistics: Theory and Methods}\/}.
\newblock \bibinfo{publisher}{John Wiley \& Sons, New York}.
%Type = Article
\bibitem[{Gottschlich \& Hinz(2014)}]{GOTTSCHLICH201452}
\bibinfo{author}{Gottschlich, J.}, \& \bibinfo{author}{Hinz, O.}
  (\bibinfo{year}{2014}).
\newblock \bibinfo{title}{A decision support system for stock investment
  recommendations using collective wisdom}.
\newblock {\it \bibinfo{journal}{Decision Support Systems}\/},  {\it
  \bibinfo{volume}{59}\/}, \bibinfo{pages}{52--62}.
%Type = Book
\bibitem[{Graham(1949)}]{Graham1949-ra}
\bibinfo{author}{Graham, B.} (\bibinfo{year}{1949}).
\newblock {\it \bibinfo{title}{The Intelligent Investor: The Definitive Book on
  Value Investing}\/}.
\newblock \bibinfo{address}{New York}: \bibinfo{publisher}{Harper Business}.
%Type = Book
\bibitem[{Grinold \& Kahn(1999)}]{Grinold1999-it}
\bibinfo{author}{Grinold, R.~C.}, \& \bibinfo{author}{Kahn, R.~N.}
  (\bibinfo{year}{1999}).
\newblock {\it \bibinfo{title}{Active Portfolio Management ({PB})}\/}.
\newblock \bibinfo{publisher}{McGraw Hill Professional, New York}.
%Type = Article
\bibitem[{Hafner \& Reznikova(2012)}]{HAFNER20123533}
\bibinfo{author}{Hafner, C.~M.}, \& \bibinfo{author}{Reznikova, O.}
  (\bibinfo{year}{2012}).
\newblock \bibinfo{title}{On the estimation of dynamic conditional correlation
  models}.
\newblock {\it \bibinfo{journal}{Computational Statistics \& Data Analysis}\/},
   {\it \bibinfo{volume}{56}\/}, \bibinfo{pages}{3533--3545}.
%Type = Article
\bibitem[{Hansen \& Lunde(2005)}]{hansen-lunde}
\bibinfo{author}{Hansen, P.~R.}, \& \bibinfo{author}{Lunde, A.}
  (\bibinfo{year}{2005}).
\newblock \bibinfo{title}{A forecast comparison of volatility models: does
  anything beat a {GARCH}(1,1)?}
\newblock {\it \bibinfo{journal}{Journal of Applied Econometrics}\/},  {\it
  \bibinfo{volume}{20}\/}, \bibinfo{pages}{873--889}.
%Type = Book
\bibitem[{Jondeau et~al.(2007)Jondeau, Poon \& Rockinger}]{Jondeau2007-eh}
\bibinfo{author}{Jondeau, E.}, \bibinfo{author}{Poon, S.-H.}, \&
  \bibinfo{author}{Rockinger, M.} (\bibinfo{year}{2007}).
\newblock {\it \bibinfo{title}{Financial Modeling Under {Non-Gaussian}
  Distributions}\/}.
\newblock \bibinfo{publisher}{Springer Science \& Business Media, New York}.
%Type = Article
\bibitem[{Lassance(2022)}]{Lassance1}
\bibinfo{author}{Lassance, N.} (\bibinfo{year}{2022}).
\newblock \bibinfo{title}{Maximizing the out-of-sample sharpe ratio}.
\newblock {\it \bibinfo{journal}{Available at SSRN}\/}, .
%Type = Article
\bibitem[{Liao \& Fan(2011)}]{Liao1}
\bibinfo{author}{Liao, Y.}, \& \bibinfo{author}{Fan, J.}
  (\bibinfo{year}{2011}).
\newblock \bibinfo{title}{High dimensional covariance matrix estimation in
  approximate factor models}.
\newblock {\it \bibinfo{journal}{Annals of Statistics}\/},  {\it
  \bibinfo{volume}{39}\/}, \bibinfo{pages}{3320--3356}.
%Type = Incollection
\bibitem[{Lichtenstein et~al.(1982)Lichtenstein, Fischhoff \&
  Phillips}]{Lichtenstein1982-fb}
\bibinfo{author}{Lichtenstein, S.}, \bibinfo{author}{Fischhoff, B.}, \&
  \bibinfo{author}{Phillips, L.~D.} (\bibinfo{year}{1982}).
\newblock \bibinfo{title}{Calibration of probabilities: The state of the art to
  1980}.
\newblock In {\it \bibinfo{booktitle}{Judgment under Uncertainty: Heuristics
  and Biases}\/} (pp. \bibinfo{pages}{306--334}).
\newblock \bibinfo{publisher}{Cambridge University Press}.
%Type = Article
\bibitem[{Makridakis et~al.(1982)Makridakis, Andersen, Carbone, Fildes, Hibon,
  Lewandowski, Newton, Parzen \& Winkler}]{MakridakisM1}
\bibinfo{author}{Makridakis, S.}, \bibinfo{author}{Andersen, A.},
  \bibinfo{author}{Carbone, R.}, \bibinfo{author}{Fildes, R.},
  \bibinfo{author}{Hibon, M.}, \bibinfo{author}{Lewandowski, R.},
  \bibinfo{author}{Newton, J.}, \bibinfo{author}{Parzen, E.}, \&
  \bibinfo{author}{Winkler, R.} (\bibinfo{year}{1982}).
\newblock \bibinfo{title}{{The accuracy of extrapolation (time series) methods:
  Results of a forecasting competition}}.
\newblock {\it \bibinfo{journal}{Journal of Forecasting}\/},  {\it
  \bibinfo{volume}{1}\/}, \bibinfo{pages}{111--153}.
%Type = Article
\bibitem[{Makridakis et~al.(1993)Makridakis, Chatfield, Hibon, Lawrence, Mills,
  Ord \& Simmons}]{Makridakis1993-bw}
\bibinfo{author}{Makridakis, S.}, \bibinfo{author}{Chatfield, C.},
  \bibinfo{author}{Hibon, M.}, \bibinfo{author}{Lawrence, M.},
  \bibinfo{author}{Mills, T.}, \bibinfo{author}{Ord, K.}, \&
  \bibinfo{author}{Simmons, L.~F.} (\bibinfo{year}{1993}).
\newblock \bibinfo{title}{The m2-competition: A real-time judgmentally based
  forecasting study}.
\newblock {\it \bibinfo{journal}{International Journal of Forecasting}\/},
  {\it \bibinfo{volume}{9}\/}, \bibinfo{pages}{5--22}.
%Type = Article
\bibitem[{Makridakis et~al.(2021)Makridakis, Fry, Petropoulos \&
  Spiliotis}]{Makridakis2021-xa}
\bibinfo{author}{Makridakis, S.}, \bibinfo{author}{Fry, C.},
  \bibinfo{author}{Petropoulos, F.}, \& \bibinfo{author}{Spiliotis, E.}
  (\bibinfo{year}{2021}).
\newblock \bibinfo{title}{{The future of forecasting competitions: Design
  attributes and principles}}.
\newblock {\it \bibinfo{journal}{{INFORMS Journal on Data Science}}\/},  {\it
  \bibinfo{volume}{1}\/}, \bibinfo{pages}{96--113}.
%Type = Article
\bibitem[{Makridakis \& Hibon(2000)}]{MAKRIDAKIS2000451}
\bibinfo{author}{Makridakis, S.}, \& \bibinfo{author}{Hibon, M.}
  (\bibinfo{year}{2000}).
\newblock \bibinfo{title}{The {M3-Competition}: results, conclusions and
  implications}.
\newblock {\it \bibinfo{journal}{International Journal of Forecasting}\/},
  {\it \bibinfo{volume}{16}\/}, \bibinfo{pages}{451--476}.
%Type = Article
\bibitem[{Makridakis et~al.(2020)Makridakis, Spiliotis \&
  Assimakopoulos}]{MAKRIDAKIS202054}
\bibinfo{author}{Makridakis, S.}, \bibinfo{author}{Spiliotis, E.}, \&
  \bibinfo{author}{Assimakopoulos, V.} (\bibinfo{year}{2020}).
\newblock \bibinfo{title}{{The M4 Competition: 100,000 time series and 61
  forecasting methods}}.
\newblock {\it \bibinfo{journal}{International Journal of Forecasting}\/},
  {\it \bibinfo{volume}{36}\/}, \bibinfo{pages}{54--74}.
%Type = Article
\bibitem[{Makridakis et~al.(2022)Makridakis, Spiliotis \&
  Assimakopoulos}]{MAKRIDAKIS20221325}
\bibinfo{author}{Makridakis, S.}, \bibinfo{author}{Spiliotis, E.}, \&
  \bibinfo{author}{Assimakopoulos, V.} (\bibinfo{year}{2022}).
\newblock \bibinfo{title}{{The M5 competition: Background, organization, and
  implementation}}.
\newblock {\it \bibinfo{journal}{International Journal of Forecasting}\/},
  {\it \bibinfo{volume}{38}\/}, \bibinfo{pages}{1325--1336}.
%Type = Book
\bibitem[{Malkiel(1973)}]{Malkiel1973-to}
\bibinfo{author}{Malkiel, B.~G.} (\bibinfo{year}{1973}).
\newblock {\it \bibinfo{title}{A Random Walk Down Wall Street: The
  {Time-Tested} Strategy for Successful Investing}\/}.
\newblock \bibinfo{address}{New York}: \bibinfo{publisher}{Norton and Company}.
%Type = Book
\bibitem[{Markowitz(1959)}]{Markowitz1959-gy}
\bibinfo{author}{Markowitz, H.} (\bibinfo{year}{1959}).
\newblock {\it \bibinfo{title}{Portfolio Selection: Efficient Diversification
  of Investments}\/}.
\newblock \bibinfo{publisher}{Wiley, {{New York}}}.
%Type = Article
\bibitem[{Molnár(2012)}]{MOLNAR201220}
\bibinfo{author}{Molnár, P.} (\bibinfo{year}{2012}).
\newblock \bibinfo{title}{Properties of range-based volatility estimators}.
\newblock {\it \bibinfo{journal}{International Review of Financial
  Analysis}\/},  {\it \bibinfo{volume}{23}\/}, \bibinfo{pages}{20--29}.
%Type = Article
\bibitem[{Montero-Manso et~al.(2020)Montero-Manso, Athanasopoulos, Hyndman \&
  Talagala}]{MONTEROMANSO202086}
\bibinfo{author}{Montero-Manso, P.}, \bibinfo{author}{Athanasopoulos, G.},
  \bibinfo{author}{Hyndman, R.~J.}, \& \bibinfo{author}{Talagala, T.~S.}
  (\bibinfo{year}{2020}).
\newblock \bibinfo{title}{{FFORMA: Feature-based forecast model averaging}}.
\newblock {\it \bibinfo{journal}{International Journal of Forecasting}\/},
  {\it \bibinfo{volume}{36}\/}, \bibinfo{pages}{86--92}.
%Type = Article
\bibitem[{Parkinson(1980)}]{ParkinsonMichael1980TEVM}
\bibinfo{author}{Parkinson, M.} (\bibinfo{year}{1980}).
\newblock \bibinfo{title}{The extreme value method for estimating the variance
  of the rate of return}.
\newblock {\it \bibinfo{journal}{The Journal of business (Chicago, Ill.)}\/},
  {\it \bibinfo{volume}{53}\/}, \bibinfo{pages}{61--65}.
%Type = Book
\bibitem[{Pedersen(2015)}]{Pedersen2015-on}
\bibinfo{author}{Pedersen, L.~H.} (\bibinfo{year}{2015}).
\newblock {\it \bibinfo{title}{Efficiently Inefficient: How Smart Money Invests
  and Market Prices Are Determined}\/}.
\newblock \bibinfo{publisher}{Princeton University Press}.
%Type = Article
\bibitem[{Petropoulos et~al.(2022)Petropoulos, Apiletti, Assimakopoulos, Babai,
  Barrow, {Ben Taieb}, Bergmeir, Bessa, Bijak, Boylan, Browell, Carnevale,
  Castle, Cirillo, Clements, Cordeiro, {Cyrino Oliveira}, {De Baets},
  Dokumentov, Ellison, Fiszeder, Franses, Frazier, Gilliland, Gönül, Goodwin,
  Grossi, Grushka-Cockayne, Guidolin, Guidolin, Gunter, Guo, Guseo, Harvey,
  Hendry, Hollyman, Januschowski, Jeon, Jose, Kang, Koehler, Kolassa,
  Kourentzes, Leva, Li, Litsiou, Makridakis, Martin, Martinez, Meeran, Modis,
  Nikolopoulos, Önkal, Paccagnini, Panagiotelis, Panapakidis, Pavía, Pedio,
  Pedregal, Pinson, Ramos, Rapach, Reade, Rostami-Tabar, Rubaszek, Sermpinis,
  Shang, Spiliotis, Syntetos, Talagala, Talagala, Tashman, Thomakos,
  Thorarinsdottir, Todini, {Trapero Arenas}, Wang, Winkler, Yusupova \&
  Ziel}]{PETROPOULOS2022705}
\bibinfo{author}{Petropoulos, F.}, \bibinfo{author}{Apiletti, D.},
  \bibinfo{author}{Assimakopoulos, V.}, \bibinfo{author}{Babai, M.~Z.},
  \bibinfo{author}{Barrow, D.~K.}, \bibinfo{author}{{Ben Taieb}, S.},
  \bibinfo{author}{Bergmeir, C.}, \bibinfo{author}{Bessa, R.~J.},
  \bibinfo{author}{Bijak, J.}, \bibinfo{author}{Boylan, J.~E.},
  \bibinfo{author}{Browell, J.}, \bibinfo{author}{Carnevale, C.},
  \bibinfo{author}{Castle, J.~L.}, \bibinfo{author}{Cirillo, P.},
  \bibinfo{author}{Clements, M.~P.}, \bibinfo{author}{Cordeiro, C.},
  \bibinfo{author}{{Cyrino Oliveira}, F.~L.}, \bibinfo{author}{{De Baets}, S.},
  \bibinfo{author}{Dokumentov, A.}, \bibinfo{author}{Ellison, J.},
  \bibinfo{author}{Fiszeder, P.}, \bibinfo{author}{Franses, P.~H.},
  \bibinfo{author}{Frazier, D.~T.}, \bibinfo{author}{Gilliland, M.},
  \bibinfo{author}{Gönül, M.~S.}, \bibinfo{author}{Goodwin, P.},
  \bibinfo{author}{Grossi, L.}, \bibinfo{author}{Grushka-Cockayne, Y.},
  \bibinfo{author}{Guidolin, M.}, \bibinfo{author}{Guidolin, M.},
  \bibinfo{author}{Gunter, U.}, \bibinfo{author}{Guo, X.},
  \bibinfo{author}{Guseo, R.}, \bibinfo{author}{Harvey, N.},
  \bibinfo{author}{Hendry, D.~F.}, \bibinfo{author}{Hollyman, R.},
  \bibinfo{author}{Januschowski, T.}, \bibinfo{author}{Jeon, J.},
  \bibinfo{author}{Jose, V. R.~R.}, \bibinfo{author}{Kang, Y.},
  \bibinfo{author}{Koehler, A.~B.}, \bibinfo{author}{Kolassa, S.},
  \bibinfo{author}{Kourentzes, N.}, \bibinfo{author}{Leva, S.},
  \bibinfo{author}{Li, F.}, \bibinfo{author}{Litsiou, K.},
  \bibinfo{author}{Makridakis, S.}, \bibinfo{author}{Martin, G.~M.},
  \bibinfo{author}{Martinez, A.~B.}, \bibinfo{author}{Meeran, S.},
  \bibinfo{author}{Modis, T.}, \bibinfo{author}{Nikolopoulos, K.},
  \bibinfo{author}{Önkal, D.}, \bibinfo{author}{Paccagnini, A.},
  \bibinfo{author}{Panagiotelis, A.}, \bibinfo{author}{Panapakidis, I.},
  \bibinfo{author}{Pavía, J.~M.}, \bibinfo{author}{Pedio, M.},
  \bibinfo{author}{Pedregal, D.~J.}, \bibinfo{author}{Pinson, P.},
  \bibinfo{author}{Ramos, P.}, \bibinfo{author}{Rapach, D.~E.},
  \bibinfo{author}{Reade, J.~J.}, \bibinfo{author}{Rostami-Tabar, B.},
  \bibinfo{author}{Rubaszek, M.}, \bibinfo{author}{Sermpinis, G.},
  \bibinfo{author}{Shang, H.~L.}, \bibinfo{author}{Spiliotis, E.},
  \bibinfo{author}{Syntetos, A.~A.}, \bibinfo{author}{Talagala, P.~D.},
  \bibinfo{author}{Talagala, T.~S.}, \bibinfo{author}{Tashman, L.},
  \bibinfo{author}{Thomakos, D.}, \bibinfo{author}{Thorarinsdottir, T.},
  \bibinfo{author}{Todini, E.}, \bibinfo{author}{{Trapero Arenas}, J.~R.},
  \bibinfo{author}{Wang, X.}, \bibinfo{author}{Winkler, R.~L.},
  \bibinfo{author}{Yusupova, A.}, \& \bibinfo{author}{Ziel, F.}
  (\bibinfo{year}{2022}).
\newblock \bibinfo{title}{Forecasting: theory and practice}.
\newblock {\it \bibinfo{journal}{International Journal of Forecasting}\/},
  {\it \bibinfo{volume}{38}\/}, \bibinfo{pages}{705--871}.
%Type = Book
\bibitem[{Prado et~al.(2021)Prado, Ferreira \& West}]{Prado2021-xl}
\bibinfo{author}{Prado, R.}, \bibinfo{author}{Ferreira, M. A.~R.}, \&
  \bibinfo{author}{West, M.} (\bibinfo{year}{2021}).
\newblock {\it \bibinfo{title}{Time Series: Modeling, Computation, and
  Inference, Second Edition}\/}.
\newblock \bibinfo{publisher}{CRC Press}.
%Type = Article
\bibitem[{Rogers \& Satchell(1991)}]{RogersL.C.G.1991EVFH}
\bibinfo{author}{Rogers, L. C.~G.}, \& \bibinfo{author}{Satchell, S.~E.}
  (\bibinfo{year}{1991}).
\newblock \bibinfo{title}{Estimating variance from high, low and closing
  prices}.
\newblock {\it \bibinfo{journal}{The Annals of applied probability}\/},  {\it
  \bibinfo{volume}{1}\/}, \bibinfo{pages}{504--512}.
%Type = Article
\bibitem[{Smyl(2020)}]{SMYL202075}
\bibinfo{author}{Smyl, S.} (\bibinfo{year}{2020}).
\newblock \bibinfo{title}{A hybrid method of exponential smoothing and
  recurrent neural networks for time series forecasting}.
\newblock {\it \bibinfo{journal}{International Journal of Forecasting}\/},
  {\it \bibinfo{volume}{36}\/}, \bibinfo{pages}{75--85}.
%Type = Book
\bibitem[{Surowiecki(2005)}]{Surowiecki2005}
\bibinfo{author}{Surowiecki, J.} (\bibinfo{year}{2005}).
\newblock {\it \bibinfo{title}{{The Wisdom of Crowds}}\/}.
\newblock \bibinfo{publisher}{Anchor Books}.
%Type = Article
\bibitem[{Swanson \& Xiong(2018)}]{Swanson1}
\bibinfo{author}{Swanson, N.~R.}, \& \bibinfo{author}{Xiong, W.}
  (\bibinfo{year}{2018}).
\newblock \bibinfo{title}{Big data analytics in economics: what have we learned
  so far, and where should we go from here?}
\newblock {\it \bibinfo{journal}{Canadian journal of Economics}\/},  {\it
  \bibinfo{volume}{3}\/}, \bibinfo{pages}{695--746}.
%Type = Article
\bibitem[{Tashman(2000)}]{TASHMAN2000437}
\bibinfo{author}{Tashman, L.~J.} (\bibinfo{year}{2000}).
\newblock \bibinfo{title}{{{Out-of-sample tests of forecasting accuracy: an
  analysis and review}}}.
\newblock {\it \bibinfo{journal}{International Journal of Forecasting}\/},
  {\it \bibinfo{volume}{16}\/}, \bibinfo{pages}{437 -- 450}.
%Type = Book
\bibitem[{Triantafyllopoulos(2021)}]{TriantafyllopoulosKostas2021BIoS}
\bibinfo{author}{Triantafyllopoulos, K.} (\bibinfo{year}{2021}).
\newblock {\it \bibinfo{title}{Bayesian Inference of State Space Models: Kalman
  Filtering and Beyond}\/}.
\newblock Springer Texts in Statistics.
\newblock \bibinfo{address}{Cham}: \bibinfo{publisher}{Springer International
  Publishing AG}.
%Type = Article
\bibitem[{West(1992)}]{West1992-ml}
\bibinfo{author}{West, M.} (\bibinfo{year}{1992}).
\newblock \bibinfo{title}{Modelling agent forecast distributions}.
\newblock {\it \bibinfo{journal}{J. R. Stat. Soc.}\/},  {\it
  \bibinfo{volume}{54}\/}, \bibinfo{pages}{553--567}.
%Type = Article
\bibitem[{West \& Crosse(1992)}]{West1992-mg}
\bibinfo{author}{West, M.}, \& \bibinfo{author}{Crosse, J.}
  (\bibinfo{year}{1992}).
\newblock \bibinfo{title}{Modelling probabilistic agent opinion}.
\newblock {\it \bibinfo{journal}{J. R. Stat. Soc.}\/},  {\it
  \bibinfo{volume}{54}\/}, \bibinfo{pages}{285--299}.
%Type = Article
\bibitem[{Yang \& Zhang(2000)}]{YangDennis2000DVEB}
\bibinfo{author}{Yang, D.}, \& \bibinfo{author}{Zhang, Q.}
  (\bibinfo{year}{2000}).
\newblock \bibinfo{title}{Drift‐independent volatility estimation based on
  high, low, open, and close prices}.
\newblock {\it \bibinfo{journal}{The Journal of business (Chicago, Ill.)}\/},
  {\it \bibinfo{volume}{73}\/}, \bibinfo{pages}{477--492}.
%Type = Article
\bibitem[{Yardley \& Petropoulos(2021)}]{YardleyP2021}
\bibinfo{author}{Yardley, E.}, \& \bibinfo{author}{Petropoulos, F.}
  (\bibinfo{year}{2021}).
\newblock \bibinfo{title}{Beyond error measures to the utility and cost of the
  forecasts}.
\newblock {\it \bibinfo{journal}{Foresight: The International Journal of
  Applied Forecasting}\/},  (pp. \bibinfo{pages}{36--45}).

\end{thebibliography}

\clearpage

\section*{Appendix A: Construction of the M6 competition universe of assets}

The 50 stocks were selected so that each GICS (Global Industry Classification Standard) sector is represented in M6 in a similar proportion with that in the S\&P500 index. In this regard, the number of assets to be sampled per sector was defined as shown in Table A1.

\begin{table}[h]
\small
\centering
\caption*{Table A1: Stocks included in the S\&P500 and the M6 investment universe per sector.}
\begin{tabular}{lrrrr}
\hline
\multirow{2}{*}{\textbf{Sector}} & \multicolumn{2}{c}{\textbf{S\&P500}} & \multicolumn{2}{c}{\textbf{M6}} \\
&\textbf{Count} & \textbf{Proportion (\%)} & \textbf{Count} & \textbf{Proportion (\%)}\\
\hline
Communication Services&27&5.3&3&6.0\\
Consumer Discretionary&63&12.5&6&12.0\\
Consumer Staples&32&6.3&3&6.0\\
Energy&21&4.2&2&4.0\\
Financial&65&12.9&7&14.0\\
Health Care&64&12.7&6&12.0\\
Industrial&74&14.7&7&14.0\\
Information Technology&74&14.7&7&14.0\\
Materials&28&5.5&3&6.0\\
Real Estate&29&5.7&3&6.0\\
Utilities&28&5.5&3&6.0\\
\hline
\end{tabular}
\end{table}

In November 2021, the 9 key features presented in Subsection \ref{sec:data} were computed for each of the stocks, taking into account the average price of the stock and its volatility, the daily and compound returns, the volatility of the returns, as well as the trading volume. These features where then used to group the stocks of each sector into diverse clusters and make sure that the selected assets would sufficiently represent the market.

To do that, the silhouette method was first employed to define the optimal number of clusters per sector. Then, based on the population size of each cluster, an appropriate number of stocks was randomly sampled per cluster. As an example, Figure A showcases that the 21 stocks of the ``Energy'' sector were grouped into two clusters of 5 and 16 stocks, respectively. Given that the examined sector should contribute 2 stocks in total to the M6 universe, the first cluster should provide $(5/21)*2=0.48\approx0$ stocks, while the second $(16/21)*2=1.52\approx2$ stocks. By randomly sampling two stocks from the second cluster, the COP and XOM stocks were included among the M6 assets.

\begin{figure}[!h]
    \centering
    \includegraphics[width=0.75\textwidth]{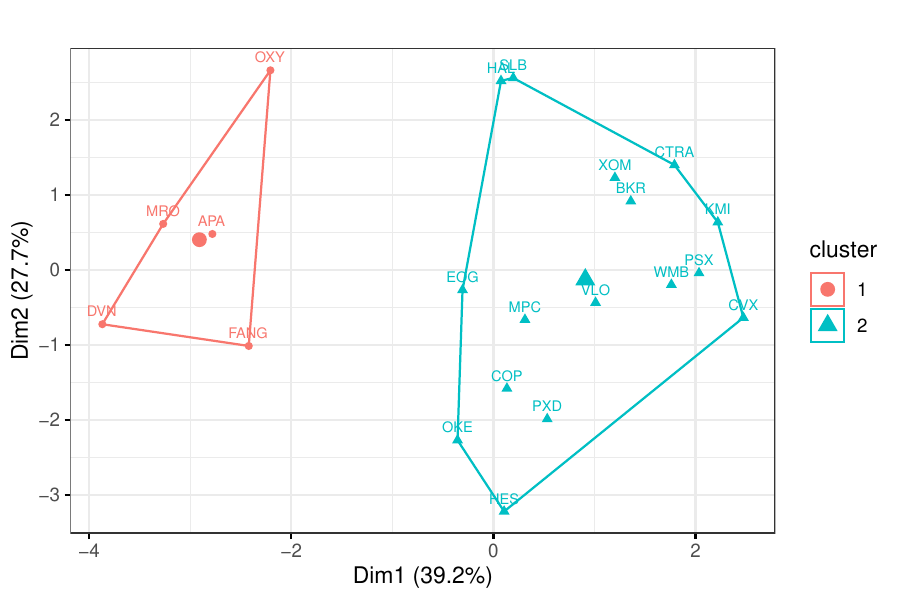}
    \caption*{Figure A: Example illustrating the clustering process taking place for the case of the ``Energy'' GICS sector.}
\end{figure}

ETFs were selected qualitatively with the objective to capture the overall returns of major markets (e.g. US, UK, Germany, Japan, China, India, Australia, Brazil), involve short- and long-term treasury and corporate bond options, allow investing in major sectors (e.g. energy, infrastructures, technology, health)
Provide precious metals and commodity options (e.g. gold and silver), and include equity style options (e.g. value, momentum and minimum volume for the United States of America and Europe). The selected ETFs can be categorized according to their type as shown in Table A2.

\begin{table}[h]
\small
\centering
\caption*{Table A2: ETFs included in the M6 investment universe per type.}
\begin{tabular}{lr}
\hline
\textbf{ETF Type} & \textbf{Count}\\
\hline
Large Cap&15\\
Small Cap&2\\
Sector&14\\
Equity Style&6\\
Government&5\\
Credit&4\\
Precious Metals&2\\
Diversified Commodities&1\\
Volatility&1\\
\hline
\end{tabular}
\end{table}

The 100 assets that were included in the M6 investment universe are presented in Table A3.

\begin{table}[h]
\small
\centering
\caption*{Table A3: Assets included in the M6 investment universe.}
\begin{tabular}{cccccccccc}
\hline
\multicolumn{10}{c}{\textbf{The M6 Investment Universe (Stocks and ETFs)}}\\
\hline
ABBV & ACN & AEP & AIZ & ALLE & AMAT & AMP & AMZN & AVB & AVY \\
AXP & BDX & BF-B & BMY & BR & CARR & CDW & CE & CHTR & CNC \\
CNP & COP & CTAS & CZR & DG & DPZ & DRE\tablefootnote{On October 3, 2022 the DRE price stopped being updated due to the company being acquired by PLD. Therefore, participants had to assume a return of zero for DRE (any investment to DRE would have no effect on portfolio returns and the rank of the asset in terms of returns would have to be forecast assuming no price change).} & DXC & EWA & EWC \\
EWG & EWH & EWJ & EWL & EWQ & EWT & EWU & EWY & EWZ & FB\tablefootnote{On June 9, 2022 the FB identifier changed into META.} \\
FTV & GOOG & GPC & GSG & HIG & HIGH.L & HST & HYG & IAU & ICLN \\
IEAA.L& IEF& IEFM.L& IEMG& IEUS& IEVL.L& IGF& INDA& IUMO.L& IUVL.L \\
IVV& IWM& IXN& JPEA.L& JPM& KR& LQD& MCHI& MVEU.L& OGN \\
PG& PPL& PRU& PYPL& RE& REET& ROL& ROST& SEGA.L& SHY \\
SLV& SPMV.L& TLT& UNH& URI& V& VRSK& VXX& WRK& XLB \\
XLC & XLE& XLF& XLI& XLK& XLP& XLU& XLV& XLY& XOM \\
\hline
\end{tabular}
\label{tab:assetsuniverse}
\end{table}

\section*{Appendix B: Connection of forecasts and investment decisions}

In order to quantify the degree to which the forecasts of the participating teams were connected with the corresponding investment decisions, we introduced a correlation measure, $r_\text{con}$, computed as follows:

\begin{itemize}[noitemsep]
\item For each submission, we multiply the investment weight of each asset with the corresponding forecasts (probability that the asset will be ranked within the first, second, third, fourth and fifth quintile). By doing that, we end up with 100 vectors, each containing 5 elements, \{Rank1, Rank2, Rank3, Rank4, Rank5\}, that represent the expected value of the investment. Elements of high positive and negative values suggest that the uncertainty around the forecast was relatively low or that the invested amount (either long or short position) was relatively large. Accordingly, elements of close to zero values suggest that the uncertainty around the forecast was relatively high or that the invested amount was relatively small.
\item We compute the average of the \{Rank1, Rank2, Rank3, Rank4, Rank5\} elements by aggregating their values across all assets and submissions made by a team throughout the competition.
\item We calculate the correlation, $r_\text{con}$, between the averages computed in the previous step and a vector containing the following integer numbers: \{1,2,3,4,5\}.
\end{itemize}

It becomes evident that if the forecasts of a team are connected with its investment decisions, $r_\text{con}$ should be close to unity (larger amounts of capital are invested at higher ranked assets and lower - or even negative - amounts of capital are invested at lower ranked assets). For instance, the following vector \{-0.036, -0.022, 0.004, 0.019, 0.073\} would score $r_\text{con}=0.96$, while the following one \{0.003,-0.019,-0.019,-0.007,0.027\} would score $r_\text{con}=0.29$  

In this regard, the submissions of the teams where classified into the following categories:

\begin{itemize}[noitemsep]
\item ``Well connected'' ($r_\text{con} \ge 0.75$);
\item ``Connected'' ($0.50 \le r_\text{con} < 0.75$);
\item ``Weekly connected'' ($0.25 \le r_\text{con} < 0.50$);
\item ``Disconnected'' ($-0.25 \le r_\text{con} < 0.25$);
\item ``Opposite connection'' ($r_\text{con} \le -0.25$);
\item ``NA'' (Teams whose forecasts were identical to those of the benchmark and, therefore, $r_\text{con}$ could not be computed).
\end{itemize}

From the 163 teams included in the ``Global'' leaderboard, 39 fell into the ``well connected'' category, 20 into the ``connected'' category, 19 were classified as ``weakly connected'', 42 as ``disconnected'', while 16 had an ``opposite connection''. This result suggests that the majority of the teams developed separate approaches for preparing their submissions for the two challenges of the M6 competition, thus often making investments that can barely be justified based on the provided forecasts. Figure B1 further supports this finding, indicating that although some teams have shorted the assets that fell according to their predictions in the worst category (Rank1) and invested more capital to the assets that fell according to their predictions in the best category (Rank5), in most of the cases, similar investments were made across all five asset classes. 

\begin{figure}[!h]
    \centering
    \includegraphics[width=0.6\textwidth]{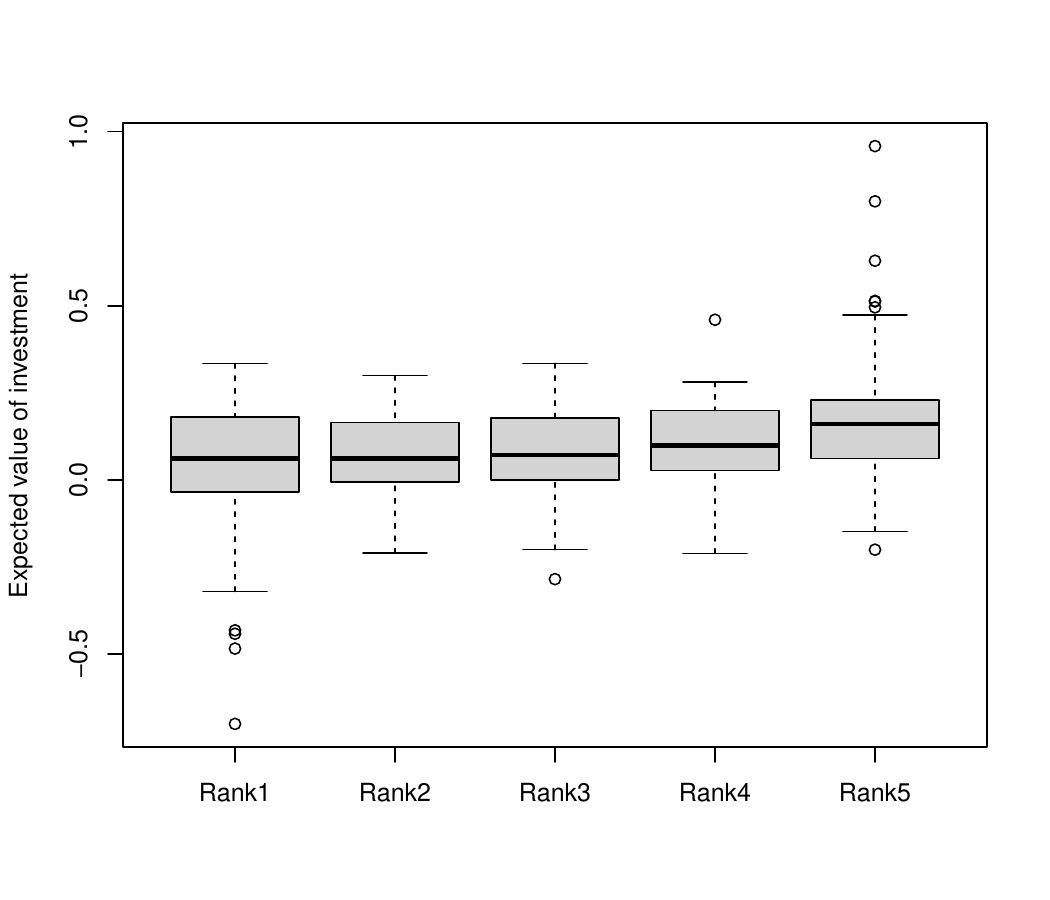}
    \caption*{Figure B1: Distribution of the values computed for the \{Rank1, Rank2, Rank3, Rank4, Rank5\} elements (expected value of investment) for the 163 teams included in the ``Global'' leaderboard.}
\end{figure}

Figure B2 provides further insights on the impact that the degree of connection between the forecasts and the investment decisions had on the performance of the teams. As seen, ``well connected'' submissions tend to involve better portfolios than the ``disconnected'' and ``opposite connection'' portfolios. However, ``weekly connected'' submissions perform similarly well and the IR values largely overlap across the individual classes. The differences in terms of performance are even smaller when it comes to RPS. In fact, we observe that ``well connected'' submissions involve some of the worst sets of forecasts, in contrast to the ``weekly connected'' submissions that perform much better on average.   

\begin{figure}[!h]
    \centering
    \includegraphics[width=0.9\textwidth]{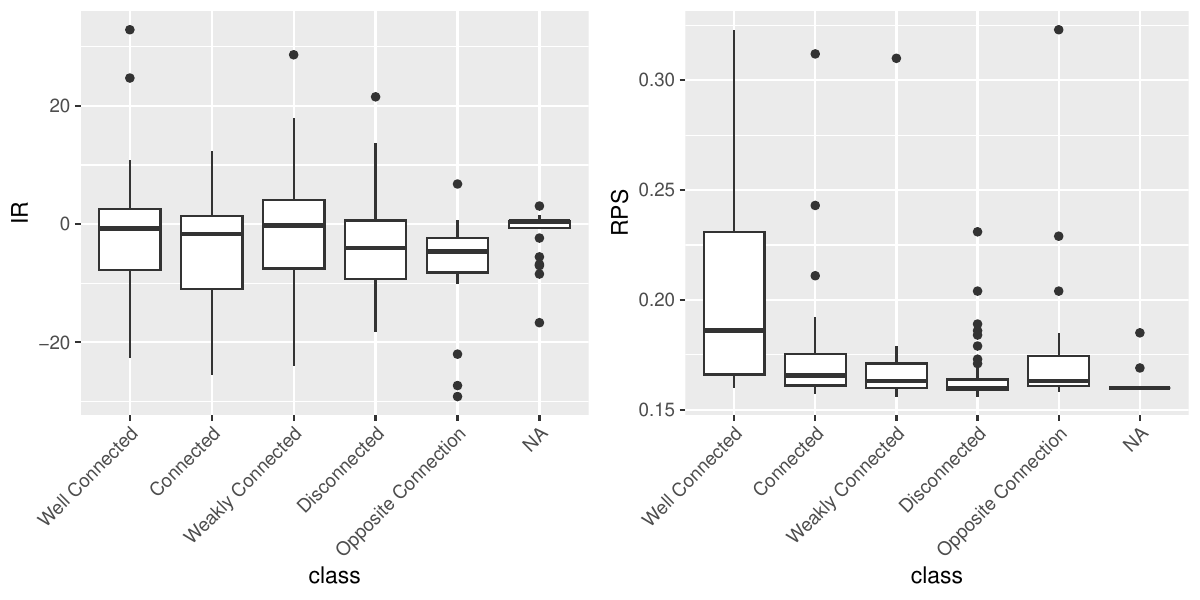}
    \caption*{Figure B2: Distribution of the IR and RPS scores achieved by the 163 teams included in the ``Global'' leaderboard, grouped based on the degree the forecasts are connected with the investment decisions.}
\end{figure}

\section*{Appendix C: Technical details of the risk model}

In this appendix, we discuss some technical details associated with the estimation of our investment risk model. Before discussing estimation of the key building blocks of the model (i.e., volatility forecasts) that are used in portfolio optimization when creating pseudo-true portfolios for comparison with M6 participant portfolios, however, we first briefly discuss the historical data set used for this purpose. Our data set includes daily prices for the cross section of M6 assets, collected from from Yahoo Finance. More specifically, we analyze 10 calendar years of daily data, ending on the start date of the competition, namely February 2, 2022. The data comprise Open, Close, High and Low prices, as well as Adjusted Prices, accounting for splits/corporate actions, dividends etc. When building our dataset, we attempted to select a broad cross section of underlying asset classes for our ETF universe, and unfortunately this meant that we are unable to obtain a balanced panel of data; several ETF securities were launched subsequent to our data start date. Additionally there are several days with missing data, due to market holidays and other reasons. We \emph{forward} filled such missing data with the last available set of prices when data were missing, effectively assigning a return of zero to market holidays, for example. 

\subsubsection*{Modeling asset level volatility}

To model asset level volatility we use Heterogeneous Exponential Realized Volatility (HEXP) models, as described in \cite{Bollerslev2018-ee}. We select these models for several reasons. First, the approach taken by the authors is to fit a `global' model for the entire panel of assets, which is inherently parsimonious and robust, as it uses the same set of parameters for each underlying asset (see above discussion). Such global models reflect a Bayesian philosophy, where the (scaled) time series of each asset is regarded as exchangeable, resulting in significant reductions in estimation error. Empirical evidence summarized in \cite{Bollerslev2018-ee} indicates that these models are appropriate for the wide range of underling asset classes, including those examined here. Second, the model can be easily fitted using OLS, making the approach accessible to competitors without sophisticated econometric knowledge and software. Third, the model offers a simple and intuitive route to account for volatility spillovers and common dynamics in volatility across asset / asset classes. Because the model uses smooth, exponentially weighted averages with several look back periods, it avoids potential variance estimation issues to which its closely related predecessor (i.e., the heterogeneous autoregression (HAR) model) is prone\footnote{For further discussion of the HAR model, see \cite{Diep1}, and the references cited therein. In addition, note that the HEXP model also has some features that draw on the MIDAS model discussed in \cite{Ghysels1}.}. Of further note is that the HEXP model described in \cite{Bollerslev2018-ee} uses a database of intra-day realized volatility estimates from a cross section of futures contracts. We do not assume that competition participants have access to such data, and instead use daily volatility estimates based on our intra-day high/low/open/close data. Evidence presented in \cite{MOLNAR201220} suggests that such estimates are reasonably interchangeable with those obtained from intra-day data.

Recall, that competitors are asked to produce portfolios which maximize the Sharpe Ratio over the 20 trading days following each submission date. We therefore build our risk model to make direct single step variance estimates with a 20 day horizon. Separately, we fit a model with a 1 day horizon as a filter prior to constructing our covariance matrix estimates. The HAR structure of our model is readily adaptable to both requirements.

To fit the variance model we begin by calculating historic variances in three ways, adopting the methodology described in \cite{YangDennis2000DVEB}. We call this the YZ estimator in the sequel. Namely, for each asset define:
\begin{align*}
\{O_t,C_t,H_t,L_t\} &= \{Open, Close, High, Low\} \text{ Prices, day t.}\\
o_t &= log(O_t)-log(C_{t-1}) & u_t &= log(H_t)-log(O_t)\\
d_t &= log(L_t)-log(O_t) & c_t &= log(C_t)-np.log(O_t)\\
V_P & = \frac{1}{T} \sum_{t=1}^T \frac{1}{4 log 2} (u_t-d_t)^2 & V_{RS} & = \frac{1}{T} \sum_{t=1}^T [u_t(u_t-c_t) + d_t(d_t-c_t)]
\end{align*}
Here, $V_P$ and $V_{RS}$ are \cite{ParkinsonMichael1980TEVM} and \cite{RogersL.C.G.1991EVFH} variance estimators. When a variance estimate is required for a given day (these values are used as a basis for the exponentially weighted moving average values which are the regressors in our model, and as dependent variables in our 1 day horizon model described below) we use the Garman-Klass estimator:
\begin{equation}\nonumber
V_{GK} = V'_O - 0.383V'_C + 1.364V_P + 0.019V_{RS},
\end{equation}
with $V'_O = o_t^2$, $V'_C = c_t^2$, and with $V_P$ and $V_{RS}$ defined as above, setting $T=1$. When a variance estimate is required for a 20 day period (these form the dependent variables in our models) we use the YZ estimator which accounts for drift in the asset price over the relevant time period. The model also requires a long-run volatility estimate for each asset. For this we utilize the YZ estimator based on an expanding window of data commencing with the first recorded price for each asset. The YZ estimator is defined as\footnote{Refer to \cite{YangDennis2000DVEB} for derivation of $V$.}:
\begin{equation}\nonumber
    V = V_O + kV_C + (1-k)V_{RS}
\end{equation}
with,
\begin{align*}
    V_O &=  \frac{1}{T-1} \sum_{t=1}^T (o_t-\bar{o})^2 & V_C &=  \frac{1}{T-1} \sum_{t=1}^T (c_t-\bar{c})^2 &
    \bar{o} & = (1/T) \sum_{t=1}^T o_t & \bar{c} & = (1/T) \sum_{t=1}^T c_t& k = \frac{.34}{1.34 + \frac{(T+1)}{(T-1)}}
\end{align*}
At each estimation date, we follow \cite{Bollerslev2018-ee}, and form 4 (centered) exponentially weighted averages of daily variance, called $ExpRV^K_T$, with centers of mass (CoM) at 1, 5, 25, and 125 days. Namely, we define:
\begin{align*}
    ExpRV^K_T &= \sum_{t=1}^T w_t RV_t, & w_t &= \frac{e ^ {- \lambda t}}{\sum e^{-\lambda t}}, & \lambda &= CoM [1,5,25,125],
\end{align*}
with $\lambda = log (1 + 1 / CoM)$. We also calculate a universe `average' volatility factor for our entire investment universe (using assets available at each historic point in time) and re-scale this back to the native scale of each asset as described in \cite{Bollerslev2018-ee}. For our $h$ step ahead forecast ($h=[1,20]$) we then stack all assets and all historic $h$ day time windows in to one vector, and run a single LS regression on the entire data set. We estimate the model as at each competition entry point. The LS model is:
\begin{align*}
    (\hat{V}_{t:t+h} - RV^{LR}_t) &= \kappa_1 (ExpRV^1_t - RV^{LR}_t) + \kappa_2 (ExpRV^5_t - RV^{LR}_t) \\ &+ \kappa_3 (ExpRV^{25}_t - RV^{LR}_t) + \kappa_4 (ExpRV^{125}_t - RV^{LR}_t) + \kappa_5 (ExpGlRV^5_t- RV^{LR}_t) + e_t.
\end{align*}

\subsubsection*{Modeling asset level volatility covariance}

There are two key components to our estimates of the covariance structure of our asset volatility. First, we assume that M6 asset returns can be described using a factor model. To do this we define factors as linear combinations of standardized returns on M6 assets. Using returns standardized using univariate volatility models is common practice in financial risk management (see \cite{Engle2002-if}). For example, \cite{AlexanderC.2002PCMf} use such returns as the basis for principal component analysis, which is used to estimate factors. We adopt a similar approach, except that we define our factors based on prior knowledge, rather than identifying them from the data. For further details, refer to \cite{Connor2019-kf}, who describe several possible approaches to building factor models. These authors note that for `asset allocation' type analysis, regression models with pre-specified factors tend to be used in practice. Building a bottom up fundamental model for a portfolio of equities would indeed require significant effort, and we assume that the data collection, estimation and testing of such a model was unrealistic for M6 participants. We therefore adopt a purely regression based approach to this problem, solely based on the use of historical returns.

More specifically, we define our factors in three hierarchical layers. The layers are fitted sequentially on residuals from the previous layer(s). This enables the use of a very simple identity prior or shrinkage target for our dynamic regression at each stage. For the top level of the hierarchy we use a single factor which we call the `M6 Market factor' (M6M), as an equally weighted combination of all 50 M6 ETF assets. In some ways this is analogous to a market portfolio for the competition, although obviously it is not market capitalization weighted or appropriate for use outside the context of M6. To construct a market capitalization weighted average we would have needed to either choose some subset of the assets, or obtain incremental data, which we chose not to do. We also decided to exclude individual equity securities, as equities already have significant underlying representation in our ETF securities.

The next layer of the factor model hierarchy is designed to capture the differential dynamics of various asset classes and style factors, which are known from the literature to have significant effects on asset pricing. We again define these using returns on various linear combinations of ETF assets form the M6 universe.

Finally, our level 3 factors are defined to be congruent with the equity market industry ETFs which form a part of the M6 Universe. At this stage, we again adopt a simple identity prior for asset loadings (i.e. we do not attempt to use prior information regarding individual equities to identify them with particular industries, although might be useful). The full set of factors are defined as in Table C1, with precise weightings available on request from the authors.

\begin{table}[h]
\begin{center}
\begin{tabular}{ c c c }
 Level & Factor & Definition \\
 \hline
 1 & M6M & M6 Market Factor \\  
 \hline
 2 & USE & US Equity factor\\
 2 & EUE & European equity factor\\
 2 & AE & Asian equity factor\\
 2 & TERM & Yield curve slope factor\\
 2 & CREDIT & Corporate debt - Govt. debt\\
 2 & MOM & Momentum factor\\
 2 & VAL & Value factor\\
 2 & SIZ & Size factor\\
 \hline
 3 & TECH & Technology factor\\
 3 & FIN & Financial factor\\
 3 & HEALTH & Healthcare factor\\
 3 & ENERGY & Energy factor\\
 3 & DISC & Consumer discretionary factor\\
 3 & IND & Industrial factor\\
 3 & COMMS & Communications factor\\
 3 & UTIL & Utilities factor\\
 3 & STAPLES & Consumer staples factor\\
 3 & MATERIAL & Materials factor\\
\end{tabular}
\end{center}
    \caption*{Table C1: Factor Definitions for M6 Risk Model Factors}
    \label{tab:factor_defs}
\end{table}

The second step of our approach is to estimate time varying loadings for each asset in these pre-specified factors. We do so by using a parsimonious variant of the BEKK model described above, with an additional shrinkage component, along the lines suggested in \cite{HAFNER20123533}. 

More specifically, to estimate asset covariance we proceed as follows. At each submission date we take the in-sample 1 day horizon fitted values from our univariate variance model for each asset. We then standardise the observed asset return for that day using the standard deviations corresponding to these variance estimates.
We then use these daily standardized returns to build the hierarchical factor model, using the BEKK approach to estimating the rolling covariance between the M6M factor and each individual asset \footnote{ We note that the M6 assets are traded both in London and in New York, and thus there is the potential for imperfect time synchronicity to affect our results.}. As the constant term in the BEKK model, we use the long run whole period covariance between each asset and M6M (estimated via an OLS regression of the standardized asset returns on the factor plus a constant). We specify our BEKK model using two \emph{scalar} parameters, namely $\gamma$, the weight on the time $t-1$ error term, and $\omega$, the weight on the long run covariance estimate. The weight on the time $t-1$ covariance estimate is defined as $(1-\omega-\gamma)$. The BEKK model used for this and the following layers of the factor model can be written as:
\begin{equation}
    \boldsymbol{\Sigma_t} = \omega \boldsymbol{\Sigma_0} + \gamma \mathbf{e}_{t-1} \mathbf{e}_{t-1}' + (1 - \omega - \gamma) \boldsymbol{\Sigma_{t-1}},
\end{equation}
where $\Sigma_t$ denotes our time varying covariance matrix and $\Sigma_0$ is our constant long run covariance matrix, the $e_{t-1}$ are previous period errors, and all other terms in the above expression are constants. We then collect the residuals from this exercise, and re-standardize them to have mean zero and standard deviation of 1.  We again use a BEKK model (with the same $\omega$ and $\delta$) to estimate time varying covariance between all the assets and each factor, with shrinkage towards the identity matrix.

The third step of our approach is to again collect the residuals from the new BEKK model, re-standardize and run the BEKK model once more with the `level 3' industry factors as regressors. 

Finally, to calculate the asset specific risks, we utilize yet another BEKK model, again using the same $\omega$ and $\gamma$ values, with the long run specific variance calculated from the entire sample period. Also, the factor covariance matrix is modeled using a similar structure, again with an identity matrix as the long run shrinkage target and the same BEKK parameters as previously\footnote{At this step, one can introduce a separate set of parameters for the factor covariance matrix, which might be expected to be more stable than the factor/asset loading process. This is left to future research, however.}. To choose $\omega$ (the weight on the previous day's cross error cross product) and $\gamma$ (the weight on the long run shrinkage target) we run a grid search, with a test data set comprising the 36 20 day trading periods ending on November 30, 2021. For each date in the period, we estimate our model and compute the log-likelihood of the entire set of M6 asset return observations for following 20 day out of sample period. Prior experience with similar models leads us to test values of $\omega$ from the set $[.030, .020, .010, .005]$ and $\gamma$ from the set $[.0025, .0050, .0075]$. The values which maximize the log-likelihood for the test data are $\omega = .01$ and $\gamma = .005$. As expected, these values place substantial weight on the shrinkage target. We use these values to estimate the covariance matrix for the rest of the competition.

\end{document}